\documentclass{article}
\usepackage[a4paper,
            bindingoffset=0.2in,
            left=0.6in,
            right=0.6in,
            top=1in,
            bottom=1in,
            footskip=.25in]{geometry}
\usepackage[utf8]{inputenc}
\usepackage{jcappub}
\usepackage{natbib}
\usepackage{fancyhdr}
\setcitestyle{square,numbers}
\usepackage{graphicx}
\usepackage{xcolor}
\usepackage{float}
\usepackage{multirow,array,longtable}
\usepackage{booktabs}
\usepackage{arydshln} % Dashed lines in array. siempre despues de 'array'
\usepackage[makeroom]{cancel}
\usepackage{amsfonts}%,mathalfa,amssymb,amsmath}
\usepackage{slashed}

\usepackage{tikz-feynman}
\usepackage{subcaption}
\usepackage{tikz}
\usepackage{amsmath}
\tikzset{
    ->-/.style={decoration={
  markings,
  mark=at position .5 with {\arrow{>}}},postaction={decorate}},
    -<-/.style={decoration={
  markings,
  mark=at position .5 with {\arrow{<}}},postaction={decorate}},
    ->/.style={decoration={
  markings,
  mark=at position .4 with {\arrow{>}}},postaction={decorate}},
}
\usepackage{amssymb}
\usepackage{longtable}
 \usepackage{lscape}

\renewcommand\eqref[1]{Eq.~(\ref{#1})}

\newcommand\figref[1]{Fig.~\ref{#1}}

\newcommand\tabref[1]{Table~\ref{#1}}

\newcommand\secref[1]{Section~\ref{#1}}
\newcommand\appref[1]{Appendix~\ref{#1}}
\newcommand{\be}{\begin{equation}}
\newcommand{\ee}{\end{equation}}
\newcommand{\bear}{\begin{eqnarray}}
\newcommand{\eear}{\end{eqnarray}}
\newcommand{\nn}{\nonumber}

\def\1loop{one-loop}

\newcommand{\mL}{\mathcal{L}}

\def\mw{m_{W}}
\def\mz{m_{Z}}

\title{Unitarity limits on HH production from VBF mediated channels}
\author[a]{A. Dobado,}
\author[b]{D.  Domenech,}
\author[b]{M.  Herrero,}
\author[c]{R. A. Morales}
\affiliation[a]{Departamento  de F\'{\i}sica Te\'orica, Instituto IPARCOS\\
Universidad Complutense de Madrid,  Plaza de las Ciencias 3, 28040 Madrid, Spain}
\affiliation[b]{Departamento de F\'{\i}sica Te\'orica and Instituto de F\'{\i}sica Te\'orica, IFT-UAM/CSIC,\\
Universidad Aut\'onoma de Madrid, Cantoblanco, 28049 Madrid, Spain}
\affiliation[c]{IFLP, CONICET - Dpto. de F\'{\i}sica, Universidad Nacional de La Plata, \\ 
C.C. 67, 1900 La Plata, Argentina}

\emailAdd{dobado@fis.ucm.es}
\emailAdd{daniel.domenech@uam.es}
\emailAdd{maria.herrero@uam.es}
\emailAdd{roberto.morales@fisica.unlp.edu.ar}

\abstract{
In this work we study the unitarity constraints on double Higgs production from Vector Boson Fusion mediated channels,  specifically  in  $f_1 f_2 \to f_3 f_4 H_5 H_6$  scattering,  where $f$ refers generically to either fermion or anti-fermion,  and both cases with quarks and leptons will be considered.  We focus on the beyond Standard Model contributions from anomalous quartic couplings of two electroweak gauge  bosons and two Higgs bosons ($VVHH$ with $VV=WW, ZZ$) from effective operators containing four derivatives.  These four derivative operators appear in the two most popular effective field theories,  SMEFT (at canonical dimension 8) and HEFT (at chiral dimension 4),  and their corresponding $VVHH$ effective vertices can be easily related for the considered processes here.  The aim of this work is to derive jointly the unitarity constraints on these related effective vertices using the peculiarities of the VBF mediated channels with four particles in the final state while keeping the propagation of the internal $VV$ pairs being off-shell as it must be for these channels.   We also discuss the relevance of the obtained  constraints for $f_1 f_2 \to f_3 f_4 H_5 H_6$  at  high energy colliders. }

\allowdisplaybreaks
\begin{document}
\begin{flushright}
%   \today\\
	IFT-UAM/CSIC-26-104
%	IPARCOS-UCM-26-YY 
\arxivnumber{2610.07194} % if you have one
\end{flushright}
\maketitle
%%%%%%%%%%%%%%%%%%%%%%%%%%%%%%%
\section{Introduction}
\label{intro}
The study of double Higgs boson production at colliders via Vector Boson Fusion (VBF)  provides one of the most efficient windows to search for Beyond Standard Model (BSM) Higgs Physics.  In particular,  the VBF-mediated channels of the type $f_1 f_2 \to f_3 f_4 H_5H_6$,  with initial fermions $f_{1,2}$  being either quarks or leptons,  are very efficient  in looking for BSM  $VVHH$ interactions of  the two internal propagating electroweak (EW) gauge bosons $VV$ (with $V$ being either $W$ or $Z$)  and the two final  Higgs bosons $HH$.  For the LHC,  at present,  the search of BSM Higgs Physics at these VBF mediated channels with initial quarks is indeed very active.  During LHC Run 2, an extensive programme was developed by ATLAS and CMS to search for Higgs-pair production via VBF channels  in several final states~\cite{CMS:2020tkr,ATLAS:2025hhd,CMS:2022cpr,ATLAS:2024lsk,CMS:2022hgz,ATLAS:2024pov},  where  they provided  important constraints on the BSM effective $VVHH$ interactions together with other effective interactions like the quartic gauge $VVVV$  and others \cite{CMS:2020gfh,  ATLAS:2024nab,  ATLAS:2025yxf, ATLAS:2026wew, CMS:2026mbq}.     

The $VVHH$ effective  interactions are the ones of our interest here.  These and other effective interactions  are treated nowadays within the framework of the two most popular Effective Field Theories (EFTs),   the Standard Model Effective Field Theory (SMEFT) and the Higgs Effective Field Theory (HEFT) (for a review see,  for instance,  \cite{Brivio:2017vri} and  \cite{Dobado:2019fxe}).   Our present study here focus on the particular bosonic effective operators in these two EFTs that contain four derivatives of scalar fields.  Concretely,  those with 4-derivatives of the doublet $\Phi$ field in the SMEFT,  generically  denoted by $(D\Phi)^4$,   and those with 4-derivatives in the HEFT,  two  acting on the singlet $H$ field and two on the non-linear $U$ field,  generically denoted by $(D H)^2 (DU)^2$.   The Wilson coefficients for the 4-derivative SMEFT effective operators have been explored by ATLAS and CMS  experiments together with other dimension 8 SMEFT effective operators,  particularly,  with those involving  the quartic gauge boson couplings  \cite{CMS:2020gfh,  ATLAS:2024nab,  ATLAS:2025yxf, ATLAS:2026wew, CMS:2026mbq}.  Usually,  they set constraints on all these SMEFT effective interactions by means of global fits.  In particular,  the quartic gauge couplings $VVVV$ are mainly constrained from double gauge boson  production via Vector Boson Scattering (VBS) (with subprocess $VV \to VV$),   and from triple gauge boson production (with subprocess $q\bar{q} \to VVV$),    whereas the $VVHH$ couplings are usually constrained from double Higgs production via VBF (with subprocess $VV \to HH$).     

The  4-derivative effective operators commented above lead to $VVHH$ effective interactions with new Lorentz structures respect to the Standard Model (SM) that grow with the external  momenta as $O(p^2)$.  The growing behaviour with energy of these effective interactions,  in turn,  lead  to perturbative unitarity violation in the EFTs predictions of scattering amplitudes for double Higgs production via VBF-mediated channels.  Therefore, for a realistic study  of the sensitivity to these new effective interactions at colliders,  the exploration must be restricted to a reduced parameter space for the EFTs coefficients where the perturbative unitarity condition is preserved.  These unitarity limits refer in general to the joint relevant variables involved in the scattering amplitudes,  which we refer in short as (energy, coefficient),  meaning the relevant energy of the studied process and the studied coefficient of the EFT.  Therefore, in practice,  the unitarity constraints for a specific coefficient really apply to the regions in the two dimensional parameter space (energy,  coefficient),  rather than to just the values of the coefficient itself.  Thus,  to derive useful limits for colliders the proper analysis must be referred to the specific process,  the specific energy and the specific coefficient  one is interested in,  rather than to study the unitarity problem in the full EFT.  

For the present study of  the $HH$ production via VBF-mediated channels,  $f_1 f_2 \to f_3 f_4 H_5H_6$,   we focus on the two coefficients involved in the  $VVHH$ effective interactions,  generically  called here $\delta$ and $\eta$.  These two  coefficients can be related  to the coefficients of the
4-derivative operators commented above by matching the corresponding EFT predictions for the scattering amplitudes in double Higgs production,   in particular those of  $VV \to HH$ (see, for instance,  \cite{Domenech:2022uud}).   Specifically,  the 4-derivative coefficients that are referred here are those of the dimension-8 SMEFT operators,  named usually $f_{S,0}$,  $f_{S,1}$ and $f_{S,2}$ (using the same notation as in \cite{Almeida:2020ylr}),  or $c_{\Phi^4 D^4}^{(1)}$,   $c_{\Phi^4 D^4}^{(2)}$ and  $c_{\Phi^4 D^4}^{(3)}$ (as in \cite{Dedes:2025oda}).  The 4-derivative coefficients within the HEFT are the two coefficients of the (chiral)dimension-4 HEFT operators,  named  $a_{ddVV1}$ and $a_{ddVV2}$ (here we use  the same notation as in  \cite{Herrero:2022krh}).  These are related with the coefficients $c_8$ and $c_{20}$ of the full list of HEFT operators in \cite{Brivio:2013pma} (see,  \cite{Herrero:2022krh} for details).     

Previous studies of the unitarity limits of these two effective parameters $\eta$ and $\delta$ were focused on the case of VBS subprocesses,  where the initial on-shell particles are the two EW vector bosons $VV$.  Some preliminary unitarity studies of the  4-derivative operators within the HEFT \cite{Delgado:2013hxa} were focused in particular on $WW \to HH$ and were performed using the Equivalence Theorem (ET),  i.e. selecting only the longitudinal vector boson polarization case where the scattering amplitudes for $W_L W_L \to HH$ were derived  from the corresponding Goldstone Boson amplitudes,   $w w \to HH$.  Indeed,  the notation used here
of $\delta$ and $\eta$ for the coefficients of the four-derivative effective operators is taken from that reference.  Other preliminary studies of unitarity for $\delta$ and $\eta$ within the HEFT in the  $WW \to HH$ channel,   this time with no use of the ET,   were done in \cite{Asiain:2021lch},   \cite{Herrero:2022krh} and
\cite{Domenech:2022uud}.  The most complete studies of unitarity limits for the VBS subprocesses,  including all channels with initial on-shell $V_1V_2$ and either final $V_3V_4$ or  $H_3H_4$,  in both HEFT and SMEFT,  were presented in \cite{Almeida:2020ylr}.  The impact of requiring unitarity preserving EFT predictions at LHC in the VBS subprocess $WZ  \to WZ$,   by using  different unitarity methods was analyzed in \cite{paperUnitarity}.   Additional  unitarity limits for the triple gauge boson production case,  $f \bar f \to V_1V_2V_3$,   were studied recently in \cite{Eboli:2026ccq}. 

The study of unitarity by means of partial wave amplitudes for the case of  multiple particles in the final state requires special treatment,  different than the usual partial wave analysis of  $2\to 2$ scattering.  Most of these studies focus on generalizations of the well-known optical-theorem for quasi-elastic relativistic scattering processes to multi-particle final states.   Previous phenomenological analysis treating the unitarity issue for multiple-Higgs production at colliders  can be found in~\cite{Kilian:2018bhs,Kilian:2021whd} and references therein.   At present,   most of the unitarity studies  for multi-particle final states use the helicity-spinor formalism to project the helicity amplitudes in terms of   partial wave amplitudes with  definite angular momentum.   In particular,  in \cite{Eboli:2026ccq} they followed the method developed in Ref.~\cite{Bresciani:2025toe} for 2$\to$3 scattering processes (see also~\cite{Degrande:2025uil}).  For our present study of unitarity with four particles in the final state we will also follow a similar procedure by using  helicity-spinor techniques to project the helicity amplitudes in terms of partial wave amplitudes with definite angular momentum. 

The main objective of this work is to perform a detailed and precise  analysis of unitarity limits for the four-derivative operators,  being  devoted for the first time specifically to the VBF-mediated processes  $f_1 f_2 \to f_3 f_4 H_5H_6$.  The scattering amplitudes of these processes with all external particles being on-shell are the ones  that are tested at colliders.  Therefore,  they are the proper ones to study the unitarity limits in the above mentioned relevant  parameter space (energy, coefficient).  Here,  energy refers to the total energy of the initial $f_1 f_2$ fermion pair,  and coefficient means $\delta$ and/or  $\eta$.  
The previous studies of unitarity refer instead to the VBS subprocesses,  with $VV$ on-shell in the initial state,  which is not really happening in the colliders.  These $VV$ pairs are intermediate states which propagate  off-shell,  i.e.  very far from the on-shell condition that  is assumed in the previous studies of unitarity in VBS subprocesses.   Our study of unitarity limits,  being focused  for the first time to our knowledge,  on the on-shell processes  $f_1 f_2 \to f_3 f_4 H_5H_6$  is,  in our opinion,  the proper one to test the sensitivity to these effective coefficients $\delta$ and $\eta$ at high energy colliders, both hadron-hadron and lepton-lepton colliders.  

One of the most appealing features of our study of unitarity at $f_1 f_2 \to f_3 f_4 H_5H_6$ scattering,  is that it provides a different avenue to test safely the coefficients $\delta$ and $\eta$ at very high energy colliders.  At present,  for instance at LHC,  to test the EFT predictions in a safe way,  i.e.  preserving unitarity,  the so-called clipping method (see, for instance \cite{Szleper:2020kcv})  is usually applied.  This method has been applied to double EW gauge boson production via VBS channels and consists in restricting the EFT prediction to phase space regions where the invariant mass of the final gauge bosons pairs is smaller than a cutoff value $m_{VV}^{\rm clip}$.  The value of $m_{VV}^{\rm clip}$ is usually derived  from the unitarity crossing value of the subprocess $V_1V_2 \to V_3V_4$,   therefore assuming this is occurring on-shell.  Next,  the constraints on the Wilson coefficients are derived from data as a function of $m_{VV}^{\rm clip}$.   In general,  after applying this clipping method to preserve unitarity,  the experimental constraints on the effective coefficients get weaker than without the unitarity cut-off (see,  for instance,
\cite{Covarelli:2024uar}).  
 In the final part of this work  we will propose to use a different clipping method  to analyze in a unitarity-safe way the sensitivity to the 4-derivative effective parameters, $\delta$ and $\eta$ by using the true on-shell scattering process 
 $f_1 f_2 \to f_3 f_4 H_5H_6$,  where the relevant invariant mass variable for the final state is $m_{f_3f_4H_5H_6} = \sqrt{s_{12}}$ rather than $m_{H_5H_6} = \sqrt{s_{56}}$.  Then,  we will finally explore how the sensitivity to the effective coefficients $\delta$ ad $\eta$  change notably by using  either clipping with $m_{f_3f_4H_5H_6}$ or with 
$m_{H_5H_6}$. 

%% poner aqui la cita al clipping method in EFT predictions.  Sobre todo, 
%%% Szleper:2020kcv y   Covarelli:2024uar%%%%

%Indeed, we will show here that the usual alternative clipping\textcolor{red}{REFr: technique (see for %instance,  the experimental analysis for diboson~\cite{CMS:2020gfh,Szleper:2020kcv} and triboson %searches~\cite{ATLAS:2024nab,ATLAS:2025yxf,CMS:2026mbq})} to provide unitarity preserving %predictions at colliders by using the VBS subprocess $VV \to HH$,  with the relevant energy being $%\sqrt{s_{VV}}=m_{HH}$ is too much restrictive,  leading to lower sensitivity to the effective coefficients.    

The paper is organized as follows.  We first review the relations of $\delta$ and $\eta$ coefficients with the corresponding 4-derivative coefficients  in the SMEFT and the HEFT in \secref{Theoretical Framework}.  Then we report  in \secref{partial-waves}  on the method used here, with spinor helicity techniques,  to set the unitarity preserving conditions for the helicity amplitudes involved with definite angular momentum. 
 The numerical results for these partial wave amplitudes for the most interesting channels will be presented in \secref{unitarity-limits}. The  two dimensional unitarity limits on the (energy,  coefficient) plane will also be presented  on \secref{unitarity-limits}.   
Finally,   the most important implications of our unitarity limits for the future high energy colliders will be reported in \secref{colliders}.  The summary of our conclusions is presented in the last section.

%%%%%%%%%%%%%%%%%%%%%%%%%%%%%%%
\section{Theoretical Framework}
\label{Theoretical Framework}
For the present  study of the anomalous couplings  $VVHH$ we work within the framework of Effective Field Theories (EFTs).  Specifically,  we choose Higgs Effective Field Theory (HEFT) and Standard Model Effective Field Theory (SMEFT) which are well known gauge invariant EFTs.  In particular,  we focus on the bosonic electroweak sector of these two theories and select the operators of our interest from their respective  $SU(2)_L \times U(1)_Y$  gauge invariant Lagrangian. 

We summarize first the HEFT Lagrangian that is relevant for this work.  This is usually organized by the chiral dimension counting,  starting in chiral dimension 2,  then chiral dimension 4, etc. 
  \begin{equation}
   \label{eqn:NLO-HEFT}
  {\mL}_{\rm HEFT}  \, = \, {\mL}_2 +{\mL}_4+...
    \end{equation}
The leading order Lagrangian  is given by:
     \begin{eqnarray}
      {\mL}_2   &=& \frac{v^2}{4}  \left(1 + 2  a \frac{H}{v} +b \frac{H^2}{v^2}+c  \frac{H^3}{v^3}+... \right)
        \text{Tr}[(D_{\mu} U)^{\dagger} D^{\mu} U] + \frac12 \partial_{\mu} H \partial^{\mu} H - V(H) \nn \\
       && - \frac{1}{2 g^2} \text{Tr}[\hat{W}_{\mu \nu} \hat{W}^{\mu \nu}] - \frac{1}{2 g'^2} \text{Tr}[\hat{B}_{\mu \nu} \hat{B}^{\mu \nu}] + {\mL}_{GF} + {\mL}_{FP},
        \label{eqn: L2}
    \end{eqnarray}
The relevant fields and quantities appearing in this Lagrangian are:  
$H$ is the Higgs field which is introduced in the HEFT as a singlet field, in contrast to the SM or the SMEFT where it is introduced as a component of the usual doublet 
$\Phi$.  Due to this singlet Nature of $H$ within the HEFT,  the first term in ${\mL}_2$ contains a polynomial function in powers of $(H/v)$. 

The $U$ field is a $2\times2$ matrix
\begin{equation}
        U \, = \, \exp \left( i \frac{{\omega_i} {\tau_i}}{v} \right),
    \end{equation}  
   that contains the three GB fields ${\omega}_i$ ($i=1,2,3$) in a non-linear representation of the $SU(2)$ symmetry group and 
   %$t_i (i=1,2,3)= \tau_i/2$ are the three generators of $SU(2)$,  
  % with 
   $\tau_i$ ($i=1,2,3$) are the three Pauli matrices.

   The EW covariant derivative of this $U$ field 
   is defined as:
   \begin{equation}
        D_{\mu} U \, = \, \partial_{\mu} U - i \hat{W}_{\mu} U + i U \hat{B}_{\mu},
    \end{equation} 
    that contains the EW gauge fields,  $\hat{W}_{\mu} = \frac{g}{2} W^i_{\mu} \tau^i$ and $\hat{B}_{\mu} = \frac{g'}{2} B_{\mu} \tau^3$ and the EW gauge couplings $g$ and $g'$.  The corresponding EW field strength tensors are given by:
       \begin{equation}
        \hat{W}_{\mu \nu} \, = \, \partial_{\mu} \hat{W}_{\nu} - \partial_{\nu} \hat{W}_{\mu} - i [\hat{W}_{\mu}, \hat{W}_{\nu}], \hspace{8mm} \hat{B}_{\mu \nu} \, = \, \partial_{\mu} \hat{B}_{\nu} - \partial_{\nu} \hat{B}_{\mu}.
    \end{equation}
    The physical gauge fields are then given,  as usual,  by:
\be
W_{\mu}^\pm = \frac{1}{\sqrt{2}}(W_{\mu}^1 \mp i W_{\mu}^2) \,,\quad
Z_{\mu} = c_W W_{\mu}^3 - s_W B_{\mu} \,,\quad
A_{\mu} = s_W W_{\mu}^3 + c_W B_{\mu} \,,
\label{eq-gaugetophys}
\ee
where we use the short notation $s_W=\sin \theta_W$ and $c_W=\cos \theta_W$,  with $\theta_W$ the weak angle.
$V(H)$ is the Higgs potential within the HEFT, which includes the triple and quartic  Higgs self-interactions:
    \begin{equation}
        V(H) \, = \, \frac12 m_H^2 H^2 +\kappa_3 \lambda v H^3 + \kappa_4 \frac{\lambda}{4} H^4.
    \label{VHEFT}    
    \end{equation}
The relations among couplings  and masses within the LO-HEFT are:   $m_H^2  = 2 \lambda v^2$,  $\mw=g v /2$ and $\mz=\mw/c_W$.   The vacuum expectation value  $v$ can be written  in terms of  the Fermi coupling constant $G_F$ by $v=(\frac{1}{\sqrt{2} G_F})^{1/2} =246$ GeV. 

Finally,  ${\mL}_{GF}$ and ${\mL}_{FP}$ are the gauge fixing  and Faddeev-Popov ${\mL}_{FP}$ terms,  respectively, whose explicit expressions in the general $R_\xi$ covariant gauges can be found in 
\cite{Herrero:2020dtv, Herrero:2021iqt,Herrero:2022krh} together with more definitions and specifications within the HEFT (for a more complete list of effective operators within the HEFT see also \cite{Brivio:2013pma}).

The  effective operators in the bosonic sector with four derivatives,  which are the ones of our interest here,  are contained in ${\mL}_{4}$ and are the following \cite{Dobado:2019fxe,Herrero:2022krh}:
 \begin{eqnarray}
 \label{eq-L4}
%\bear
{\mL}_{4} \, &=& a_{ddVV1} \frac{\partial^\mu H\,\partial^\nu H}{v^2} 
{\rm Tr}\Big[ (D_\mu  U)^{\dagger} D_\nu U \Big] 
+a_{ddVV2} \frac{\partial^\mu H\,\partial_\mu H}{v^2} {\rm Tr}\Big[ (D^\nu U)^{\dagger} D_\nu  U \Big] +... \nn
\end{eqnarray}  

The SM can be reached from the HEFT by setting the HEFT parameters to $a=b=\kappa_3=\kappa_4=1$ and $c=0$ in ${\mL}_2$ and the rest of parameters set to zero,  i.e.  $a_i's=0$ in 
${\mL}_4$.

Regarding the SMEFT,  the set of  EW  bosonic effective operators  with four derivatives that are relevant for the present work are contained in the dimension 8 set of operators defining the SMEFT Lagrangian:
\begin{eqnarray}
  \label{eqn:NLO-SMEFT}
  {\mL}_{\rm SMEFT}  \, &=& \, {\mL}_{\rm SM} + {\mL}_6  +{\mL}_8+...
 \end{eqnarray}
    where the leading order SMEFT Lagrangian,  with canonical dimension 4,  is the SM Lagrangian,   ${\mL}_{\rm SM}$.   For the next to leading order contributions,  
${\mL}_6$ contains the SMEFT operators with canonical dimension 6,    ${\mL}_8$  those of canonical dimension 8,  etc. .   We use here the basis of operators and notation as defined in \cite{Grzadkowski_2010}. 
 The notation for the SM Lagrangian is set by:   
\bear 
\label{SMlag4}
 {\mL}_{\rm SM}   &=&
                   -\frac{1}{4} W_{\mu\nu}^I W^{I\mu\nu}
                   -\frac{1}{4} B_{\mu\nu}   B^{\mu\nu} +  {\mL}_{GF}^{\rm SM} + {\mL}_{FP}^{\rm SM} \nn \\
&&   + \left( D_\mu \Phi \right)^\dagger \left( D^\mu \Phi \right) 
   + \mu^2 \Phi^\dagger \Phi - \lambda \left( \Phi^\dagger \Phi \right)^2  ,
\eear
where $\Phi$ is the SM complex scalar doublet,  
\bear 
\Phi & =& \left( \begin{array}{c} \phi^+ \\ \frac{1}{\sqrt{2}} (v +
    H+ i \phi^0) \end{array} \right) \;, \quad
\label{doubletdef}
\eear 
the normalization chosen  here is as in the HEFT,  $m_H^2= 2 \lambda v^2$,  and the conventions for the covariant derivative and SM electroweak field strengths are set here as in  \cite{Grzadkowski_2010}:
\bear
  D_\mu &=&  \partial_\mu+i g' B_\mu Y + i g W^{I}_\mu T^{I} \,\,, \, \, I= 1,2,3   \nn\\
  W_{\mu\nu}^I &=& \partial_\mu W^{I}_\nu-\partial_\nu W^I_\mu -g \epsilon^{IJK} W^J_\mu W^K_\nu \nn\\
   B_{\mu\nu} &=& \partial _\mu B_\nu -\partial_\nu B_\mu
\label{SMdefs}   
\eear

\begin{figure}[!t]
    \centering
        \includegraphics[scale=0.25]{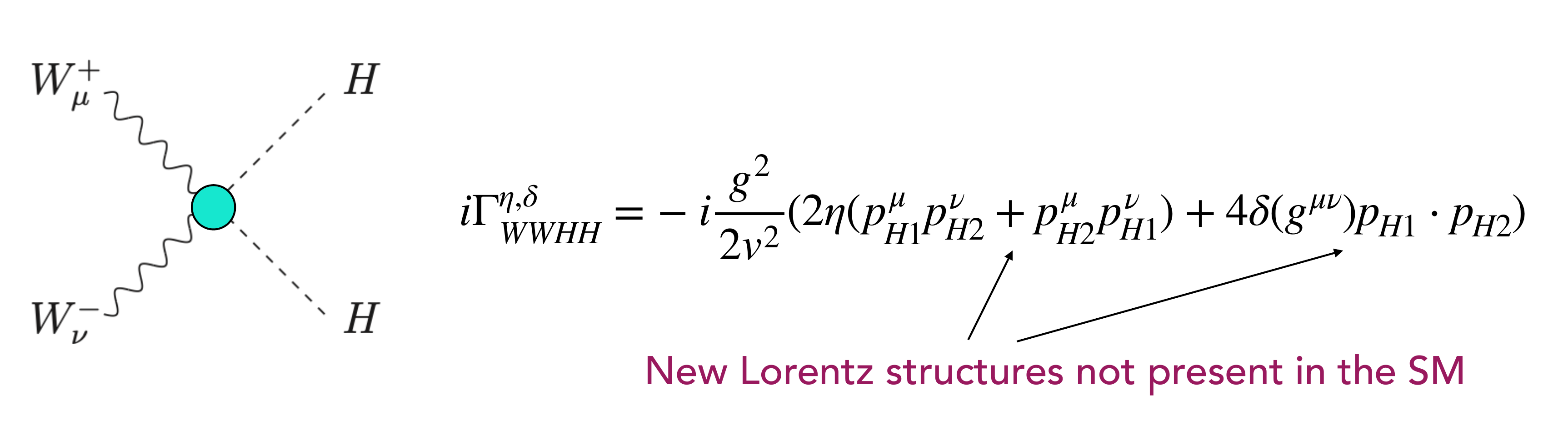}
    \caption{Feynman rule for the BSM vertex $WWHH$ from the 4-derivative  operators within HEFT, in \eqref{eq-L4},  and within the SMEFT, in \eqref{dim8},  written in terms of $\eta$ and $\delta$ according to the relations in  \eqref{coeffs}.  This is in contrast to the SM Feynman rule given by  $i\Gamma^{SM}_{WWHH}=i\frac{g^2}{2}g^{\mu \nu}$.  The corresponding Feynman rule for the BSM vertex $ZZHH$ is similar to the $WWHH$ one,  by just replacing $g^2$ by $g^2/\cos^2 \theta_W$.}
    \label{FR}
    \end{figure}
    
The definitions of the gauge fixing Lagrangian in the covariant $R_\xi$ gauges,   ${\mL}_{GF}^{\rm SM}, $ and the Fadeev-Popov Lagrangian,  ${\mL}_{FP}^{\rm SM}$,  are the usual ones in the SM.  
%It should be noticed that the usual normalization in the SMEFT for the Higgs self-coupling in \eqref{SMlag4} is different than its usual normalization in the HEFT in   \eqref{VHEFT}.  Specifically,  $\lambda_{\rm SMEFT} =2 \lambda_{\rm HEFT}$.  

The relevant set of operators for this work,  with four derivatives $(D\Phi)^4$,  within the SMEFT are given by:
\bear
\label{dim8}
 {\mL}_{8}&=&
 \frac{c_{\Phi^4 D^4}^{(1)}}{\Lambda^4} ((D_\mu\Phi)^\dagger D_\nu\Phi)((D^\nu\Phi)^\dagger D^\mu\Phi) \nn \\
 &+&  \frac{c_{\Phi^4 D^4}^{(2)}}{\Lambda^4} ((D_\mu\Phi)^\dagger D_\nu\Phi)((D^\mu\Phi)^\dagger D^\nu\Phi) \nn \\
  &+&  \frac{c_{\Phi^4 D^4}^{(3)}}{\Lambda^4} ((D_\mu\Phi)^\dagger D^\mu\Phi)((D_\nu\Phi)^\dagger D^\nu\Phi)  +...
\eear
This notation for the 4-derivative operators and the corresponding coefficients can be related easily to those of Ref.~\cite{Almeida:2020ylr} by:
\begin{equation}
c_{\Phi^4 D^4}^{(1)}= f_{S,2} \, ; \, c_{\Phi^4 D^4}^{(2)}= f_{S,0} \,; \, c_{\Phi^4 D^4}^{(3)}= f_{S,1}
\end{equation}

The relations among the previous coefficients of the 4-derivative operators in the two EFT's,  HEFT and SMEFT,  have been found in references~\cite{Domenech:2022uud,Domenech:2025gmn}  by the procedure of matching the set of on-shell scattering amplitudes for double and triple Higgs production from $VV$ with $VV=WW, ZZ$.  The correspondence found is the following:
\begin{eqnarray}
\label{coeffs}
 a_{ddVV1}&=&\frac{v^4}{4\Lambda^4} (c_{\Phi^4 D^4}^{(1)}+c_{\Phi^4 D^4}^{(2)})  \equiv \eta \nn \\
  a_{ddVV2}&=&\frac{v^4}{4\Lambda^4} (c_{\Phi^4 D^4}^{(3)}) \equiv \delta 
\end{eqnarray} 

The Feynman rules of the relevant EFT $VVHH$ vertices  are summarized in \figref{FR} in terms of the parameters $\eta$ and $\delta$ which will be the ones used from now on to refer jointly to the previous coefficients of both EFT's according to \eqref{coeffs}.

\section{Partial waves computation for $f_1f_2 \to f_3 f_4 H_5 H_6$}
\label{partial-waves}

\begin{figure}[!t]
    \centering
        \includegraphics[scale=0.3]{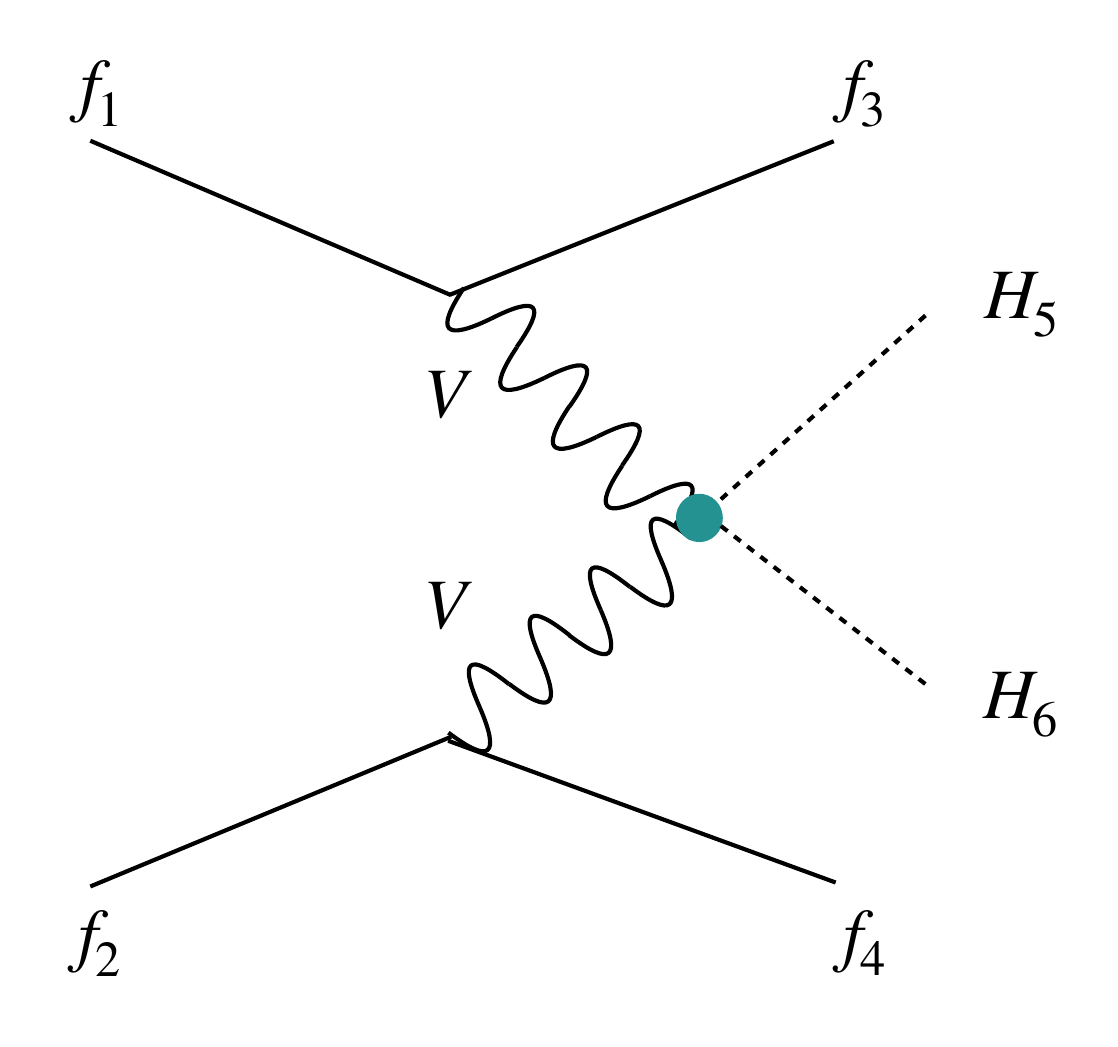}
    \caption{Generic diagram for VBF mediated channels $f_1 f_2 \to f_3 f_4 H_5 H_6$.}
    \label{VBF}
    \end{figure}

In this section we present the procedure to get the unitarity constraints on $\eta$  and $\delta$ from the study of the partial waves for the generic scattering process $f_1 f_2 \to f_3 f_4 H_5 H_6$ where  the subscripts refer to the momenta of the particles.  We assume here $p_1$ and $p_2$ incoming and $p_3$, $p_4$,  $p_5$ and  $p_6$ outgoing. This generic process (with $f$ refereeing jointly to either fermion or  anti-fermion)  is very sensitive to $\eta$ and $\delta$,  specially  at high energies in the TeV range and above.  At these high energies this scattering process is well known to be dominated by the so-called Vector Boson Fusion (VBF)  mediated  channels,  which are represented schematically in \figref{VBF}.  Since  the final state must contain two Higgs bosons,  the two vector bosons $VV$ involved in the intermediate states of these channels,  and depending on the particular process,  can be either $WW$ ($WW \equiv W^+W^-$) or $ZZ$.  Examples of particular processes of this generic type with interest for the future colliders will be presented and commented later. 

Since this process has four particles in the final state,   the standard partial wave analysis  of two-to-two  scattering processes is not applicable.  Instead,  we use here the 
massless spinor-helicity formalism (for a pedagogical introduction see,  for instance,  \cite{Dixon:2013uaa}) and apply the particular techniques described in \cite{Bresciani:2025toe}.  For the presentation in this section we follow closely the presentation and notation of the partial waves in \cite{Eboli:2026ccq}, where they analyzed the different process $f_1 f_2 \to V_3 V_4 V_5$ with three EW gauge bosons in the final state.   

In order to get the partial waves for the process of our interest,  $f_1f_2 \to f_3 f_4 H_5 H_6$,  we project the corresponding helicity amplitude 
${\cal M}(f_{1,\lambda_1} f_{2, \lambda_2} \to f_{3, \lambda_3} f_{4, \lambda_4} H_{5,\lambda_5} H_{6, \lambda_6})$ 
(or ${\cal M}_{\lambda_1, \lambda_2;  \lambda_3, \lambda_4, \lambda_5, \lambda_6}$ in short) into a basis of kinematic polynomials $\{{\cal B}^{J,a}_{\lambda_1,\lambda_2;\lambda_3, \lambda_4, \lambda_5,\lambda_6}\}$ with definite total angular momentum $J$.  The index $a$ usually refers to additional quantum numbers needed to completely characterize the states besides helicities and total angular momentum.  In the present case it may  also represent  different possibilities for the total angular momentum of the various pairs of particles in the final state.  The partial-wave amplitudes  of our interest,  
$T^{J,a}(f_{1,\lambda_1} f_{2, \lambda_2} \to f_{3, \lambda_3} f_{4, \lambda_4} H_{5,\lambda_5} H_{6, \lambda_6})$ are then obtained by expanding the scattering amplitude in this basis:
\begin{eqnarray}
\hspace{-2cm}
{\cal M}(f_{1,\lambda_1} f_{2, \lambda_2} \to f_{3, \lambda_3} f_{4, \lambda_4} H_{5,\lambda_5} H_{6, \lambda_6}) =& \nn \\
 \sum_{J,a} 
T^{J,a}(f_{1,\lambda_1} f_{2, \lambda_2} \to f_{3, \lambda_3} f_{4, \lambda_4} H_{5,\lambda_5} H_{6, \lambda_6}) \,\, \,\,
{\cal B}^{J,a}_{\lambda_1,\lambda_2;\lambda_3, \lambda_4, \lambda_5,\lambda_6} \,\,.  \,\,&
\end{eqnarray}

We normalize the partial-wave basis as in \cite{Bresciani:2025toe, Eboli:2026ccq}:

\begin{eqnarray}
\int d\Phi_{f_1 ,f_2} d \Phi_{f_3, f_4,H_5,H_6} ({\cal B}^{J,a}_{\lambda_1,\lambda_2;\lambda_3, \lambda_4, \lambda_5,\lambda_6})({\cal B}^{J',b}_{\lambda_1,\lambda_2;\lambda_3, \lambda_4, \lambda_5,\lambda_6})^* =
 (2J+1) \delta^{J,J'} \delta^{a,b},
\end{eqnarray}
where $d\Phi_S$ is the phase space for state $S$.  Notice that with this normalization above,  the integrations are performed over the initial state 
($d \Phi_{\rm in}=d \Phi_{f_1,f_2}$) and the final state ($d\Phi_{\rm out}=d \Phi_{f_3, f_4,H_5,H_6}$). 
Then finally,  the partial-wave amplitudes can be written as:

\begin{eqnarray}
T^{J,a}(f_{1,\lambda_1} f_{2, \lambda_2} \to f_{3, \lambda_3} f_{4, \lambda_4} H_{5,\lambda_5} H_{6, \lambda_6})=& \nn \\
\frac{1}{2J+1}
\int d\Phi_{f_1 ,f_2} d \Phi_{f_3, f_4,H_5,H_6} {\cal M}_{\lambda_1, \lambda_2;  \lambda_3, \lambda_4, \lambda_5, \lambda_6} \,\,( {\cal B}^{J,a}_{\lambda_1,\lambda_2;\lambda_3, \lambda_4, \lambda_5,\lambda_6})^* \,\,. \,\,&
\label{TJa-def}
\end{eqnarray}

All the previous formulas of the helicity amplitudes of our interest here refer to scattering of two-to-four  particles,  with labels referring to the momenta of the particles,  i.e.  $1,2 \to 3, 4,5,6$.  The possible helicities $\lambda_i$,  $(i=1,...,6)$, for the present case of  $f_1f_2 \to f_3 f_4 H_5 H_6$ are $\pm 1/2$,  for fermions $1$,  $2$,  $3$ and $4$ and $0$ for scalars $5$ and $6$.  Then,  the various helicity channels  refer to the 16  possibilities a priori for the particle helicities,  i.e.  $(\pm, \pm; \pm, \pm, 0, 0)$ using the short notation.

In order to extract  the kinematic polynomials we also use the massless spinor-helicity techniques.   In particular,   we follow the general steps described in  
\cite{Bresciani:2025toe} where they focused in the case of $2 \to 3$ processes.  We present here the case of $2 \to 4$ processes with effective $VVHH$ interactions derived from the 4-derivative operators of both HEFT and SMEFT,  contributing via the $\delta$ and $\eta$ as in \figref{FR}.  Notice that these $VVHH$ effective interactions translate into contributions to the helicity amplitudes for the  process  
$f_1f_2 \to f_3 f_4 H_5 H_6$,  via the VBF diagram in \figref{VBF},   which contain two internal propagators of EW gauge bosons $VV$,  therefore  our computation of partial waves here must apply to amplitudes of non-contact type.  
Since, to our knowledge,  the case of $2\to 4$ processes  is not available in the literature, we have done the full excersice of deriving the relevant polynomials for the present two-to-four case.  For this exercise,  we have used the Mathematica code provided in Ref.~\cite{Degrande:2025uil}, which is based on the package ABC4EFT \cite{Li:2022tec,  Li:2020gnx,  Li:2020xlh}.  It implements the Casimir operator method and sistematically constructs both non-orthogonal ${\cal P}$ and orthonormal ${\cal B}$ bases.     
These outputs are in all-incoming particles convention. Therefore,  to go to the physical notation, which  is the one used in the partial waves computation with ${1,2}$ incoming and ${3,4,5,6}$ outgoing, one must analytically extend the definition of the spinor variables using the forthcoming \eqref{inout}.

We summarize in  \appref{A1}  the obtained results for the relevant polynomials in the present study of unitarity limits to the BSM $\eta$ and $\delta$ in  $f_1f_2 \to f_3 f_4 H_5 H_6$.   The corresponding  matrices ${\cal R}$ that transfer from the non-orthogonal ${\cal P}$ to the orthonormal basis ${\cal B}$ are also collected in this \appref{A1}.

%%%%%%
Notice that the polynomials  ${\cal P}^{J,a}$ in \tabref{table-samehelicities},  \tabref{table-opposhelicitiesmp} and \tabref{table-opposhelicitiespm},  
 are normalized to one,  but they are not yet orthogonal for the different values of the `quantum number $a$'. 
  Thus,  they provide a different set of partial waves denoted by $\tilde T^{J,b}$  defined as:

 \begin{eqnarray}
 \label{Ttilde}
\tilde T^{J,b}(f_{1,\lambda_1} f_{2, \lambda_2} \to f_{3, \lambda_3} f_{4, \lambda_4} H_{5,\lambda_5} H_{6, \lambda_6})=& \nn \\
\frac{1}{2J+1}
\int d\Phi_{f_1 ,f_2} d \Phi_{f_3, f_4,H_5,H_6} {\cal M}_{\lambda_1, \lambda_2;  \lambda_3, \lambda_4, \lambda_5, \lambda_6} \,\,( {\cal P}^{J,b}_{\lambda_1,\lambda_2;\lambda_3, \lambda_4, \lambda_5,\lambda_6})^* \,\,.\,\,&
\end{eqnarray} 

 We next obtain the corresponding normalized and orthogonal basis of polynomials,   $\mathcal{B}^{J,a}$,  by applying the proper transfer matrices to the non-orthogonal polynomials  
 $\mathcal{P}^{J,a}$:
\begin{eqnarray}
\mathcal{B}^{J,a}_{\lambda_1,\lambda_2;\lambda_3, \lambda_4, \lambda_5,\lambda_6} =\sum_b \mathcal{R}^J_{a,b} \,\,
\mathcal{P}^{J,a}_{\lambda_1,\lambda_2;\lambda_3, \lambda_4, \lambda_5,\lambda_6} \,\, .
\label{transfer}
\end{eqnarray}
The values of these transfer matrices for the relevant cases here are collected in  \appref{A1}.  We include the cases with same sign helicities and with opposite sign helicites.  For the same sign helicities case,  we study the projections to $J=0$ and $J=1$,  where  the transfer matrices 
 are  $7 \times 7$ and $9 \times 9$  respectively.  For the opposite sign helicities case,  there are projections with $J=1$ and $J=2$.  In this case we select to study the $J=2$ projection,  which is not present in the same sign helicities case and it is simpler having a lower number of  polynomials.  The corresponding $3 \times 3$ transfer matrices and the 3 polynomials for this $J=2$ case are also collected in \appref{A1}.

To  extract finally  the partial-wave amplitudes $T^{J}$,   the corresponding linear combination of the $\tilde T^{J}$'s is used:
\begin{eqnarray}
\label{TJa}
T^{J,a}(f_{1,\lambda_1} f_{2, \lambda_2} \to f_{3, \lambda_3} f_{4, \lambda_4} H_{5,\lambda_5} H_{6, \lambda_6})=\sum_b (\mathcal{R}^J_{ab})^* \tilde T^{J,b}(f_{1,\lambda_1} f_{2, \lambda_2} \to f_{3, \lambda_3} f_{4, \lambda_4} H_{5,\lambda_5} H_{6, \lambda_6})\,\,.
\end{eqnarray} 

These last partial waves are the ones that must fulfill the generalized optical theorem: 
\begin{eqnarray}
T^{J,a}_{i\to f}-(T^{J,a}_{i\to f})^* = i \sum_X T^{J,a}_{i \to X}(T^{J,a}_{f \to X})^* \,\,,\,\,
\end{eqnarray}
leading to the perturbative unitary condition that we are looking for:
\begin{eqnarray}
|T^{J,a}(f_{1,\lambda_1} f_{2, \lambda_2} \to f_{3, \lambda_3} f_{4, \lambda_4} H_{5,\lambda_5} H_{6, \lambda_6})|<1 .
\end{eqnarray}

The details of the 4-body phase space integration defining $\tilde T^{J}$ in \eqref{Ttilde} are provided in \appref{A2}.  The focus point of our proposal is to implement the spherical parametrization of  Ref.~\cite{Degrande:2025uil},  introduced in \cite{Larkoski:2020thc},  applied to our present case and to perform the non-trivial integrations for the non-contact amplitudes containing denominators (from the $V$ propagators) numerically by means of VEGAS integrator~\cite{Lepage:2020tgj}.  The details of this parametrization and the integration are given in \appref{A2}.
    
%%%%%%%%%
Regarding the explicit expressions for the helicity amplitudes of the processes  $f_1f_2 \to f_3 f_4 H_5 H_6$ in \figref{VBF}, they all have a similar general structure:

\begin{eqnarray}
i{\cal M}&=&i{\cal M}^{\rm SM}+i{\cal M}^{\rm BSM} \nn \\
&=&(i \Gamma^{\mu'}_{V_1ff'})\left(\frac{-ig_{\mu'\mu}}{q_1^2-m_V^2}\right) (i\Gamma^{\nu'}_{V_2ff'})    \left(\frac{-ig_{\nu'\nu}}{q_2^2-m_V^2}\right)
(i \Gamma^{\mu \nu}_{V_1V_2HH}) \,\,, \,\,
\label{generic}
\end{eqnarray}
where the $V_1V_2$ intermediate EW bosons can be either $WW$ and/or $ZZ$ depending on the specific process.  The contributions with intermediate $WW$ and $ZZ$ are called WBF and ZBF mediated channels,  respectively.   Notice also that only the $g_{\mu \nu}$ part of the $\Delta^{V}_{\mu \nu}$ propagator contributes here,  since we are assuming massless fermions.  

The various vertices appearing in \eqref{generic} can be written in a very compact form, using the spinor-helicity techniques,  as follows: 
\begin{eqnarray}
i\Gamma^{\mu}_{Wff'}&=&i \frac{g}{\sqrt{2}} J^\mu_L \,\, ; \,\, i\Gamma^{\mu}_{Zff'}=i \frac{g}{c_W}( g^f_L J^\mu_L + g^f_R J^\mu_R)  \,\,; \,\, 
  g^f_L=T^f_3-Q^f s_W^2  \,\,; \,\, g^f_R= -Q^f s_W^2  \,\,,\,\, \nn \\
 && J^\mu_L= \bar u_{-} \gamma^\mu P_L u'_{-}= \bar u_{-} \gamma^\mu P_L v'_{+}=\bar v_{+} \gamma^\mu P_L u'_{-}=\bar v_{+} \gamma^\mu P_L v'_{+}= [p|\gamma^\mu |p'\rangle  \,\,,\,\, \nn \\
&&   J^\mu_R= \bar u_{+} \gamma^\mu P_R u'_{+}= \bar u_{+} \gamma^\mu P_R v'_{-}=\bar v_{-} \gamma^\mu P_R u'_{+}=\bar v_{-} \gamma^\mu P_R v'_{-}= \langle p|\gamma^\mu |p']  \,\,,\,\,\nn \\
i\Gamma^{\mu \nu}_{WWHH}&=&i \frac{g^2}{2}(g^{\mu\nu}- \frac{4 \delta}{v^2}g^{\mu \nu} 
p_{H1} .  p_{H2}-\frac{2 \eta}{v^2}
(p^\mu_{H1} p^\nu_{H2}+p^\nu_{H1} p^\mu_{H2})) \,\,,\,\, \nn \\
i\Gamma^{\mu \nu}_{ZZHH}&=&i \frac{g^2}{2c_W^2}(g^{\mu\nu}- \frac{4 \delta}{v^2}g^{\mu \nu} 
p_{H1} .  p_{H2}-\frac{2 \eta}{v^2}
(p^\mu_{H1} p^\nu_{H2}+p^\nu_{H1} p^\mu_{H2}))\,\,,\,\, 
\label{vertices} 
\end{eqnarray}
where,  in the massless case,  the four-components spinors with momentum $p$ are written in terms of bi-spinors as $u_-=v_+=(|p\rangle\,0)^T$,  $u_+=v_-=(0\,|p])^T$,  $\bar u_-= \bar v_+=(0 \, [p|)$,  and $\bar u_+=\bar v_-=(\langle p| \, 0)$. 

Notice that the first terms in the vertices $i\Gamma^{\mu \nu}_{WWHH}$ and $i\Gamma^{\mu \nu}_{ZZHH}$ correspond to the SM contributions,  whereas the last  terms correspond to the BSM contributions containing the anomalous couplings $\eta$ and $\delta$,  as  given in \figref{FR}. 
 Notice also that,  as already said,  we are working with massless fermions in using the helicity-spinor techniques,  but we have included the mass of the corresponding intermediate EW gauge boson 
 $m_V$  in the denominators of the propagators.  This is necessary to regulate the involved integrals for phase space integration defining the partial waves in \eqref{Ttilde}.  As we will see in the next section  with these $m_V$  included in the denominators  these integrals are finite and well behaved. 
 
Then,  the final result for ${\cal M}^{\rm BSM}$ can be written in a very compact form:
\begin{eqnarray}
{\cal M}^{\rm BSM}&=& \frac{{\cal C}_V}{(q_1^2-m_V^2)(q_2^2-m_V^2)} J^\mu_1 J^\nu_2 \left(\frac{-1}{v^2}\right) (4 \delta g_{\mu\nu}  p_5.p_6+ 
2 \eta  (p_{5 \mu} p_{6\nu}+p_{6\mu} p_{5\nu})),
\label{ABSM}
\end{eqnarray}
which should be compared with the corresponding amplitude of the SM:
\begin{eqnarray}
{\cal M}^{\rm SM}&=& \frac{{\cal C}_V}{(q_1^2-m_V^2)(q_2^2-m_V^2)} J^\mu_1 J^\nu_2 (g_{\mu\nu}) \,\,.
\label{ASM}
\end{eqnarray}
 The coefficients ${\cal C}_V$ in the global factors above are: ${\cal C}_W=\frac{g^4}{4}$ for WBF mediated channels,   and ${\cal C}_Z=\frac{g^4}{2c_W^4}$ for ZBF mediated channels.  The mass in the propagators are $m_V=m_W$ and $m_V=m_Z$, respectively.  The two currents above,  $J_1^\mu$ and $J_2^\nu$,   refer to the fermionic currents in the two  interaction points with the $V_1$ and $V_2$ bosons respectively. They carry  inside the remaining factors of the corresponding couplings,  like $g_L$,  $g_R$ etc in the $V=Z$ case.  

It should also be noted that in all these results in Eqs. (\ref{vertices})-(\ref{ASM})  we have used the convention of all-momenta  incoming.  Thus,  in order to get the wanted results in the physical convention for the process $f_1 f_2 \to f_3 f_4 H_5 H_6$ with $p_{1,2}$  incoming and $p_{3,4,5,6}$ outgoing, the following changes should be 
done to the final states ($ i=3,4,5,6$) momenta and helicities:
\bear 
\label{inout}
&&p_i \to -p_i;  |-p_i \rangle \to - |p_i \rangle; |-p_i] \to |p_i];  \langle -p_i| \to - \langle p_i|;   [-p_i| \to [p_i|;\, \lambda_i \to -\lambda_i\,.
\eear
From all the expressions above and using the proper Fierz Identities,  
one can get the final result written exclusively in terms of  products of $\langle ij \rangle $'s  and $[ij]$'s.  In particular,  for the products of the two currents $J_1$ and $J_2$ involved in \eqref{ABSM} and \eqref{ASM} the following identities are applied:
\begin{equation}
[p|\gamma^\mu |p'\rangle [q|\gamma_\mu |q'\rangle =  2 \langle p'q'\rangle [qp] \,\,, \,\, {\rm in} \,\,J_L J_L  \,\,,
\end{equation}
\begin{eqnarray}
[p| \gamma^\mu |p'\rangle \langle q|\gamma_\mu |q'] &=&  2 \langle p'q \rangle [q'p] \,\,, \,\, {\rm in} \,\,J_L J_R \,\,, 
\end{eqnarray}
\begin{eqnarray}
\langle p|\gamma^\mu |p'][q|\gamma_\mu |q' \rangle &=&  2 \langle pq'\rangle [qp'] \,\,, \,\, {\rm in} \,\,J_R J_L \,\,,
\end{eqnarray}
\begin{eqnarray} 
\langle p|\gamma^\mu |p'] \langle q|\gamma_\mu |q']&=&  2 \langle pq\rangle [q'p'] \,\,, \,\, {\rm in} \,\,J_R J_R \,\, .
\end{eqnarray}

%%%%%%   
\begin{table}[h!]
    \centering
    \begin{tabular}{cccc}
       \hline
        \hline
        \multicolumn{4}{c} {Factors in helicity amplitudes of relevant VBF mediated processes $f_1 f_2 \to f_3 f_4 H_5 H_6$}  \\
        \hline
       Process   & Helicity Amplitude &  Global Factor ($G$)   \\
        $ u d \to d u HH $ & $ {\cal M}^{WW}_{-, -; -,-, 0,0}$ &  ${\cal C}_W$  \\
         & $ {\cal M}^{ZZ}_{-, -; -,-, 0,0}$ &  ${\cal C}_Zg^u_L g^d_L$    \\
         & $ {\cal M}^{ZZ}_{+ +; +,+, 0,0}$ &  ${\cal C}_Zg^u_R g^d_R$   \\
         & $ {\cal M}^{ZZ}_{-, +; +,-, 0,0}$ &  ${\cal C}_Z g^u_L g^d_R$  \\
          & $ {\cal M}^{ZZ}_{+, -; -,+, 0,0}$ &  ${\cal C}_Z g^u_R g^d_L$   \\  
         $ u u \to u u HH $ & $ {\cal M}^{ZZ}_{-, -; -,-, 0,0}$ &${\cal C}_Zg^u_L g^u_L$   \\
          & $ {\cal M}^{ZZ}_{+ +; +,+, 0,0}$ &  ${\cal C}_Zg^u_R g^u_R$  \\
            & $ {\cal M}^{ZZ}_{-, +; -,+, 0,0}$ &  ${\cal C}_Z g^u_L g^u_R$   \\
             & $ {\cal M}^{ZZ}_{+, -; +,-, 0,0}$ &  ${\cal C}_Z g^u_R g^u_L$  \\
         $ d d  \to d d  HH $ & $ {\cal M}^{ZZ}_{-, -; -,-, 0,0}$ &${\cal C}_Zg^d_L g^d_L$ \\
            & $ {\cal M}^{ZZ}_{+ +; +,+, 0,0}$ &  ${\cal C}_Zg^d_R g^d_R$   \\
            & $ {\cal M}^{ZZ}_{-, +; -,+, 0,0}$ &  ${\cal C}_Z g^d_L g^d_R$   \\
             & $ {\cal M}^{ZZ}_{+, -; +,-, 0,0}$ &  ${\cal C}_Z g^d_R g^d_L$   \\
           $ e^-e^+ \to \nu_e \bar \nu_e H H$ & $ {\cal M}^{WW}_{-, +; -,+, 0,0}$ &  ${\cal C}_W$   \\
           $e^-e^+ \to e^-e^+ HH $ & $ {\cal M}^{ZZ}_{-, -; -,-, 0,0}$ &${\cal C}_Zg^e_L g^e_R$   \\
           & $ {\cal M}^{ZZ}_{+ +; +,+, 0,0}$ &  ${\cal C}_Zg^e_R g^e_L$   \\
            & $ {\cal M}^{ZZ}_{-, +; -,+, 0,0}$ &  ${\cal C}_Z g^e_L g^e_L$   \\
             & $ {\cal M}^{ZZ}_{+, -; +,-, 0,0}$ &  ${\cal C}_Z g^e_R g^e_R$   \\
         \hline
        \hline
\end{tabular}
    \caption{\label{table-Gfactors}  Most relevant VBF mediated processes involved at high energy colliders that provide the best unitarity limits.  The coefficients involved are:  ${\cal C}_W=\frac{g^4}{4}$,   ${\cal C}_Z=\frac{g^4}{2c_W^4}$,  $g^f_L=T_3^{f_L}-Q^{f} s_W^2$, and $g^f_R=-Q^f s_W^2$.}
    \end{table}

Looking at these products of currents,  one can immediately conclude that the resulting non-vanishing helicity amplitudes for the VBF mediated channels of interest here contain  an even number of same sign helicities.  Using the short notation,  $(\pm, \pm;\pm,\pm, 0,0)$ with just the helicities signs in the fermion helicities $\lambda_{1,2,3,4}$'s and setting  $\lambda_{5,6}=0$ for the two final Higgs bosons,  the non-vanishing helicity amplitudes are,  generically,  of these six types: $(--;--00)$, $(++;++00)$,  $(-+;-+00)$,  $(+ -; +-00)$, $(+ -; - + 00)$ and  $(-+;+-00)$.   Any other combination of helicities give vanishing amplitudes.

We provide in \tabref{table-Gfactors} a list of specific processes $f_1f_2 \to f_3 f_4 H_5 H_6$ which are the most relevant ones for the setting of unitarity limits at future  high energy colliders,  including both  $pp$ and $e^-e^+$ (also $\mu^- \mu^+$) options.   To simplify,  we take only first generation fermions and assume no-quark mixing,  
i.e $V_{\rm CKM}=1$.  In the case of quarks we show only those channels that involve valence quarks in the $pp$ option.  The global factors here refer to the global coefficients  that multiply $(-1/v^2)$ in \eqref{ABSM}.

%%%%%
As can be seen in \tabref{table-Gfactors} some of these processes are mediated by just WBF ($ e^-e^+ \to \nu_e \bar \nu_e H H$),    some by just ZBF ( $ u u \to u u HH $,  $ dd  \to dd HH $,  $e^-e^+ \to e^-e^+ HH $)  and some  by both WBF and ZBF ($ u d \to d u HH $).  In the latter case,  one must obviously add the two contributions when computing the helicity amplitudes and, therefore, in  the forthcoming partial waves computation.  Notice also that the vector boson mass involved in the denominators of these two VBF cases are $m_W$ for WBF channels and $m_Z$ for ZBF channels. 
 
To end this section we summarize the full analytical results of the  helicity amplitudes for three selected processes,  1) $ u d \to d u HH $,   2) $uu \to uuHH$  (Idem $dd \to dd HH$,  changing $u \leftrightarrow d$)  and 3) $e^-e^+ \to e^-e^+ HH $ (Idem $\mu^- \mu^+ \to \mu^- \mu^+ HH$,  changing $e \leftrightarrow \mu$),  which can serve as reference cases for the two types,  respectively: `WBF+ZBF', 1),  and  `only ZBF', 2) and 3).  All these amplitudes contain the SM result (in the first term) and the BSM result (in the last terms).  The BSM amplitudes with the contributions from $\delta$ and $\eta$ are the ones of our interest here for the following study of unitarity.
The SM contribution is included here just for comparison.
%For shortness, the SM contribution of the contact $VVHH$ interaction is included here just for comparison (the contributions with three internal propagators are negligible for this unitarity study but are considered in the numerical analysis of \secref{colliders}).
These results are the following:
 
 \textcolor{white}{NECESITAMOS ESPACIO PARA QUE LAS TABLAS NO CORTEN LAS REFERENCIAS}
\vspace{200cm}
 
 1) $ u_1 d_2 \to d_3 u_4 H_5H_6 $:
  \begin{eqnarray}
 {\cal M}(ud;duHH) &=&  {\cal M}^W+{\cal M}^Z \nn \\
 {\cal M}_{(--;--00)}^W &=&\Big( {\cal C}_W  \Big) \frac{2}{(t_{13}-m_W^2)(t_{24}-m_W^2)} \Big(    \langle 1 2 \rangle [43]  \nn \\
 & &-\frac{1}{v^2}(2\delta s_{56}\langle 1 2 \rangle [43]
 + \eta (\langle 1 5 \rangle [53] \langle 26  \rangle [64]+\langle 1 6 \rangle [63] \langle 25  \rangle [54] )) \Big)  \nn \\
 {\cal M}^W&=& 0 \rm{\,\,\,\,\,\,for\,\, other\,\, helicities} \nn \\ 
  {\cal M}_{(--;--00)}^Z &=&\Big( {\cal C}_Z g^u_L g^d_L  \Big) \frac{2}{(t_{14}-m_Z^2)(t_{23}-m_Z^2)} \Big(    \langle 1 2 \rangle [34]  \nn \\
 & &-\frac{1}{v^2}(2\delta s_{56}\langle 1 2 \rangle [34]
 + \eta (\langle 1 5 \rangle [54] \langle 26  \rangle [63]+\langle 1 6 \rangle [64] \langle 25  \rangle [53] )) \Big)  \nn \\
   {\cal M}_{(++;++00)}^Z &=&\Big( {\cal C}_Z g^u_R g^d_R \Big) \frac{2}{(t_{14}-m_Z^2)(t_{23}-m_Z^2)} \Big(    \langle 43 \rangle [21]  \nn \\
 & &-\frac{1}{v^2}(2\delta s_{56}\langle 43 \rangle [21]
 +\eta (\langle 4 5 \rangle [51] \langle 36  \rangle [62]+\langle 4 6 \rangle [61] \langle 35  \rangle [52] )) \Big)  \nn \\
    {\cal M}_{(-+;+-00)}^Z &=&\Big( {\cal C}_Z g^u_L g^d_R  \Big) \frac{2}{(t_{14}-m_Z^2)(t_{23}-m_Z^2)} \Big( -   \langle 13 \rangle [24]  \nn \\
 & &+\frac{1}{v^2}(2\delta s_{56}\langle 13 \rangle [24]
 + \eta (\langle 15 \rangle [54] \langle 36  \rangle [62]+\langle 1 6 \rangle [64] \langle 35  \rangle [52] )) \Big)  \nn \\
     {\cal M}_{(+-;-+00)}^Z &=&\Big( {\cal C}_Z g^u_R g^d_L \Big) \frac{2}{(t_{14}-m_Z^2)(t_{23}-m_Z^2)} \Big( -   \langle 42 \rangle [31]  \nn \\
 & &+\frac{1}{v^2}(2\delta s_{56}\langle 42 \rangle [31]
 +\eta (\langle 45 \rangle [51] \langle 26  \rangle [63]+\langle 4 6 \rangle [61] \langle 25  \rangle [53] )) \Big)  
 \label{ud2duHH}
 \end{eqnarray}
 
  2) $ u_1u_2 \to u_3u_4 H_5H_6 $ (Idem for $dd \to dd HH$,  changing $u \leftrightarrow d$):
  %%%%%%%%%
    \begin{eqnarray}
 {\cal M}(uu;uuHH) &=& {\cal M}^Z  \nn \\
  {\cal M}_{(--;--00)}^{Z} &=&\Big[\Big( {\cal C}_Z g^u_L g^u_L  \Big) \frac{2}{(t_{13}-m_Z^2)(t_{24}-m_Z^2)} \Big(    \langle 1 2 \rangle [43]  \nn \\
 & &-\frac{1}{v^2}(2\delta s_{56}\langle 1 2 \rangle [43]
 + \eta (\langle 1 5 \rangle [53] \langle 26  \rangle [64]+\langle 1 6 \rangle [63] \langle 25  \rangle [54] )) \Big) \Big]  \nn \\
 &-& \Big[ 3 \leftrightarrow 4 \Big] \nn \\
   {\cal M}_{(++;++00)}^{Z} &=& \Big[\Big( {\cal C}_Z g^u_R g^u_R \Big) \frac{2}{(t_{13}-m_Z^2)(t_{24}-m_Z^2)} \Big(    \langle 34 \rangle [21]  \nn \\
 & &-\frac{1}{v^2}(2\delta s_{56}\langle 34 \rangle [21]
 +\eta (\langle 3 5 \rangle [51] \langle 46  \rangle [62]+\langle 3 6 \rangle [61] \langle 45  \rangle [52] )) \Big) \Big] \nn \\
 &-& \Big[ 3 \leftrightarrow 4  \Big]  \nn \\
%%% rober llama IIa a lo siguiente
    {\cal M}_{(-+;-+00)}^{Z} &=&\Big[\Big( {\cal C}_Z g^u_L g^u_R  \Big) \frac{2}{(t_{13}-m_Z^2)(t_{24}-m_Z^2)} \Big( -   \langle 14 \rangle [23]  \nn \\
 & &+\frac{1}{v^2}(2\delta s_{56}\langle 14 \rangle [23]
 + \eta (\langle 15 \rangle [53] \langle 46  \rangle [62]+\langle 1 6 \rangle [63] \langle 45  \rangle [52] )) \Big) \Big]  \nn \\
     {\cal M}_{(+-;+-00)}^{Z} &=& \Big [\Big( {\cal C}_Z g^u_R g^u_L \Big) \frac{2}{(t_{13}-m_Z^2)(t_{24}-m_Z^2)} \Big( -   \langle 32 \rangle [41]  \nn \\
 & &+\frac{1}{v^2}(2\delta s_{56}\langle 32 \rangle [41]
 +\eta (\langle 35 \rangle [51] \langle 26  \rangle [64]+\langle 3 6 \rangle [61] \langle 25  \rangle [54] )) \Big) \Big]  \nn \\
 {\cal M}_{(-+;+-00)}^{Z} & = &{\cal M}_{(-+;-+00)}^{Z} \left[3 \leftrightarrow 4 \right] \,;\, 
 {\cal M}_{(+-;-+00)}^{Z} = {\cal M}_{(+-;+-00)}^{Z} \left[ 3 \leftrightarrow  4 \right]
\label{uu2uuHH}
  \end{eqnarray}

  3) $ e^-_1 e^+_2 \to e^-_3 e^+_4 H_5H_6 $ (Idem for $\mu^-\mu^+ \to \mu^- \mu^+ HH$, changing $ e \leftrightarrow  \mu$):
  \begin{eqnarray}
 {\cal M}(e^-e^+;e^-e^+HH) &=&  {\cal M}^Z \nn \\
 {\cal M}_{(--;--00)}^Z &=&\Big( {\cal C}_Z g^e_L g^e_R  \Big) \frac{2}{(t_{13}-m_Z^2)(t_{24}-m_Z^2)} \Big(    \langle 1 2 \rangle [43]  \nn \\
 & &-\frac{1}{v^2}(2\delta s_{56}\langle 1 2 \rangle [43] 
 + \eta (\langle 1 5 \rangle [53] \langle 26  \rangle [64]+\langle 1 6 \rangle [63] \langle 25  \rangle [54] )) \Big)  \nn \\
   {\cal M}_{(++;++00)}^Z &=&\Big( {\cal C}_Z g^e_R g^e_L  \Big) \frac{2}{(t_{13}-m_Z^2)(t_{24}-m_Z^2)} \Big(    \langle 34 \rangle [21]  \nn \\
 & &-\frac{1}{v^2}(2\delta s_{56}\langle 34 \rangle [21]
 +\eta (\langle 3 5 \rangle [51] \langle 46  \rangle [62]+\langle 3 6 \rangle [61] \langle 45  \rangle [52] )) \Big)  \nn \\
    {\cal M}_{(-+;-+00)}^Z &=&\Big( {\cal C}_Z g^e_L g^e_L  \Big) \frac{2}{(t_{13}-m_Z^2)(t_{24}-m_Z^2)} \Big( -  \langle 14 \rangle [23]  \nn \\
 & &+\frac{1}{v^2}(2\delta s_{56}\langle 14 \rangle [23]
 + \eta (\langle 15 \rangle [53] \langle 46  \rangle [62]+\langle 1 6 \rangle [63] \langle 45  \rangle [52] )) \Big)  \nn \\
     {\cal M}_{(+-;+-00)}^Z &=&\Big( {\cal C}_Z g^e_R g^e_R \Big) \frac{2}{(t_{13}-m_Z^2)(t_{24}-m_Z^2)} \Big( -   \langle 32 \rangle [41]  \nn \\
 & &+\frac{1}{v^2}(2\delta s_{56}\langle 32 \rangle [41]
 +\eta (\langle 35 \rangle [51] \langle 26  \rangle [64]+\langle 3 6 \rangle [61] \langle 25  \rangle [54] )) \Big)  
 \label{ee2eeHH}
 \end{eqnarray}
 
 In the previous results we have used the short notation $s_{ij}=(p_i+p_j)^2$ and $t_{ij}=(p_i-p_j)^2$.  These variables can be written in the massless spinor-helicity  formalism as $s_{ij}=\langle  ij \rangle [ji]$ and $t_{ij}=-\langle  ij \rangle [ji]$.  Besides,   one can always rewrite the previous results in different order of the indices inside a bi-spinor product by using the antisymmetry properties of these products,  i.e.,  $\langle ij \rangle =  -\langle ji \rangle$ and   $[ij]= -[ji]$.

\section{Unitarity constraints on $\delta$ and $\eta$ versus  $\sqrt{s_{12}}$} 
\label{unitarity-limits}

In this section we present the numerical results for the partial-waves amplitudes $\tilde T^{J}$, defined  in  \eqref{Ttilde},  and for $T^{J}$,  defined in \eqref{TJa-def}.  The practical procedure here is,  first computing the $\tilde T^{J}$'s  and then obtaining from them the $T^{J}$'s  by the use of \eqref{TJa} with the proper transfer matrices.  
As said previously,  to perform the phase space integration needed to get  $\tilde T^{J}$ we use the parametrization given in \appref{A2}, which is based on \cite{Degrande:2025uil}.  

The results in this section will be presented exclusively for the processes selected previously:  1) $ u_1 d_2 \to d_3 u_4 H_5H_6 $, 
2a) $ u_1u_2 \to u_3u_4 H_5H_6 $ 2b)  $d_1d_2 \to d_3d_4 H_5H_6$ and 3) $ e^-_1 e^+_2 \to e^-_3 e^+_4 H_5H_6 $.
The reason for this particular choice is because these processes are expected to be the most relevant ones in setting constraints to  the $\delta$ and $\eta$ parameters which,  at the same time,  may have relevant implications for very high energy colliders,  above 10 TeV,  like  HL-LHC,  \cite{ATLAS:2025eii},   HE-LHC \cite{FCC:2018bvk},  future circular hadron-hadron colliders like FCC-pp \cite{FCC:2018vvp},  and  future lepton-lepton colliders like the International Muon collider \cite{Casarsa:2024kct}.

In this sense,  since the unitarity constraints come from the behaviour of helicity amplitudes at high energies,  those channels with initial  valence proton quarks ($u$ and/or $d$) will be the most relevant ones for $pp$ colliders.  Due to the properties of the  PDF's,  the valence quarks are expected to reach the highest available energies at $pp$ colliders with higher probability than the sea quarks and anti-quarks.  In particular,  the VBF-mediated amplitudes initiated with valence quarks are expected to be the most relevant ones for the high energy $pp$ colliders,  like the LHC  and the future circular colliders FCC-pp.  The class of events with two Higgs bosons plus two jets having VBF topologies are usually selected by tagging these two final jets with the proper VBF cuts.  More specifically,  the usual requirements for these two jets are that they are placed in opposite hemispheres with large differences in rapidities,  large  transverse momentum and large joint  invariant mass, $m_{jj}$.  These requirements are very efficient to enlarge the VBF-mediated contributions versus the non-VBF contributions.  Indeed the non-VBF contributions at these large energies are highly suppressed  with these VBF cuts and provide no significant rates to the final $HHjj$ events.  

Regarding the $e^-e^+$  colliders (and,  similarly,  for  $\mu^- \mu^+$ colliders) we select the cases with detectable particles in the final state in addition to the two Higgs bosons,  i.e.  those with $e^-e^+$ in the final state.  These are  the only ones that provide the proper tagging techniques to analyze the VBF topologies by the proper VBF cuts on these final $e^-$ and $e^+$,  in a similar way as its done for $pp$ colliders with the two final jets.  In contrast,  the case with (undetectable) neutrinos and anti-neutrinos  in the final state  does not provide these specific tagging techniques and therefore the VBF topologies (WBF in this case)  will be more difficult to get separated from the non-VBF topologies.  In consequence,  the $e^-e^+ \to \nu \bar \nu HH$ case is not an efficient processes to test the unitarity constraints. 

For the chosen processes here,  the lowest allowed values of the total angular momentum $J$ in the partial waves $T^J$  are $J=0$,  $J=1$ and $J=2$.   In particular,  for the processes 1) and 2)  the helicity channels with same sign helicities  (in short notation) $(--; --00)$ and $(++;++00)$ contribute  to $J=0$,  and to $J=1$,  whereas the helicity channels with opposite helicities
$(\pm \mp; \mp \pm 00)$ in 1) and $(\pm \mp; \pm \mp 00)$ in 2) contribute to $J=1$ and $J=2$.  Similarly,  for the process 3) the  helicity channels $(--; --00)$ and $(++;++00)$ contribute  to $J=0$,  and to $J=1$,  whereas the helicity channels  
$(\pm \mp; \pm \mp 00)$ contribute to $J=1$ and $J=2$.  

For the numerical evaluations in each process and in each helicity channel we have chosen seven input energy values $E_{12} ({\rm TeV})=\sqrt{s_{12}}=0.5, 1,3, 14/3, 10, 20,50$,  covering a wide range of future  collider energies.   All the following results are displayed separately for $\delta$ and $\eta$.  Whenever we give results for one of these two parameters,  the other parameter is set to zero,  for simplicity.  We report on our findings for the various $J$ cases in the following.

\begin{figure}[h!]
\hspace{-0.5cm}
   % \centering
  \includegraphics[width=1.0\textwidth]{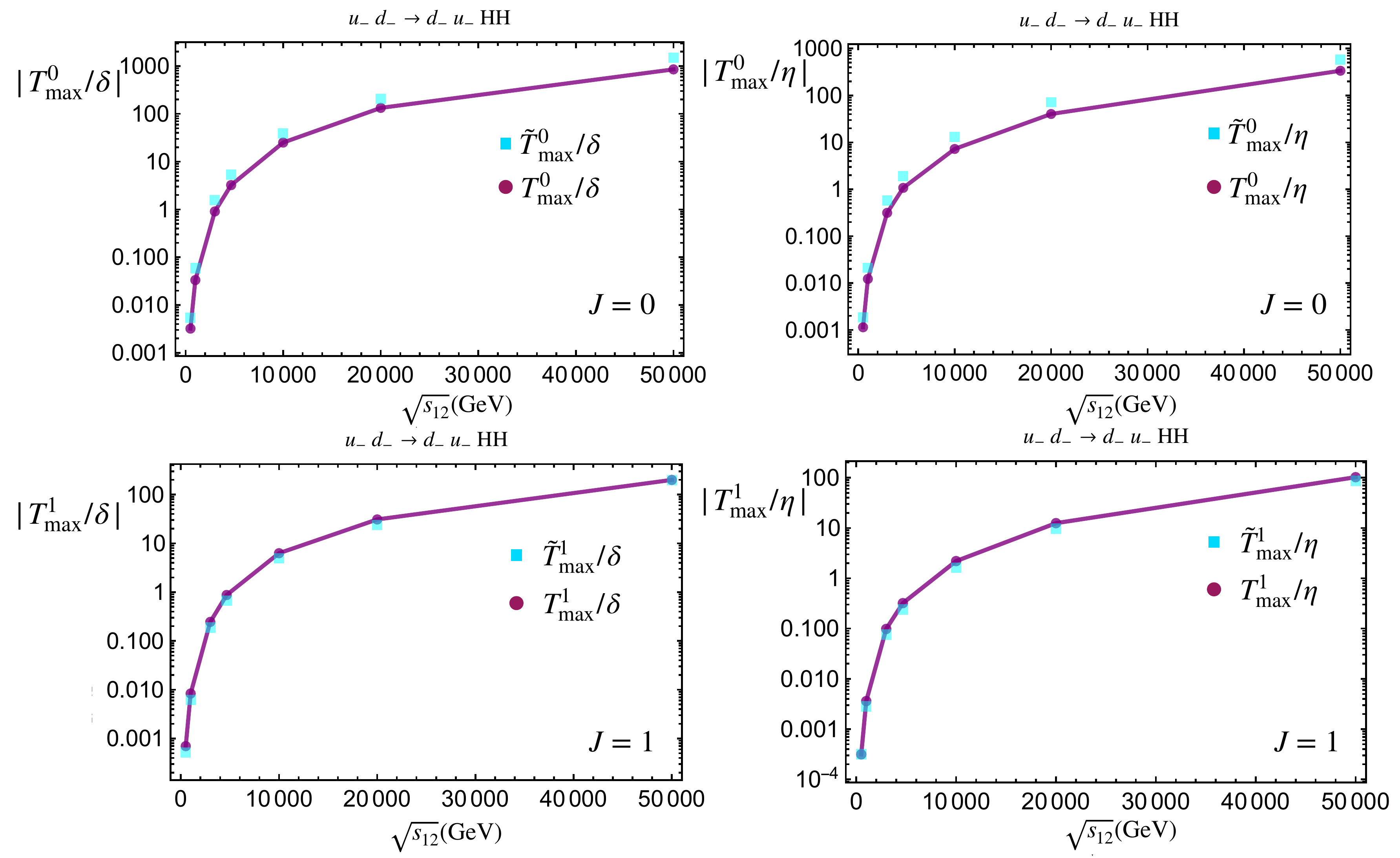} 
  \includegraphics[width=1.0\textwidth]{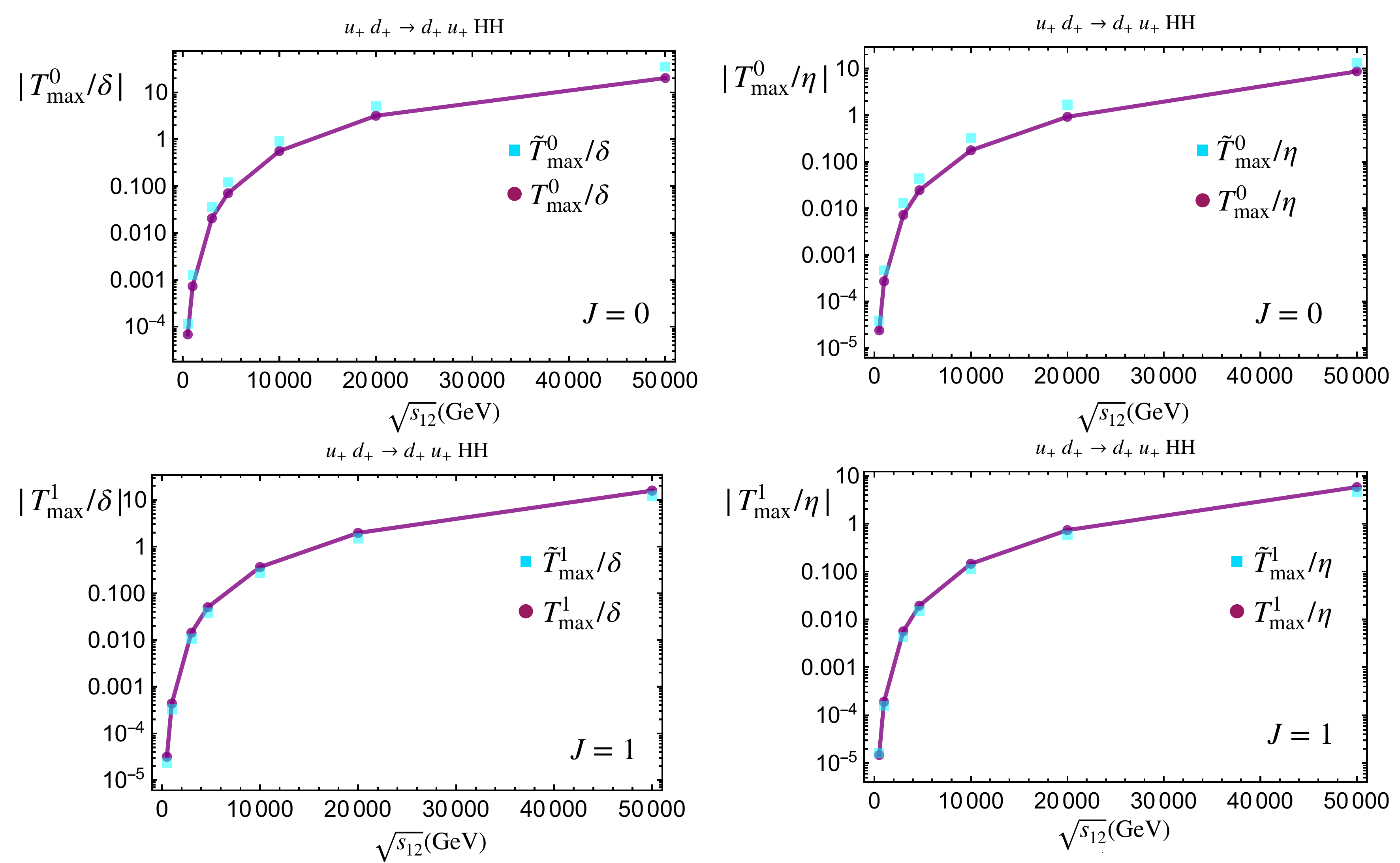} 
    \caption{Results for $|T_{\rm max}^{J}(u d \to d u HH)|$ normalized to $\delta$ (plots on the left) and to $\eta$ (plots on the right) versus the total initial energy $E_{12}= \sqrt{s_{12}}$.  The helicities are $(- - ; - - 00)$  on the two first rows and  $(+ + ; + + 00)$ in the two last rows.  The two cases  $J=0$ and $J=1$ are displayed.  The colored points are the predictions for the seven selected values of $E_{12}({\rm TeV})=0.5, 1, 3, 14/3,  10, 20,50$.   Blue squared points are for $|\tilde T^J_{\rm max}|$,  purple round points are for $|T^J_{\rm max}|$.  The lines join the $|T^J_{\rm max}|$ predictions.}
\label{J0J1-udtoduHH}    
\end{figure}

%\newpage

%%%%%%%%%%%
\vspace {0.3cm}
{\bf PARTIAL WAVES WITH  $J=0$}

\vspace {0.1cm}
We start reporting the results for $J=0$.  In this case,  only the helicity channels with the same sign helicites: $(- -; - -00)$ and  
$(+ +; + +00)$ contribute.  The helicity channels with opposite helicities do not contribute to $J=0$.  The numerical results for this $J=0$ case are organized as follows:

\vspace{0.5cm}
 $\tilde T^0(u d \to duHH)/\delta$   is summarized in \tabref{tab:udJ0delta} and (together with $T^0/\delta$) in \figref{J0J1-udtoduHH}.

 $\tilde T^0(u d \to duHH)/\eta$   is  summarized in \tabref{tab:udJ0eta} and (together with $T^0/\eta$) in \figref{J0J1-udtoduHH}.

 $\tilde T^0(uu \to uu HH)/\delta$  is summarized in \tabref{tab:uuJ0delta}  and (together with $T^0/\delta$) in \figref{J0J1-uutouuHH}. 

 $\tilde T^0(uu \to uu HH)/\eta$   is summarized in \tabref{tab:uuJ0eta}  and  (together with $T^0/\eta$) in \figref{J0J1-uutouuHH}. 

 $\tilde T^0(dd \to dd HH)/\delta$ is summarized in \tabref{tab:ddJ0delta}  and (together with $T^0/\delta$) in \figref{J0J1-ddtoddHH}.

$\tilde T^0(dd \to dd HH)/\eta$  is summarized in \tabref{tab:ddJ0eta}  and (together with $T^0/\eta$) in \figref{J0J1-ddtoddHH}.

\begin{figure}[h!]
\hspace{-0.5cm}
   % \centering
  \includegraphics[width=1.0\textwidth]{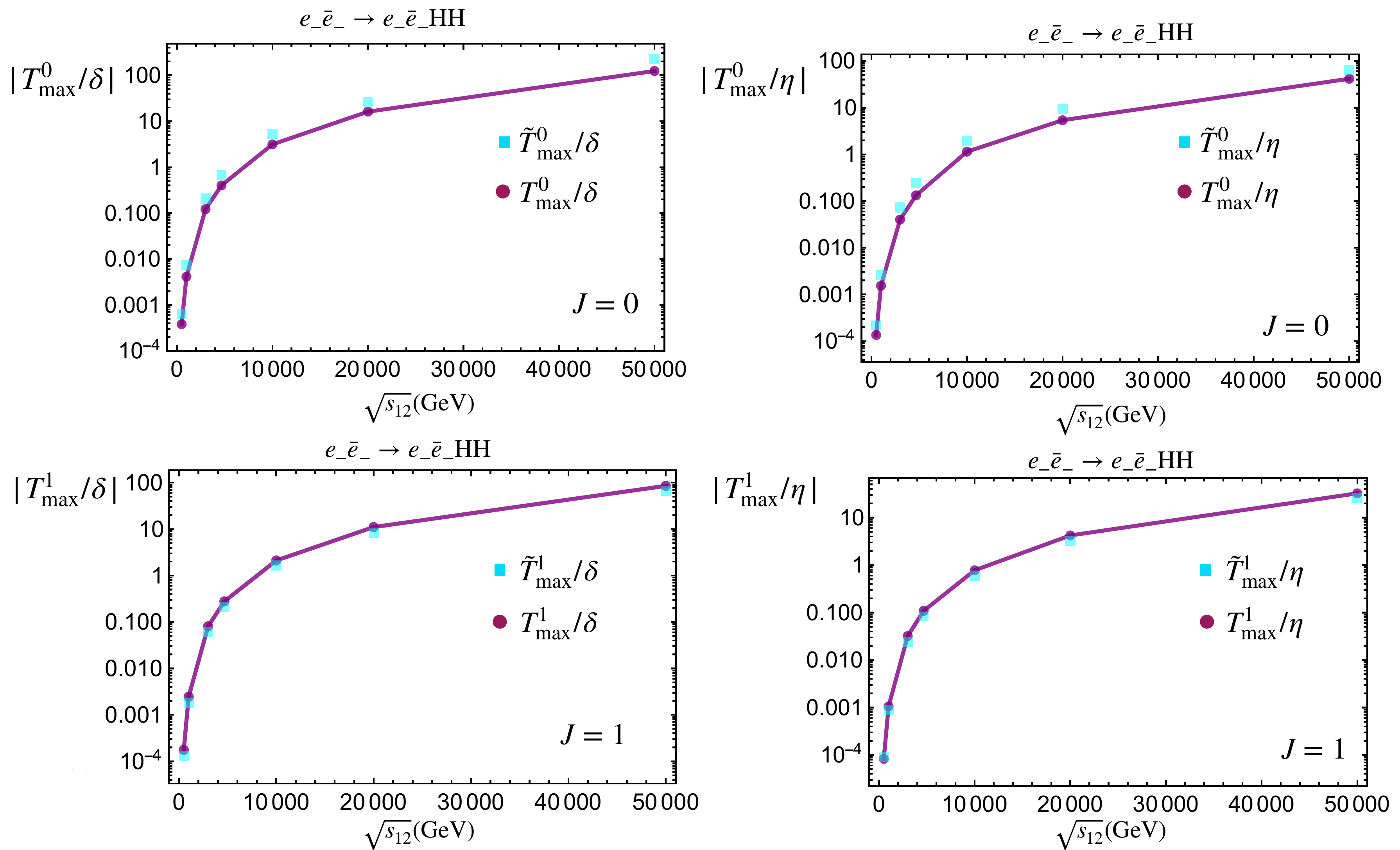} 
    \caption{Results for $|T_{\rm max}^{J}(e \bar e  \to e \bar e HH)|$ normalized to $\delta$ (plots on the left) and to $\eta$ (plots on the right) versus the total initial energy $E_{12}= \sqrt{s_{12}}$.  The helicities are $(- - ; - - 00)$.  The two cases  $J=0$ and $J=1$ are displayed.  The colored points are the predictions for the seven selected values of $E_{12}({\rm TeV})=0.5, 1, 3, 14/3,  10, 20,50$.   Blue squared points are for $|\tilde T^J_{\rm max}|$,  purple round points are for $|T^J_{\rm max}|$.  The lines join the $|T^J_{\rm max}|$ predictions.  The case of positive helicities $(+ +; + + 00)$ (not shown) is identical to the $(- - ; - - 00)$ case.}
\label{J0J1-eebtoeebHH}    
\end{figure}

$\tilde T^0(e^-e^+ \to e^- e^+ HH)/\delta$  is  summarized in \tabref{tab:eeJ0delta} and (together with $T^0/\delta$) in \figref{J0J1-eebtoeebHH}.

$\tilde T^0(e^-e^+ \to e^- e^+ HH)/\eta$ is summarized in \tabref{tab:eeJ0eta} and (together with $T^0/\eta$) in \figref{J0J1-eebtoeebHH}.

\vspace{0.5cm}
The most important features that we learn from these mentioned tables and figures,  for the $J=0$ case,  are reported in the following items. 
\begin{itemize}
\item[1.-]
For each process and  each helicity channel,  there are 7 solutions for the partial waves ${\tilde T}^{J,a}$ corresponding to $a=1,...,7$.  
The results for all ${\tilde T}^{J,a}$ are real numbers,   including positive and negative solutions.  The size of all $|{\tilde T}^{J,a}|$ 
grow with $E_{12}$.
\item[2.-]
For each quark-quark process,  each energy $E_{12}$,   and  each $a$,  the rates of ${\tilde T}^{J,a}$ for $(- -; - -00)$ are always larger than for $(+ +; + +00)$. 
Therefore the helicity channels  $(- -; - -00)$  are  more constraining than the $(+ +; + +00)$ ones.   For the electron-positron case these $(- - ;- - 00)$ and $(+ +; + +00)$  are identical,  therefore equally constraining. 
\item[3.-]
For each process,  each energy $E_{12}$,   and  each $a$,  the rates of ${\tilde T}^{J,a}/\delta$ are always larger than those of ${\tilde T}^{J,a}/\eta$.
Therefore,  the constraints for $\delta$ are always more restrictive than for $\eta$.
\item[4.-]
The process and helicity channel that provide the largest rates for the partial waves with $J=0$ is $u d \to du HH$ with helicities $(- -; - -00)$.  Therefore,  the tightest constraints for both $\delta$ and $\eta$ will come from this case.  The reason for the dominance of rates in this particular case is because  it is the only one where the WBF diagram contributes and this always dominates over the contribution from the ZBF topology.  Indeed,  for the $J=0$ case,  we have checked that the ratio $|{\tilde T}^0_{\rm WBF}/{\tilde T}^0_{\rm ZBF}|$  is within the interval (2-2.3) which can be explained by the ratio of the corresponding couplings involved $|{\cal C}_W/({\cal C}_Z g_L^u g_L^d)|$.  We have also checked that in this particular case both contributions to ${\tilde T}^0$, from WBF and ZBF,  go with the same sign,  therefore,  the final rates shown for ${\tilde T}^0$, containing WBF+ZBF,  are enhanced respect to the separate contributions from WBF and ZBF. 
\item[5.-]
In all the studied cases,  the behaviour of ${\tilde T}^{J,a}$ with the total energy, $E_{12}=\sqrt{s_{12}}$,  when  normalized to either $\delta$ or $\eta$,  can be (qualitatively and quantitatively) described by the following generic function of $s_{12}$ given by:
\begin{equation}
{\tilde T}^J(s_{12})/{\delta, \eta}= K \times \frac{s_{12}}{v^2} \times ( k_1 \log {\frac{\sqrt{s_{12}}}{m_V}}+ k_2 ), 
\label{scaling}
\end{equation}
where the coefficients $ K$,  $k_1$ and $ k_2$ are constant with energy.  Their particular values depend on the particular process,  the  specific  helicities,  the parameters $\delta$ and $\eta$,  and also on the angular momentum value $J$.  The global factor $K$ can be derived from the product of the various factors involved in the computation of ${\tilde T}^J$ as follows:
\begin{equation}
 K=G \times  S \times  \frac{1}{(2J+1)}, 
 \end{equation}
 
%\newpage

\begin{figure}[h!]
\hspace{-0.5cm}
   % \centering
  \includegraphics[width=1.0\textwidth]{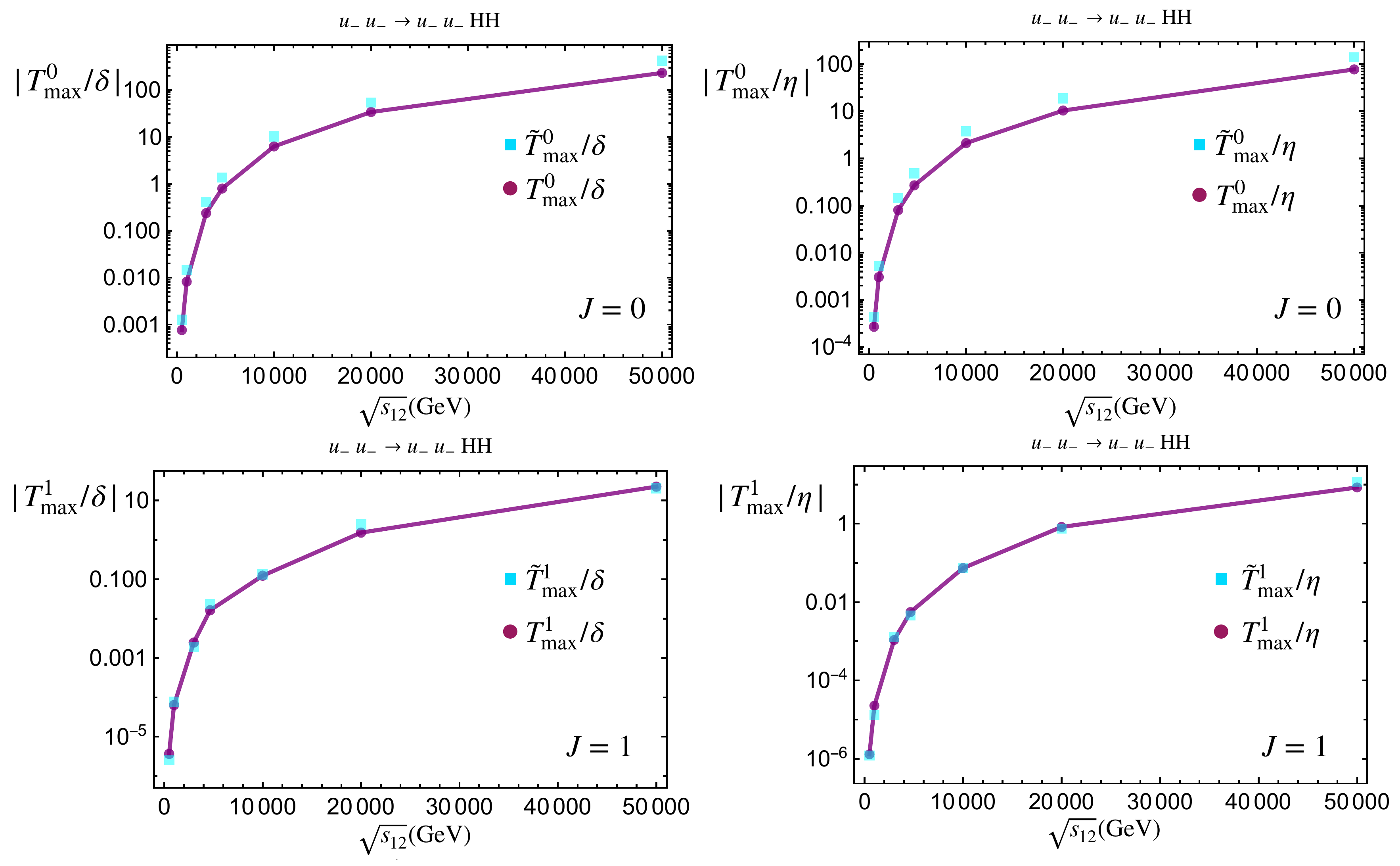} 
  \includegraphics[width=1.0\textwidth]{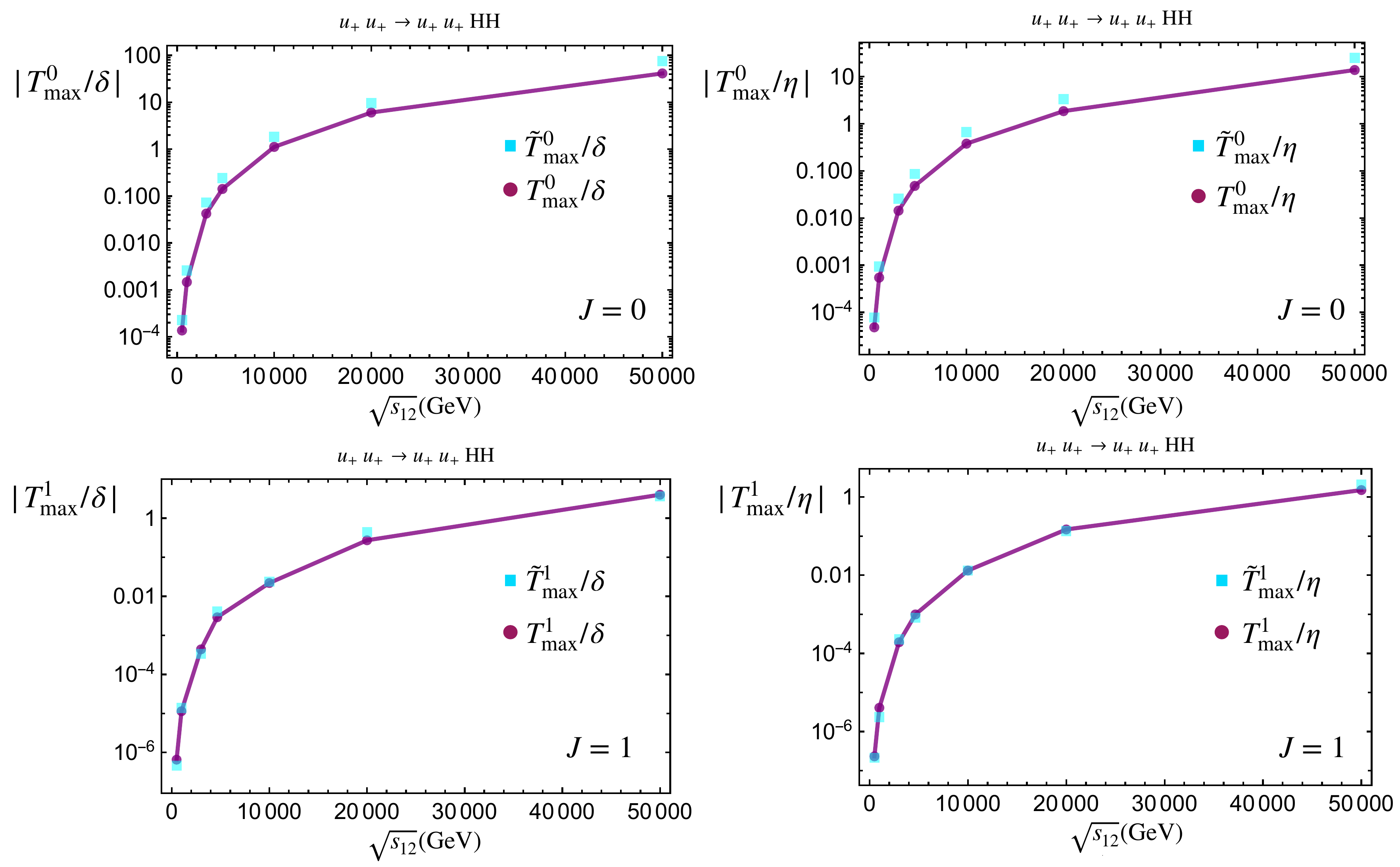} 
    \caption{Results for $|T_{\rm max}^{J}(u u \to u u HH)|$ normalized to $\delta$ (plots on the left) and to $\eta$ (plots on the right) versus the total initial energy $E_{12}= \sqrt{s_{12}}$.  The helicities are $(- - ; - - 00)$  on the two first rows and  $(+ + ; + + 00)$ in the two last rows.  The two cases  $J=0$ and $J=1$ are displayed.  The colored points are the predictions for the seven selected values of $E_{12}({\rm TeV})=0.5, 1, 3, 14/3,  10, 20,50$.   Blue squared points are for $|\tilde T^J_{\rm max}|$,  purple round points are for $|T^J_{\rm max}|$.  The lines join the $|T^J_{\rm max}|$ predictions. }
\label{J0J1-uutouuHH}    
\end{figure}

%\newpage
 
\begin{figure}[h!]
\hspace{-0.5cm}
   % \centering
  \includegraphics[width=1.0\textwidth]{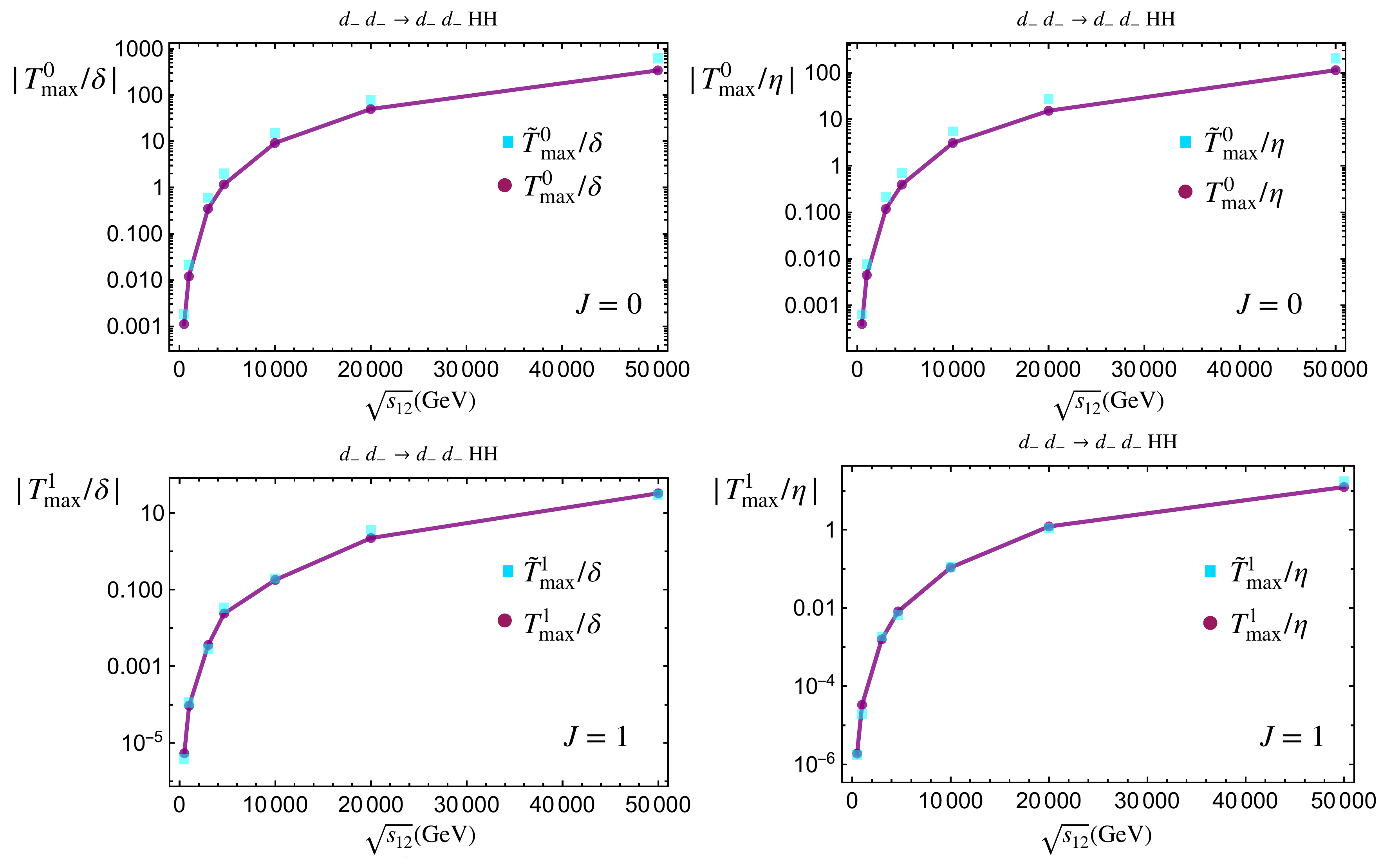} 
  \includegraphics[width=1.0\textwidth]{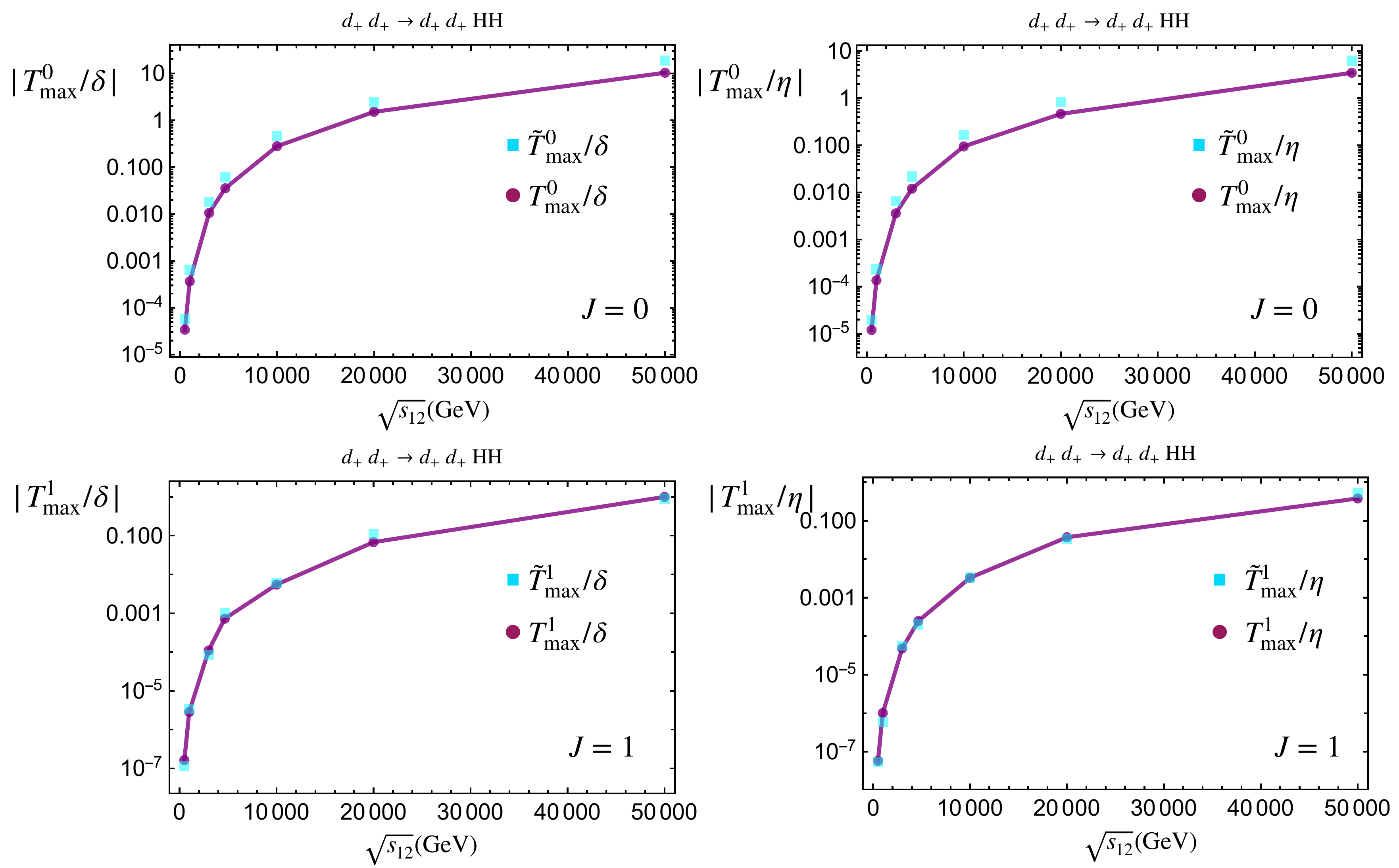} 
    \caption{Results for $|T_{\rm max}^{J}(d d \to d d HH)|$ normalized to $\delta$ (plots on the left) and to $\eta$ (plots on the right) versus the total initial energy $E_{12}= \sqrt{s_{12}}$.  The helicities are $(- - ; - - 00)$  on the two first rows and  $(+ + ; + + 00)$ in the two last rows.  The two cases  $J=0$ and $J=1$ are displayed.  The colored points are the predictions for the seven selected values of $E_{12}({\rm TeV})=0.5, 1, 3, 14/3,  10, 20,50$.   Blue squared points are for $|\tilde T^J_{\rm max}|$,  purple round points are for $|T^J_{\rm max}|$.  The lines join the $|T^J_{\rm max}|$ predictions. }
\label{J0J1-ddtoddHH}    
\end{figure}

%\newpage

%%%%%% 
where $G$ is the global factor in the helicity amplitudes,  $\cal M$,  as given in \tabref{table-Gfactors}.  The factor $S$ is a symmetry factor that must be introduced in the integrand defining the partial wave amplitudes,  and it can be derived from the products of $\frac{1}{\sqrt{N_{\rm id}!}}$ for each group of $N_{\rm id}$ identical particles, both in the initial and in the final states.  For the studied cases,  these are:
 \begin{eqnarray} 
\label{Sfactor}
S_{ud \to duHH}&=&\frac{1}{\sqrt{2}}\,\,, \,\, S_{uu \to uuHH} = S_{dd  \to dd HH} =\frac{1}{\sqrt{2}} \frac{1}{\sqrt{2}} \frac{1}{\sqrt{2}}\,\,, \,\,
S_{e \bar e \to e \bar e HH}=\frac{1}{\sqrt{2}} \,\,. 
\end{eqnarray}
The numerical values of the $ k_1$ and $k_2$  coefficients are also different  for each process and helicity channel and for  the two  cases 
${\tilde T}^J/\delta$ and ${\tilde T}^J/\eta$.
%$\frac{{\tilde T}^J(s_{12})}{\delta}$ and  $\frac{{\tilde T}^J(s_{12})}{\eta}$.  
They can be derived numerically from the results of the integration with VEGAS.

This general  behaviour with $E_{12}$ in \eqref{scaling} can be easily inferred from the scaling of the various factors involved in the integral.   
The BSM helicity amplitudes  themselves scale as $\sim s_{12}^0$,  the relevant Polynomials,  ${\cal P}^J$,  scale as $\sim s_{12}^{-1}$,  the 4-body phase space,  $\rm {dLips_4}$  scales as $\sim s_{12}^{2}$,  and the integration with the propagators of the two intermediate $VV$ bosons provides the extra logarithmic dependence, scaling as $\sim \log {\frac{\sqrt{s_{12}}}{m_V}}$,  where $m_V$ is the mass of the propagating EW gauge boson. 

\item[6.-]
Regarding the hierarchy among the size of $|{\tilde T}^J|$  for the various channels,  we find that their rates  follow roughly a similar  hierarchy as expected from the ratios for the corresponding $G$-factors.  Some examples are:
\begin{eqnarray}
|R^{\rm WBF/ZBF}_{uddu}|& \simeq & |{\cal C}_W/({\cal C}_Z (g_L^u g_L^d))| \simeq 2 \nn\\
|R^{- - - -/++++}_{uuuu}| & \simeq & |(g_L^u g_L ^u)/(g_R^u g_R ^u)| \simeq 5.6 \nn \\
|R^{- - - -/++++}_{dddd}| & \simeq & |(g_L^d g_L ^d)/(g_R^d g_R ^d)| \simeq 33 \nn \\
|R^{- - - -/++++}_{e \bar e e \bar e }| & \simeq & |(g_L^e g_R ^e)/(g_R^e g_L ^e)| = 1 \nn \\
|R^{- - - -}_{uuuu/dddd}| & \simeq &  |(g_L^u g_L ^u)/(g_L^d g_L ^d)| \simeq 0.7 \nn \\
|R^{++++}_{uuuu/dddd}| & \simeq &  |(g_R^u g_R ^u)/(g_R^d g_R ^d)| \simeq 4 
\label{ratios}
\end{eqnarray}
\item[7.-]
The main  results for the partial waves respect to the orthonormal basis,  $T^J$,  are summarized  in Figures \ref{J0J1-udtoduHH}-\ref{J0J1-eebtoeebHH}.   For shortness, instead of reporting on all the cases,  as we did in Tables \ref{tab:udJ0delta}-\ref{tab:eeJ0eta}   for $a=1,...,7$,  we now display only the values of the largest solutions for $|T^J|$ and include,  for comparison in each case,  the largest solution for $|{\tilde T}^J|$ (which in this case correspond to the $a=2$ solutions in the tables).  These figures are the most relevant ones to get the constraints on $\delta$ and $\eta$ that we are looking for.  They show clearly the announced functional behavior with energy given in \eqref{scaling},  for all the studied cases:  $|{\tilde T}^J_{\rm max}/\delta|$,  $|{T}^J_{\rm max}/\delta|$,  $|{\tilde T}^J_{\rm max}/\eta|$ and 
$|{T}^J_{\rm max}/\eta|$. 
The summary of the numerical predictions for $|{T}^0_{\rm max}|$,  in  the most relevant cases of quark-quark and electron-positron scattering with same sign helicities,  for the seven selected energies,  $E_{12}({\rm TeV})=(0.5, 1,  3, 14/3, 10, 20,50)$ are (correspondingly) as follows:

 \textcolor{white}{NECESITAMOS ESPACIO PARA QUE LAS TABLAS NO CORTEN LAS REFERENCIAS}
\vspace{200cm}
\begin{eqnarray}
|T^0_{\rm max}/\delta (u_{-}d_{-} \to u_{-} d_{-} HH)| &=&
(0.003, 0.033, 0.91, 3.23, 25.1, 133, 852)  \nn \\
|T^0_{\rm max}/\eta (u_{-}d_{-} \to u_{-} d_{-} HH)| &=&
(0.001, 0.012, 0.31, 1.07, 7.23, 40.3, 336)  \nn \\
|T^0_{\rm max}/\delta (e_{-} \bar e_{-} \to e_{-} \bar e_{-} HH)| &=& 
( < 10^{-3},  0.004,  0.12,  0.4,  3.13,  16.14, 124) \nn \\
|T^0_{\rm max}/\eta (e_{-} \bar e_{-} \to e_{-} \bar e_{-} HH)| &=&
(<10^{-3}, 0.002, 0.04,  0.13, 1.14,  5.4, 41.4)  
\end{eqnarray}
\end{itemize}
In summary,   in this $J=0$ case,  we conclude that the tightest unitarity constraints for $\delta$ and $\eta$ in the studied processes of quark-quark scattering and electron-positron scattering  can be extracted from the following inequalities,  
\begin{eqnarray}
|T^0_{\rm max} (\delta, \eta) (u_{-}d_{-} \to u_{-} d_{-} HH)| < 1 & \nn \\
|T^0_{\rm max} (\delta, \eta) (e_{-} \bar e_{-} \to e_{-} \bar e_{-} HH)| < 1 &
\end{eqnarray}

%%%%%%%%
%\newpage

%%%%%%%%
To get these inequalities in functional form we use some extrapolating functions which we have proven to provide a very good approximation to the  previous numerical results for $|T^0|$.  We assume the following specific $|T^0_{\rm max}(\delta, \eta)|$ functions,  following the announced behavior with $s_{12}$ in \eqref{scaling}:
\begin{eqnarray}
|T^0_{\rm max} (\delta, 0)(u_{-}d_{-} \to u_{-} d_{-} HH)| &= & |\delta| \,\, | K^{uddu} \frac{ s_{12}}{v^2} \Big( k^{uddu}_{1, \delta} \ln (\sqrt{s_{12}}/m_W)+
k^{uddu}_{2,\delta} \Big) |  \nn \\
|T^0_{\rm max}(0, \eta)(u_{-}d_{-} \to u_{-} d_{-} HH)| &= & |\eta| \,\,| K^{uddu} \frac{ s_{12}}{v^2} \Big( k^{uddu}_{1, \eta} \ln (\sqrt{s_{12}}/m_W)+
k^{uddu}_{2,\eta} \Big) |  \nn \\
|T^0_{\rm max}(\delta,0) (e_{-} \bar e_{-} \to e_{-} \bar e_{-} HH)| &=& |\delta|\,\, |K^{e \bar e e \bar e} \frac{ s_{12}}{v^2} \Big( k^{e \bar e e \bar e}_{1, \delta} \ln (\sqrt{s_{12}}/m_Z)+
k^{e \bar e e \bar e}_{2,\delta} \Big)|   \nn \\
|T^0_{\rm max}(0,\eta) (e_{-} \bar e_{-} \to e_{-} \bar e_{-} HH)| &=&|\eta| \,\,  |K^{e \bar e e \bar e} \frac{ s_{12}}{v^2} \Big( k^{e \bar e e \bar e}_{1, \eta} \ln (\sqrt{s_{12}}/m_Z)+
k^{e \bar e e \bar e}_{2,\eta} \Big) | \,\,,  
\label{Fit-Functions}
\end{eqnarray}
where, we have set the global normalization factors  as:
\begin{eqnarray}
|K^{uddu}|&=&|{\cal C}_W (1.5) \frac{1}{ \sqrt{2}}|=0.0484,  \nn \\ 
|K^{e\bar e e \bar e}|&=& |{\cal C}_Z (g_L^e g_R ^e) \frac{1}{\sqrt{2}} |=0.0066,  
\label{Fit-K} 
 \end{eqnarray}
 which are inspired in the corresponding couplings for $uddu$ and $e\bar e e \bar e$ respectively times the $(1/\sqrt{2})$ symmetry factor due to the  two identical final Higgs bosons.  In the $uddu$ case we have taken effectively into account the two contributions from WBF and ZBF by approximating ZBF to 0.5 WBF (justifying  the 1.5 factor),  according to our previous numerical results for the ratio WBF/ZBF.

\begin{figure}[h!]
\hspace{-0.5cm}
   % \centering
  \includegraphics[width=1.0\textwidth]{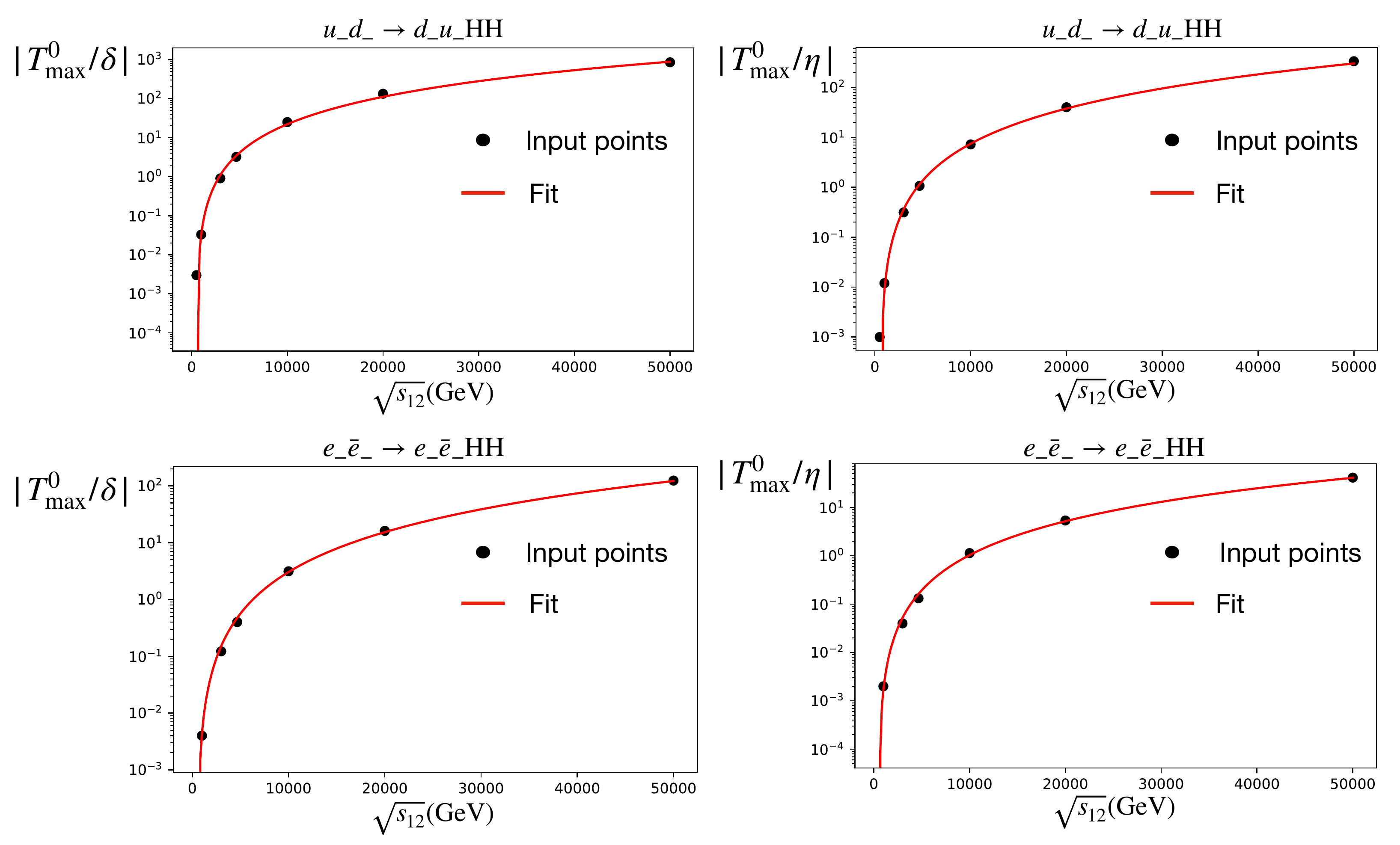} 
    \caption{Fits with the extrapolating functions in Eqs. (\ref{Fit-Functions})-(\ref{Fit-k1k2}) for $|T^{0}_{\rm max} /\delta|$ (left plots) and $|T^{0}_{\rm max} /\eta|$ (right plots) in the most constraining cases with helicities $(- -; - - 00)$.}
\label{Fits}    
\end{figure}

 Then,  from the fits of these functions to the reported  data for the seven energy values we  get finally the following values for the corresponding $k_{1,2}$: 
\begin{eqnarray}
&&k^{uddu}_{1, \delta}=0.102 \,;\, k^{uddu}_{2,\delta}=-0.203\,;\,  k^{uddu}_{1, \eta}=0.0358 \,; \, k^{uddu}_{2,\eta}= -0.0749  \nn \\
&& k^{e\bar e e \bar e}_{1, \delta}=0.103\,;\,  k^{e\bar e e \bar e}_{2, \delta}=-0.206 \,;\,   k^{e\bar e e \bar e}_{1, \eta}=0.0341 \,;\,
k^{e\bar e e \bar e}_{2, \eta}=-0.0661,  
\label{Fit-k1k2}
 \end{eqnarray}
 These extrapolating functions above,  turn out to provide very good fits, as can be seen in \figref{Fits}. 
  At this point,  we would like to comment that we have tried simpler functions,  like the one with just the logarithmic term in the parenthesis (i.e., setting $k_2=0$) and,  although we have checked that it also fits well to the high energy points,  the quality of the total fit turns to be worse than the previous one.

 \vspace {0.3cm}
{\bf PARTIAL WAVES WITH  $J=1$}

\vspace {0.1cm}
Next we report shortly on  the $J=1$ case. 
We  analyze  the same sign helicities case,  i.e.  for $(- - ; - - 00)$ and $(+ +; + + 00)$,  which can be compared with the previous $J=0$ case.  For shortness,   in this $J=1$ case,   instead of reporting on all the seven solutions for $a=1,...,7$,  we just report on the results for the largest rates of 
$|{\tilde T}^1|$ (which occur again for a=2) and $|{T}^1|$,  named  $|{\tilde T}^1_{\rm max}|$ and $|{T}^1_{\rm max}|$ respectively,   in 
Figures \ref{J0J1-udtoduHH}-\ref{J0J1-eebtoeebHH}.  The numerical results for these same sign helicities case are organized as follows:

\vspace{0.3cm} 
 $|{\tilde T}^1_{\rm max}/(\delta,\eta)|$ and $|{T}^1_{\rm max}/(\delta, \eta)|$ for $ud \to du HH$ are displayed  in \figref{J0J1-udtoduHH}.
 
 $|{\tilde T}^1_{\rm max}/(\delta,\eta)|$ and $|{T}^1_{\rm max}/(\delta, \eta)|$ for $uu \to uu HH$ are displayed  in \figref{J0J1-uutouuHH}.
 
 $|{\tilde T}^1_{\rm max}/(\delta,\eta)|$ and $|{T}^1_{\rm max}/(\eta, \delta)|$ for $dd \to dd HH$ are displayed  in \figref{J0J1-ddtoddHH}.
 
 $|{\tilde T}^1_{\rm max}/(\delta,\eta)|$ and $|{T}^1_{\rm max}/(\eta, \delta)|$ for $e \bar e \to e \bar e HH$ are displayed  in \figref{J0J1-eebtoeebHH}.

 \vspace {0.3cm} 
From these figures we learn the following:
\begin{itemize}
\item[1.-] In all cases,   the $|T^1|$  rates are always lower than the corresponding $|T^0|$ rates.  This is mainly because  of the global factor $1/(2J+1)$ which reduces the $J=1$ rates  by approximately $(1/3)$ respect to the previous $J=0$ rates.  Therefore,  the partial waves $|T^1|$ provide always milder  constraints for $\delta$ and $\eta$ than the partial waves $|T^0|$. 

\item[2.-] In all cases,  the behaviour with $s_{12}$ follow the same functional form as previously, namely,  those given  by the corresponding  logarithmic 
function,  as provided  in \eqref{scaling} with the proper $K$, $k_1$, $k_2$ factors.  We do not provide the corresponding fits in this $J=1$ case,  for shortness. 
\item[3.-] The ratios of the corresponding global $G$-factors explain approximately the hierarchy of the various $|T^1|$ rates for the different studied cases.  Notice that the ratios of the $G$-factors do not depend on the $J$ value.  Then,  the ratios among these $|T^1|$ rates follow a similar pattern as those in the previous $|T^0|$ case,  as  given in   \eqref{ratios}.  
\item[4.-]  We have not studied the $J=1$ case for the channels with opposite helicities  because they are more involved requiring the analysis of a too large system with many more polynomials ($a=1,...,13$) than in the same helicities case.  However,  by looking at the ratios of the corresponding 
$G$-factors,  and taking into account the reducing factor $(1/3)$ for $J=1$,  we expect that the corresponding $|T^J|$ rates for $J=1$ in this opposite sign helicities case will also provide milder constraints than for the dominant $|T^0|$ case.  
\end{itemize}

 \vspace {0.3cm}
{\bf PARTIAL WAVES WITH  $J=2$}

\begin{figure}[h!]
\hspace{-0.5cm}
   % \centering
  \includegraphics[width=1.0\textwidth]{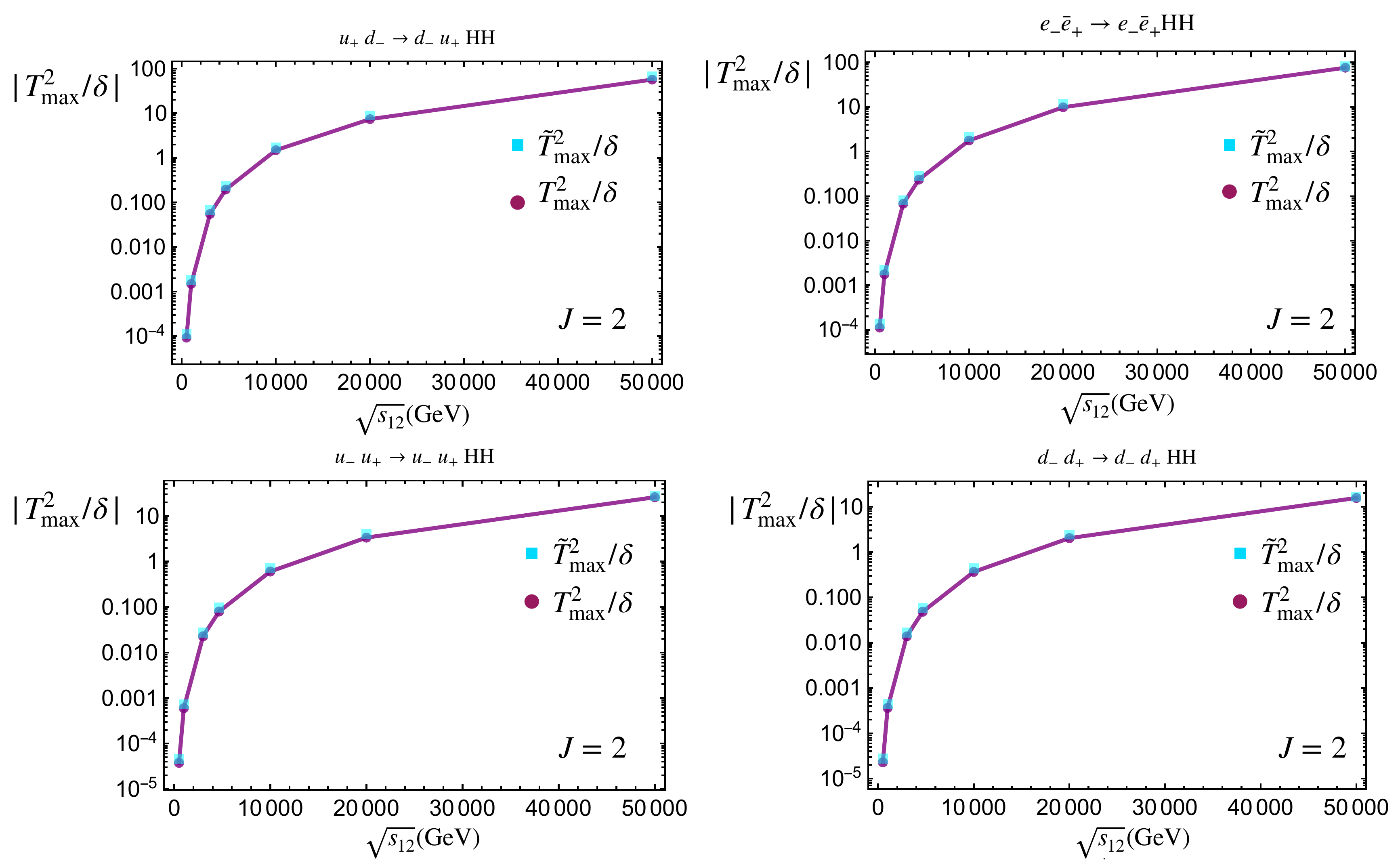} 
    \caption{Results for $|T_{\rm max}^{J}|$ with $J=2$  normalized to $\delta$ versus the total energy $E_{12}= \sqrt{s_{12}}$.  The selected processes having  helicities with opposite sign are:  $ud \to du HH$ with helicities $(-+;+ - 00)$ (upper-left plot),  $e \bar e  \to e \bar e HH$ with helicities $(- +; - +00)$ (upper-right plot),  $uu \to uu HH$ with helicities $(- +; - +00)$ (lower-left plot),  and $dd \to dd HH$ with helicities $(- +; - +00)$ (lower-right plot). The colored points are the predictions for $E_{12}({\rm TeV})=0.5, 1, 3, 14/3,  10, 20,50$.   Blue squared points are for $|\tilde T^J_{\rm max}|$,  purple round points are for $|T^J_{\rm max}|$.  The lines join the 
$|T^J_{\rm max}|$ predictions. }
\label{J2}    
\end{figure}

Next we report shortly on the $J=2$ case.   Now we have selected just the cases with the largest rates.  Specifically,  these cases correspond to  the partial waves with opposite helicities for the processes  $ud \to du HH$,  $uu \to uu HH$,  $dd \to dd HH$ and $e \bar e \to e \bar e HH$.  As in the previous $J=1$ case,  for shortness,   we only show the results for the largest rates,  named $|{\tilde T}^2_{\rm max}|$ and $|{T}^2_{\rm max}|$ respectively.  The summary of results for the largest  partial waves for $J=2$,  which correspond to $\delta$,   are summarized in  \figref{J2}. 
From this figure we learn the following:

\begin{itemize}
\item[1.-] In all cases,   the  $|{T}^2_{\rm max}|$  rates are always lower than the corresponding $|{T}^0_{\rm max}|$ rates.  This is mainly justified by the reducing factor $1/(2J+1)$  which is $1/5$ for the $J=2$ case.  
\item[2.-] Comparing the various processes and helicity channels shown in this figure we find that the following hierarchy holds in this $J=2$ case:  
$|T^2(e_{-} \bar e_{+} \to e_{-} \bar e_{+} HH)|$ $>$ 
$|T^2(u_{+} d_{-} \to d_{-} u_{+} HH)|$ $>$ $|T^2(u_{-}u_{+} \to u_{-}u_{+} HH)|$  $>$ $|T^2(d_{-}d_{+}\to d_{-}d_{+} HH)|$
\item[3.-]  We have finally confirmed for $J=2$ that  the non-VBF configuration appearing in the process $e \bar e \to e \bar e HH$ leads to negligible rates,  compared to the VBF configuration.   This non-VBF contribution  also contains the effective vertex $ZZHH$ involving $\delta$ and $\eta$ and participates via the diagram with a virtual $Z$-boson propagating in a `s-type' channel,  $e^+e^- \to Z^* \to e^+e^- HH$.   Besides,  we have also checked  the different behaviour with $s_{12}$.   This non-VBF contribution does not contain any logarithmic dependence but scales instead just linearly with $s_{12}$.  In any case it is totally irrelevant for the constraints of $\delta$ and $\eta$ we are looking for.  
\end{itemize}

Overall, in the complete analysis of the partial waves for $J=0$, $ J=1$ and $J=2$,  we find that the $|T^0_{\rm max}|$ rates for the $(- - ; - - 00)$ case clearly provide the tightest constraints on the $\delta$ and $\eta$ parameters.
These unitarity limits are  functions of  the $\sqrt{s_{12}}$ total energy,   and they can be derived easily from the condition $|T^0_{\rm max}|<1$ by using the results in  Eqs. (\ref{Fit-Functions})-(\ref{Fit-k1k2}).  They can be written jointly as follows:  
\begin{eqnarray}
|\delta| & < &  |K  \frac{ s_{12}}{v^2} \Big( k_{1, \delta} \ln (\sqrt{s_{12}}/m_W)+
k_{2,\delta} \Big)|^{-1} \\
|\eta| & < &  |K  \frac{ s_{12}}{v^2} \Big( k_{1, \eta} \ln (\sqrt{s_{12}}/m_W)+
k_{2,\eta} \Big)|^{-1}\,,\
\end{eqnarray}
where the corresponding $K$ and $k_{1,2}$ factors  for the two processes $ud \to du HH$ and $e^-e^+\to e^-e^+ HH$ are given in \eqref{Fit-K} and \eqref{Fit-k1k2}.  

From these results,  we see that for each energy $E_{12}$ the constraints on $\delta$ are always tighter than the constraints on $\eta$.  We also see that for each energy the constraints for $ud \to d u HH$ are always tighter than the constraints for $e^- e^+ \to  e^- e^+ HH$.   It should also be noticed that all the results obtained for $e^- e^+ \to  e^- e^+ HH$ apply identically for $\mu^- \mu^+ \to \mu^- \mu^+ HH$ since all the fermion masses have been neglected in the whole analysis  of unitarity in this work. 

\begin{figure}[h!]
\hspace{-0.5cm}
\includegraphics[width=0.8\textwidth]{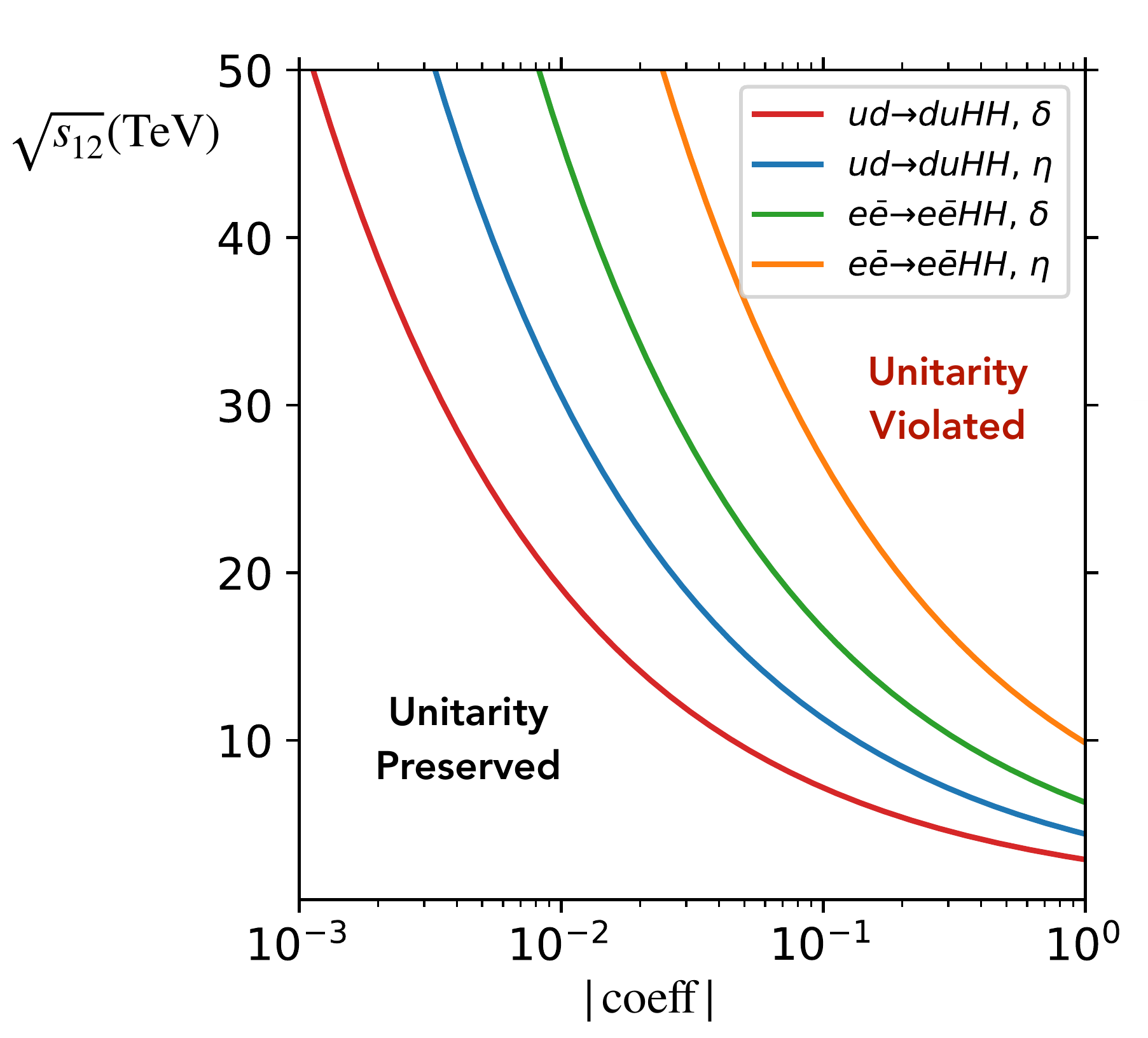}
    \caption{Most relevant unitarity limits for $f_1 f_2 \to f_3 f_4 H_5H_6$ as a function of the module of the coefficient involved and the relevant energy $\sqrt{s_{12}}=m_{f_3f_4H_5H_6}$.  Each line refer to the particular process and coefficient as specified in the legend.  The other coefficient,  not being specified in each line,  is set to zero. The region to the right above each line is where unitarity is violated.  The region to the left below each line is where unitarity is preserved}
\label{s12versuscoeff}    
\end{figure}

The summary of all these most relevant unitarity limits for the most constraining VBF mediated $f_1 f_2 \to f_3 f_4 H_5H_6$ processes is  represented graphically  in \figref{s12versuscoeff}.   In this case,  the constraints are on  the $(\sqrt{s_{12}},  |{\rm coeff}|)$ plane, with  $ \sqrt{s_{12}}$ the total initial energy and 
$|{\rm coeff}|$ the module of the coefficient being studied, i.e.  either $|\delta|$ (setting $\eta=0$) or 
$|\eta|$ (setting $\delta=0$).   Only perturbative values of the coefficients,  i.e.  with $|{\rm coeff}|<1$ are shown.  There are four lines,  two for $ud \to d u HH$ (one for $\delta$ in red,  the other one for 
$\eta$ in blue),   and two for  $e^- e^+ \to  e^- e^+ HH$   (one for $\delta$ in green,  the other one for $\eta$ in orange).  In this figure,  each line define two areas: the area to the right and above the line is where perturbative unitarity is violated and the area to the left and below the line is where perturbative unitarity is preserved.   It is also clear that the line for $ud \to d u HH$ in the $\delta$ case  (red line) provides  the largest excluded region by unitarity limits. 
For instance,   at $E_{12}=10\, {\rm  TeV}$,  the best constraints for  $ud \to d u HH$ are  $|\delta | < 4 \times 10^{-2}$ and 
$|\eta| < 1.5 \times 10^{-1}$,  whereas  for $e^- e^+ \to  e^- e^+ HH$ these are $|\delta | < 3.5 \times 10^{-1}$ and $|\eta| < 1$ respectively. 

\begin{figure}[h!]
    \centering
    \includegraphics[width=1\textwidth]{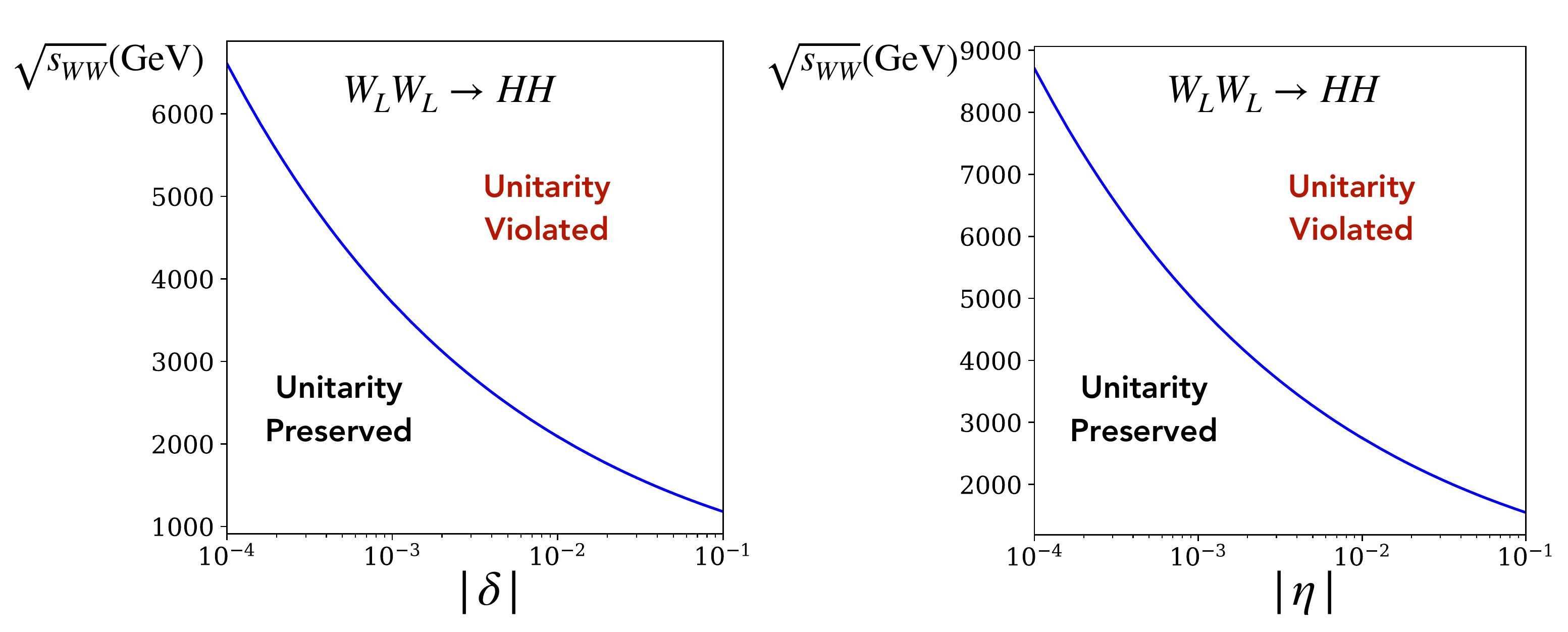}
    \caption{Most relevant unitarity limits for $W W  \to HH$,  coming from $W_L W_L \to HH$,  as a function of the module of the coefficient involved and the relevant energy $\sqrt{s_{WW}}=m_{HH}$.  Each line refer to the particular coefficient specified in the legend.  The other coefficient,  not being specified in each line,  is set to zero. The region to the right above each line is where unitarity is violated.  The region to the left above each line is where unitarity is preserved.}
\label{fig:unitarity-WW2HH}
\end{figure}

Finally, it is also interesting to mention that all these unitarity limits found for VBF-mediated processes  $f_1 f_2 \to f_3 f_4 H_5H_6$ turn out to be looser than those obtained for the on-shell Vector Boson Scattering (VBS) processes of the $VV \to HH$ type.  For comparison,  we show in \figref{fig:unitarity-WW2HH} the unitarity limits for the most constraining case in VBS which is $W_L W_L \to HH$.  The most relevant partial wave amplitude $a_J$ in this case (using the usual definition for the 2 $\to$ 2 case)  is for $J=0$,  where the simplified prediction is $|a_0(\delta, \eta)| =|\frac{1}{16 \pi v^4} (\delta+ \frac{1}{3} \eta) s_{WW}^2|$.     Notice that in this case the constraint  $|a_0| <1$ acts  on the $(\sqrt{s_{WW}},  |{\rm coeff}|)$ plane,  where the initial energy is $\sqrt{s_{WW}}=m_{HH}$ which is obviously different than the relevant energy for $f_1 f_2 \to f_3 f_4 H_5H_6$ which,  as said repeatedly,  is $\sqrt{s_{12}}$.  
This comparison  also indicates that,  in practice,  the proper unitarity limits to be applied for high energy $pp$ and $e^- e^+$ colliders (or,  equivalently $ \mu^- \mu^+$)  are those from  $f_1 f_2 \to f_3 f_4 H_5H_6$ that have been presented here, rather than those obtained from $VV \to HH$ since these VBS process do not really occur at colliders with on-shell $VV$ intermediate pairs.   This issue will be discussed  in the following section.

\section{Implications for  high energy  colliders}
\label{colliders}

In this section we study the most important implications for high energy colliders of our previous results on unitarity limits for the VBF-mediated processes $f_1 f_2 \to f_3 f_4 H_5H_6$.  In view of the summary in Fig. \ref{s12versuscoeff} we can already conclude that for the future planned electron-positron colliders ILC \cite{ILC:2019gyn} and CLIC \cite{Aicheler:2018arh} there are not relevant  implications of our unitarity limits since the projected energies are up to 3 TeV at most.  Neither for the ee option of the Future Circular Collider FCC-ee \cite{FCC:2018evy} with energy up to 365 GeV there will be any impact of our unitarity limits.  Regarding lepton-lepton colliders,  it is only in the International Muon Collider with projected energy 10 TeV  \cite{Casarsa:2024kct},   where there could be a small impact of our unitarity limits.  Whereas $\eta$ is not constrained at this energy of 10TeV,  the future  studies of $\delta$ at this collider will be limited by the unitarity constraint $\delta< 0.35$.

The situation is very different for high energy hadron-hadron colliders with energies above 10 TeV,  like HL-LHC,  with $ \sqrt{s}=14$ TeV (for a review of the HL-LHC Highlights,  see for instance,  \cite{ATLAS:2025eii}),  the HE-LHC with projected energy   $\sqrt{s}=27$ TeV \cite{FCC:2018bvk}, 
and for the hadron-hadron version of the Future Circular Collider FCC-hh,  with planned energy of up to $ \sqrt{s}=100$ TeV \cite{ FCC:2018vvp,FCC:2018byv}.
For the study in this section we focus on the most actual case,  the LHC collider. 
Specifically,  we focus on the predictions  of the cross sections for the process $pp \to HHjj$ with two jets and two Higgs bosons in the final state,  and where the $jj$ pairs here means the two jets that are produced from the two final quarks occurring in the partonic process of our interest here,  $q_1 q_2 \to q_3 q_4 H_5 H_6$.  To simplify,  we will not consider here real jets, i.e.  we will not do any showering nor hadronization process,  but just compute with initial and final quarks  $q_{1,2,3,4}$ (quarks here refereeing to either quarks or anti-quarks), so whenever we refer here to the final $jj$ we really mean the  $q_3 q_4$ in the final partonic state.

Regarding the predictions for the LHC (14 TeV)  we compute all the cross sections of our interest here,  by   generating events with MadGraph5 (MG5~\cite{Alwall:2014hca}),  with the following VBF selected cuts which are applied at generation level:

\begin{equation}
    p_T(j) > \text{20 GeV},\quad m_{jj} > \text{500 GeV},\quad 2<|\eta_j|<5,\quad \eta_{j_1}\cdot \eta_{j_2} < 0,\quad \Delta R_{jj} > 0.4
\label{VBF-cuts}    
\end{equation}
This selection favors the VBF configuration over other possible configurations,  typically by requiring  two forward-backward detectable jets,  localized in different hemispheres,  with large differences in rapidity and with high joint invariant mass $m_{jj}$.  

For our predictions of BSM $HHjj$ events at LHC,  we have implemented our own model-file in MG5 with the setting of the $VVHH$ effective interactions (for both cases,  $VV=WW$ and $ZZ$) as described generically in Fig. \ref{FR}, where the dependence of the predictions with $\delta$ and $\eta$ is introduced.

\begin{figure}[h!]
    \centering
        \includegraphics[width=1.0\textwidth]{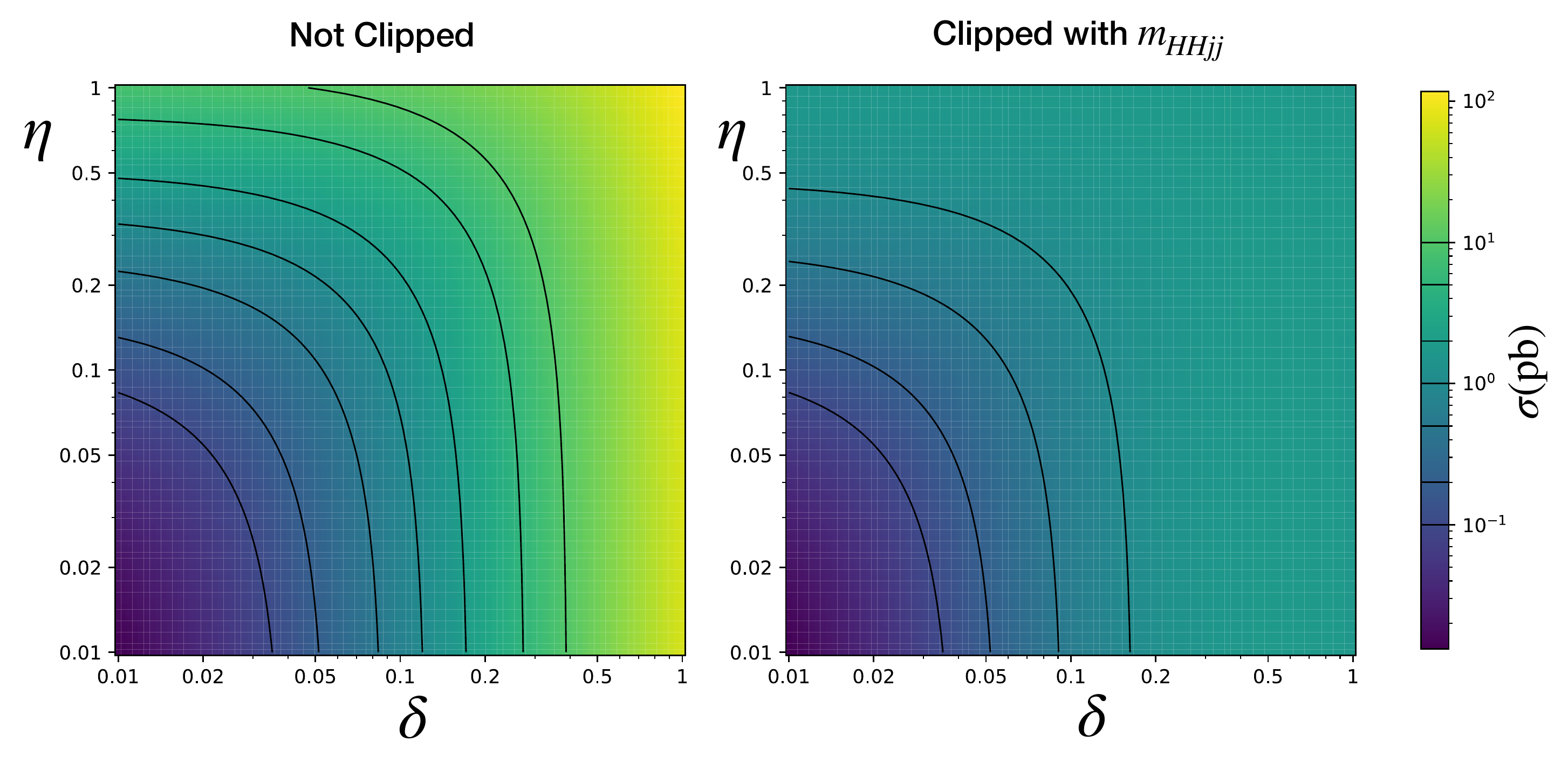}
    \caption{Predictions for  $\sigma(pp\rightarrow HHjj)({\rm pb})$ at the LHC(14TeV), with VBF cuts of \eqref{VBF-cuts}, in the 
    $(\delta, \eta)$ plane.  The plot in the left, named `Not clipped',  is with no unitarity cut-offs. The plot on the right,  named `Clipped with $m_{HHjj}$'  is with the unitarity cut-off provided by $m_{HHjj}$. }
\label{Clipped2DmHHjj}    
\end{figure}

\begin{figure}[h!]
    \centering
      \includegraphics[width=1.0\textwidth]{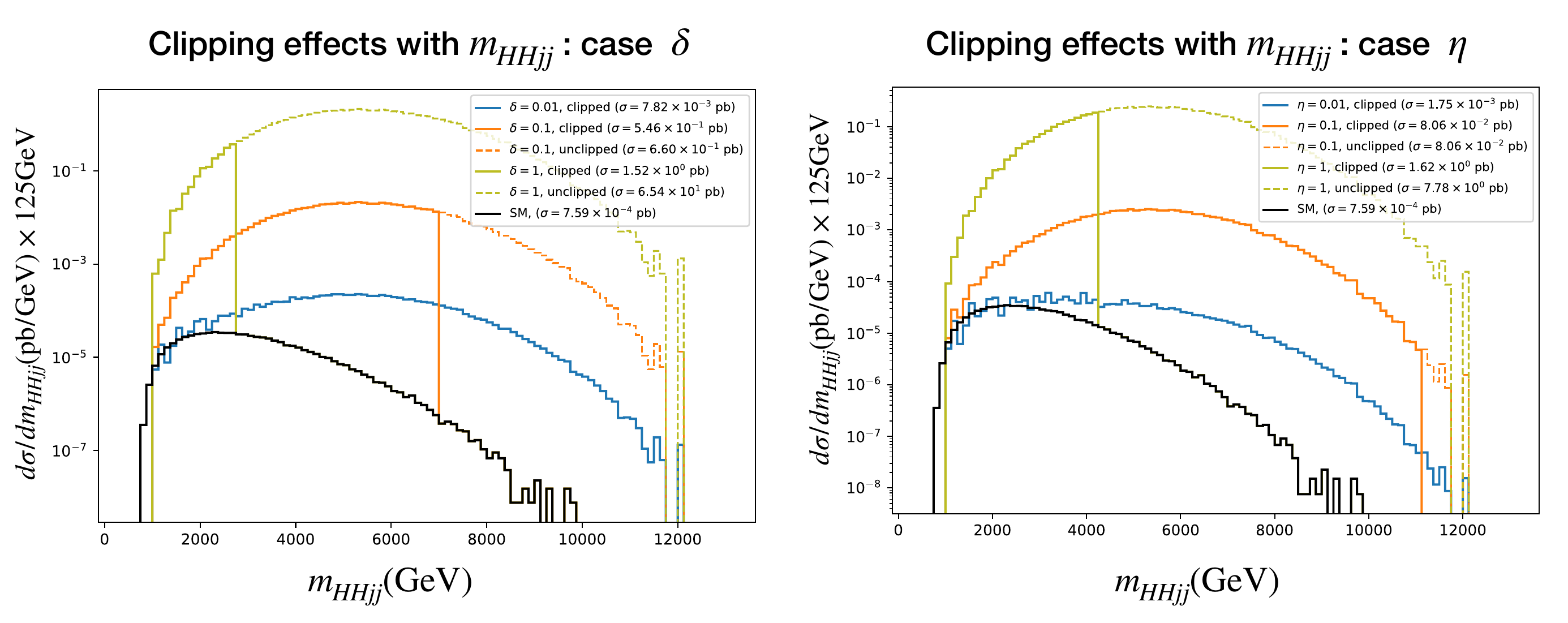}
    \caption{Predictions for the cross sections distributions 
    $d\sigma(pp\rightarrow HHjj)/dm_{HHjj}\,({\rm pb/GeV})$ at the LHC(14TeV) for several choices of $\delta$ (left plot, setting $\eta=0$) and of $\eta$ (right plot, setting $\delta=0$).  The effects of clipping in  $m_{HHjj}$  are displayed by the dashed lines.  Events  predicted in these dashed lines  are unitarity violating above $m_{HHjj}^{\rm clip}$ and must be removed for a safe prediction of the total cross section. }
   \label{ClippedHist} 
\end{figure}

\begin{figure}[h!]
    \centering
        \includegraphics[width=1.0\textwidth]{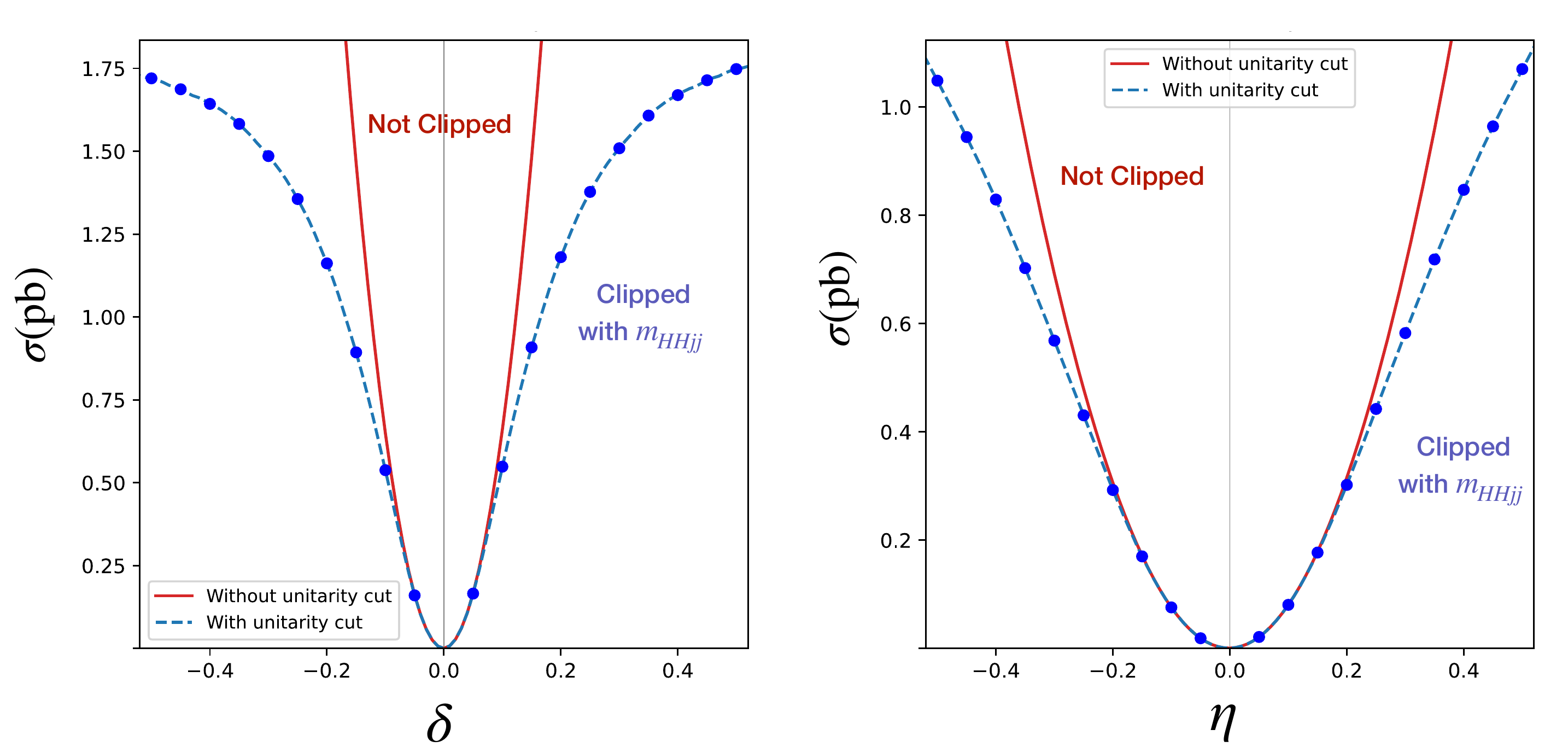}
    \caption{Predictions for  $\sigma(pp\rightarrow HHjj)({\rm pb})$ at the LHC(14TeV), with VBF cuts of \eqref{VBF-cuts}, versus 
    $\delta$ (left plot,  setting $\eta=0$) and versus $\eta$ (right plot,  setting $\delta=0$). The lines in red, named `Not Clipped',  are with no unitarity cut-offs.  The lines in blue,  named `Clipped with $m_{HHjj}$',  are with the unitarity cut-off   provided by  $m_{HHjj}$.}
\label{Clipped1DmHHjj}    
\end{figure}

\begin{figure}[h!]
    \centering
        \includegraphics[width=1.0\textwidth]{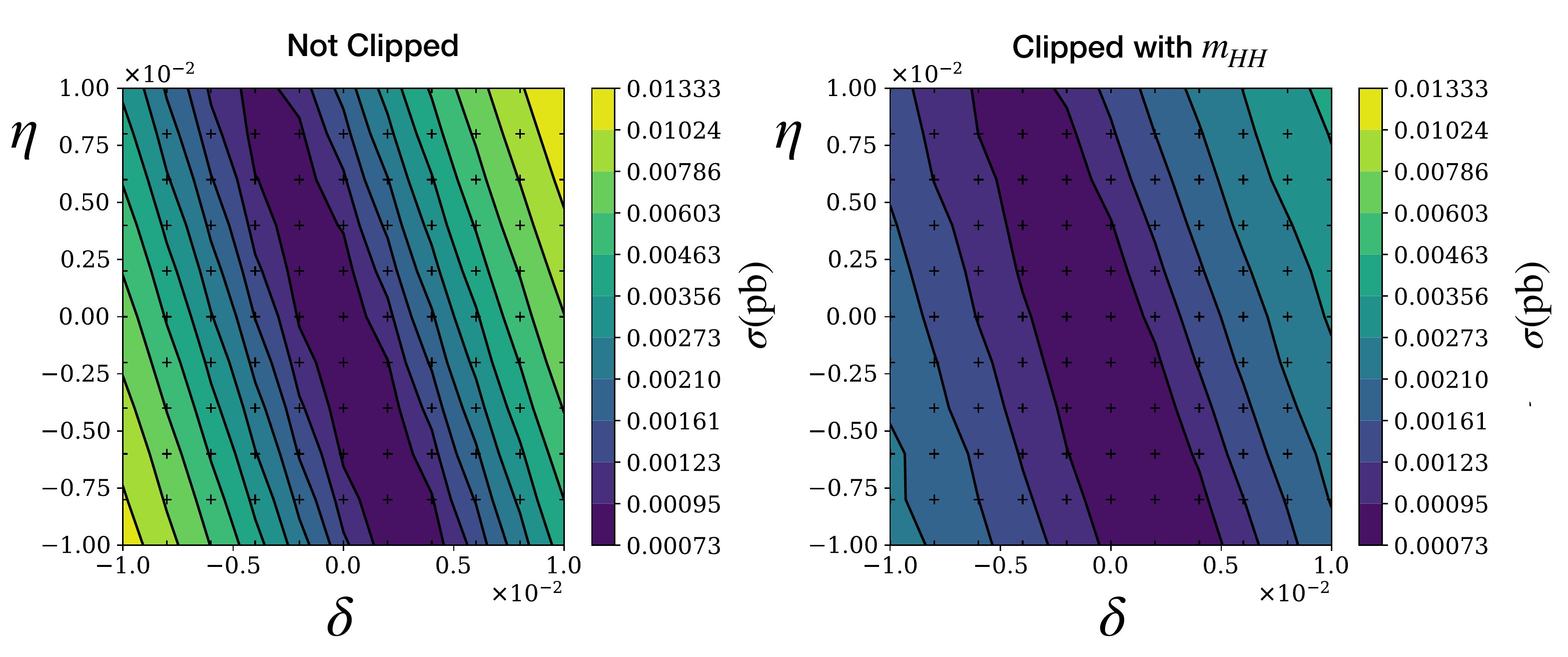}
    \caption{Predictions for  $\sigma(pp\rightarrow HHjj)({\rm pb})$ at the LHC(14TeV), with VBF cuts of \eqref{VBF-cuts}, in the 
    $(\delta, \eta)$ plane.  The plot in the left, named `Not clipped',  is with no unitarity cut-offs. The plot on the right,  named `Clipped with $m_{HH}$'  is with the unitarity cut-off provided by $m_{HH}$. }
\label{Clipped2DmHH}    
\end{figure}

\begin{figure}[h!]
    \centering
      \includegraphics[width=1.0\textwidth]{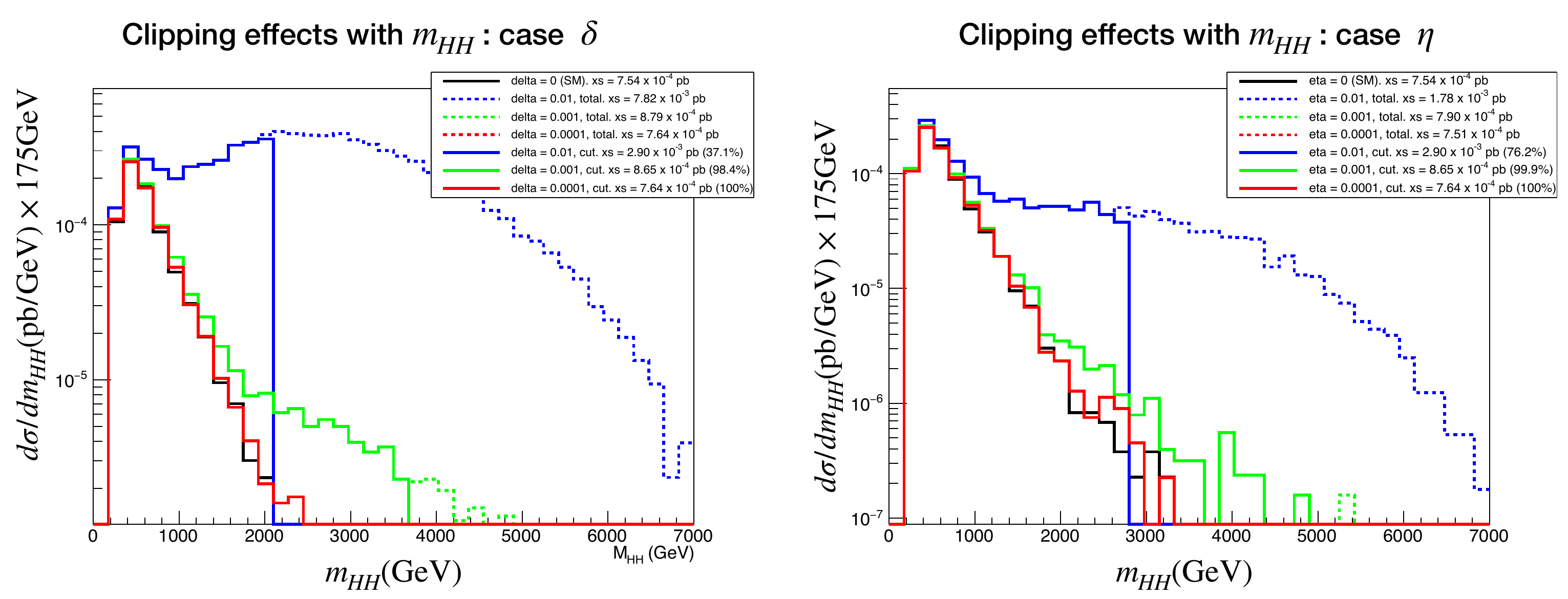}
    \caption{Predictions for the cross sections distributions 
    $d\sigma(pp\rightarrow HHjj)/dm_{HH}\,({\rm pb/GeV})$ at the LHC(14TeV) for several choices of $\delta$ (left plot, setting $\eta=0$) and of $\eta$ (right plot, setting $\delta=0$).  The effects of clipping with  $m_{HH}$  are displayed by the dashed lines.  Events  predicted in these dashed lines  are unitarity violating above $m_{HH}^{\rm clip}$ and must be removed for a safe prediction of the total cross section. }
   \label{ClippedHistmHH} 
\end{figure}

\begin{figure}[h!]
    \centering
        \includegraphics[width=1.0\textwidth]{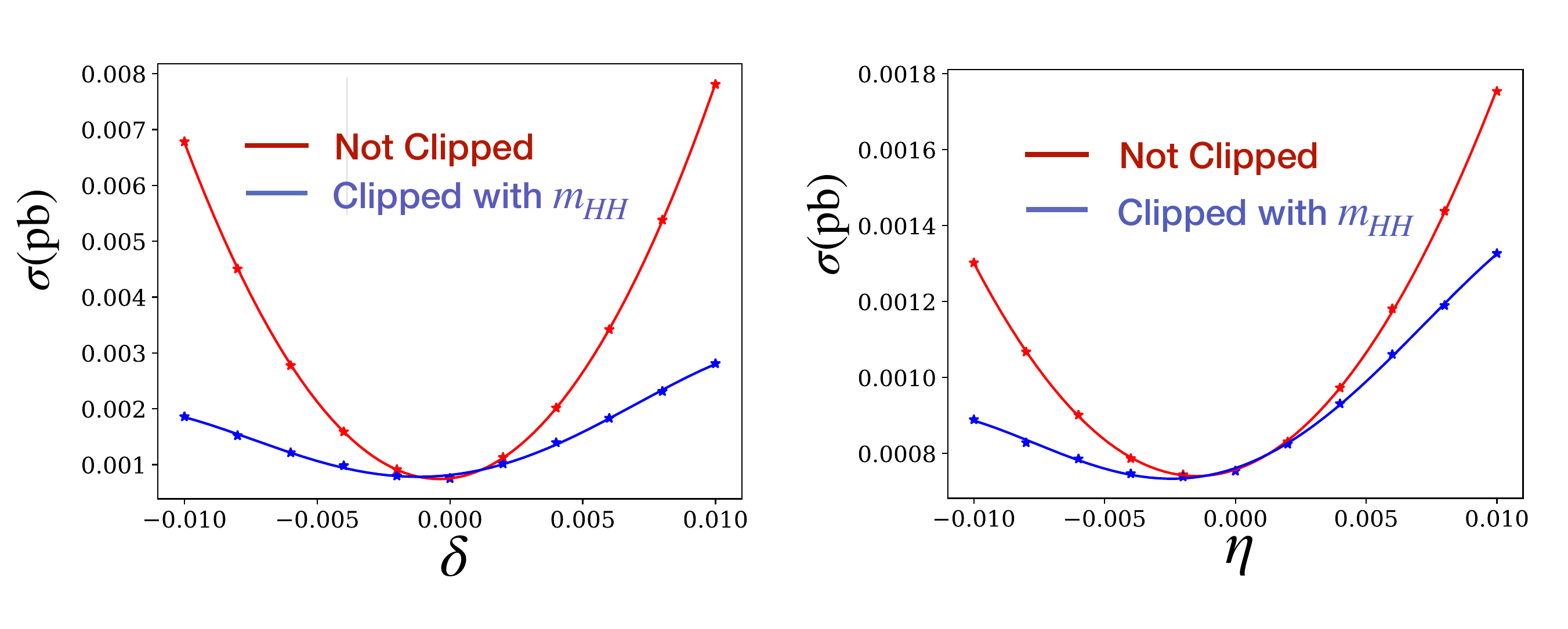}
\caption{Predictions for  $\sigma(pp\rightarrow HHjj)({\rm pb})$ at the LHC(14TeV), with VBF cuts of \eqref{VBF-cuts}, versus 
    $\delta$ (left plot,  setting $\eta=0$) and versus $\eta$ (right plot,  setting $\delta=0$). The lines in red, named `Not Clipped', are  with no unitarity cut-offs.  The lines in blue,  named `Clipped with $m_{HH}$',  are with the unitarity cut-off  provided by  $m_{HH}$.}
\label{Clipped1DmHH}    
%    \caption{Fitted total cross section: $ \sigma(\delta) = 65.408753\delta^2 + 0.052885\delta + 0.000755$ (pb). Fitted clipped cross section: $ \sigma_{clip}(\delta) -70250.827532\delta^4 - 55.726918 \delta^3 + 22.131522\delta^2 + 0.053106\delta + 0.000815$ (pb)}
\end{figure}

The results are summarized in Figures \ref{Clipped2DmHHjj} to \ref{Clipped1DmHH}.  We first apply the method of clipping with $m_{HHjj}$ as extracted from the unitarity cut-off $\sqrt{s_{12}}$,  obtained  from our study in the present work of $f_1f_2 \to f_3f_4 H_5H_6$,  and then we compare with the method of clipping with $m_{HH}$ as extracted from the unitarity cut-off 
$\sqrt{s_{WW}}$ from $WW \to HH$.   

\figref{Clipped2DmHHjj} shows the predictions for  $\sigma(pp\rightarrow HHjj)$ at the LHC, with VBF cuts of \eqref{VBF-cuts}, in the $(\delta, \eta)$ plane.  It displays the differences between the rates with no clipping method applied (therefore unitarity violating rates) and the unitarity safe rates with the clipping method applied to the variable $m_{HHjj}$. 

\figref{ClippedHist} shows the histograms of the cross sections distributions 
 $d\sigma(pp\rightarrow HHjj)/dm_{HHjj}$ for several choices of $\delta$ and $\eta$ parameters.  It displays the effects on the rates when using the clipping method in the variable  $m_{HHjj}$.
 
\figref{Clipped1DmHHjj} shows the effects on the total cross section of clipping with $m_{HHjj}$ in one dimensional plots, i.e.  with just one parameter at each time. 
\vspace{1mm}

For the comparison with these results above we also include here the results for the case of clipping with the variable $m_{HH}$ as extracted from the unitarity-cut off in $W_LW_L \to HH$ scattering. 
 
\figref{Clipped2DmHH} shows the predictions for  $\sigma(pp\rightarrow HHjj)$ at the LHC, with VBF cuts of \eqref{VBF-cuts}, in the $(\delta, \eta)$ plane.  It displays the differences between the rates with no clipping method applied (therefore unitarity violating rates) and the unitarity safe rates with the clipping method applied to the variable $m_{HH}$.  Notice that the scale in the coefficients values is smaller than in the previous case.  

\figref{ClippedHistmHH} shows the histograms of the cross sections distributions 
 $d\sigma(pp\rightarrow HHjj)/dm_{HH}$ for several choices of $\delta$ and $\eta$ parameters.  It displays the effects on the rates when using the clipping method with the variable  $m_{HH}$.

\figref{Clipped1DmHH} shows the effects on the total cross section of clipping with $m_{HH}$ in one dimensional plots, i.e.,  with just one parameter at each time. 
\vspace{2mm}

%We present the full cross section $\sigma(\eta, \delta)$ as well as the cross section computed by integrating the differential cross section 
%$\frac{\text{d}\sigma}{\text{d}M_{HH}} (\eta, \delta, M_{HH})$ up to $M_{HH}\equiv \sqrt{\hat{s}} = \sqrt{\hat{s}_{max}}(\eta, \delta)$ obtained %with the loose criteria.

\textcolor{white}{NECESITAMOS ESPACIO PARA QUE LAS TABLAS NO CORTEN LAS REFERENCIAS}
\vspace{200cm}

\textcolor{white}{NECESITAMOS ESPACIO PARA QUE LAS TABLAS NO CORTEN LAS REFERENCIAS}
\vspace{200cm}

\textcolor{white}{NECESITAMOS ESPACIO PARA QUE LAS TABLAS NO CORTEN LAS REFERENCIAS}
\vspace{200cm}

From all these figures we learn the following: 
\begin{itemize}
\item[1.-] The sensitivity at LHC to the parameters $\delta$ and $\eta$ gets reduced when a clipping method to ensure unitarity safe predictions is applied.  By reduced sensitivity here we mean that the cross sections  predicted for a given value of 
$\delta$ or $\eta$ get lower values when applying a unitarization method, i.e.  when applying  a clipping method.  Therefore the constraints on $\delta$ and $\eta$ (presented usually as CL allowed intervals for these parameters) that will be presumably derived from the comparison of these predictions with real data will be weaker (i.e.  wider allowed intervals) than without using any unitarity cut-off.  

This reduction in the sensitivity to $\delta$ and $\eta$ when applying a clipping method can be seen clearly in Figs. \ref{Clipped1DmHHjj} and \ref{Clipped1DmHH} by noticing that the curves in blue with a unitarity cut-off set in (i.e.  with a clipping method applied) get the two branches (for positive and negative coefficients, respectively) more open and lower  than the curves in red where no unitarity cut-off is set in (i.e with no clippling method applied).  
%Concretely, for $\sigma=1.8$ pb in the left plot of \figref{Clipped1DmHHjj}, the maximum allowed $|\delta|$ value is 0.1 without unitarity cut but 0.2 clipping with $m_{HHjj}$. Regarding left plot of \figref{Clipped1DmHH} for $\sigma=1.8\cdot10^{-3}$ pb, the allowed range is (-0.004,0.003) with no unitarity cut-off but it is (-0.009,0.006) clipping with $m_{HH}$.
We also see in these two figures that  the size  of the coefficients $\delta$  and  $\eta$  where the two different clipping  methods  act is different.  The effects of the cut-off by  $m_{HHjj}^{\rm clip}$ start being noticed  at larger values of the coefficients,  say in the range $0.01-0.1$,  whereas the effects of the cut-off by $m_{HH}^{\rm clip}$ start being noticed already at lower values of the coefficients,  say in the range $0.001-0.01$.  

\item[2.-] These different range values of the coefficients where the two clippling methods start acting can also be seen in Figs. \ref{ClippedHist} and   \ref{ClippedHistmHH}.  The dashed lines in these two histograms  represent the areas where unitarity is violated and where the predictions must be removed for a unitarity-safe computation of the total cross section.  The values of the clipping invariant mass $m^{\rm clip}$ depend on the values of the coefficients and are precisely at the points where the solid lines change into dashed lines.  In the case of  $m_{HH}$ we see that values of the coefficients larger than around $0.001$  are affected,  whereas in the case of $m_{HHjj}$  only  values  larger than around $0.01$ are affected.  This implies that for values below $0.01$ the analysis with real data using the cut-off with $m_{HHjj}^{\rm clip}$ will not change the future constraints for these parameters respect to the studies with no unitarity cut-off.  However, for the case of using the cut-off with  $m_{HH}^{\rm clip}$ it will change the future constraints on the parameters providing wider allowed intervals than without the unitarity cut-off.

\item[3.-]  The sensitivity to $\delta$ and $\eta$ is generically larger in $WW \to HH$ scattering than in $q_1 q_2 \to q_3 q_4 HH$ scattering,  basically due to the different dependence with the relevant energy of the amplitude involved in each case,  which is $s_{WW}^2$  versus  $s_{12}\, {\rm log} s_{12}$,  respectively.   However,  since the clipping method with $m_{HH}$ assumes on-shell $VV$ pairs in the intermediate states,  which is not really happening at LHC,  for a realistic future study of these parameters confronting real data,  we recommend to use instead the more proper clipping method using the variable $m_{HHjj}$.  
\end{itemize}
 
%\newpage

%%%%%%%%%%%%%%
\section{Conclusions}
\label{conclu}
%\textcolor{red}{Put all solutions together on one equation and all correlations together in another equation.}
%\newpage
In the present work we have analyzed in a detailed and precise way the unitarity limits on double Higgs production from VBF mediated channels of the type  $f_1 f_2 \to f_3 f_4 H_5H_6$,  with  fermions $f_{1,2}$  being either quarks or leptons.   We have focused our study on the BSM effective interactions of two EW gauge bosons and two Higgs bosons  which are involved in those scattering processes via the internal propagating EW gauge boson $VV$ pairs.  Concretely we have studied the BSM effective interactions $VVHH$ introduced from the effective operators with four derivatives,  both within the HEFT and SMEFT,  which we have jointly parametrized  by two parameters $\delta$ and $\eta$. 
The crucial point of our novel study is precisely considering that these internal $VV$ pairs are usually off-shell at high energy colliders,  so that one should  not work the unitarity issue at the subprocess $VV \to HH$ level,  but rather at the on-shell  $f_1 f_2 \to f_3 f_4 H_5H_6$ process level.   Since these processes involve 4 particles in the final state we have used the spinor helicity techniques to perform properly the analysis of the projections onto partial wave amplitudes $T^J$ with fixed total angular momentum $J$ (considering massless final particles).  We have considered in our analysis projections onto $J=0$,  $J=1$ and $J=2$. This has allowed us to follow with precision the unitarity constraints on these partial waves,  by requiring the perturbative unitarity condition $|T^J|< 1$.    

In our detailed study of unitarity limits,  we have explored the most relevant  cases for future high energy colliders,  including both hadron-hadron and lepton-lepton colliders.  For this objective,  we  have selected the most interesting cases,  with either quarks or leptons in the external legs.  In particular,  we have fully explored the following scattering on-shell processes $u d \to du HH$,  $uu \to uu HH$, $ dd \to dd HH$, 
and $e^- e^+ \to e^- e^+ HH$ (with similar results for $\mu^- \mu^+ \to \mu^- \mu^+HH$).  In the first case there are two types of VBF channels contributing,  WBF and ZBF, whereas in the other three cases only ZBF participates.  
The main outcome from our detailed analysis of the partial $T^J$ waves in the various helicity channels 
explored,   considering both possibilities having  either same sign helicities or opposite sign helicities, is the conclusion that the most constraining helicity channels are those for $J=0$ with  same sign negative helicities, i.e.  for the $(- - ; - - 00)$  helicity channels.
The summary of our best unitarity limits for these most constraining cases is displayed in Figure \ref{s12versuscoeff}.  We have also concluded that  the best  constraining channels corresponds to $ud \to du HH$ scattering where the WBF participates and the rates are consequently larger.  This concluding plot also tells clearly the two different regions in the relevant two dimensional (energy, parameter) plane.  The most constraining cases  also show that these two different regions separating the allowed-disallowed by unitarity limits areas will be most relevant for future hadron-hadron colliders than for lepton-lepton colliders.  

In the particular case of LHC at 14 TeV,  we have studied the phenomenological consequences of limiting the future data analysis from the setting of a unitarity cut-off.  For that purpose,  we have compared the two clipping methods,  one with the variable $m_{HHjj}$ and the other one with the variable $m_{HH}$.  The unitarity cut-off by means of $m_{HHjj}$ is  extracted from our detailed  analysis  of unitarity in $q_1 q_2 \to q_3 q_4 H_5 H_6$ processes.  The unitarity cut-off by means of $m_{HH}$ is extracted from $W_LW_L \to HH$ scattering.  From this study,  we have concluded  the following.  
First,  the sensitivity at LHC to the parameters $\delta$ and $\eta$ gets reduced when a clipping method is applied to ensure unitarity safe predictions.  By reduced sensitivity here we mean that the cross sections  predicted for a given value of 
$\delta$ or $\eta$ get lower values when applying a unitarization method, i.e.  when applying  a clipping method.  Therefore the constraints on $\delta$ and $\eta$ (presented usually as CL allowed intervals for these parameters) that will be presumably derived from the comparison of these predictions with real data will be weaker (i.e.  wider allowed intervals) than without using any unitarity cut-off.   Second,  our proposed clipping method with the variable $m_{HHjj}$ ensures a unitarity safe analysis of future data and,  therefore,  it will provide a safe determination of the CL allowed by data intervals for these parameters $\delta$ and $\eta$. 
 Third,  we have checked that the sensitivity to $\delta$ and $\eta$ is generically larger in $WW \to HH$ scattering than in $q_1 q_2 \to q_3 q_4 H_5H_6$ scattering,  basically due to the different dependence with the relevant energy of the amplitude involved in each case,  which is $s_{WW}^2$  versus  $s_{12}\, {\rm log} s_{12}$,  respectively.   However,  since the clipping method with $m_{HH}$ assumes on-shell $VV$ pairs in the intermediate states,  which is not really happening at LHC,  for a realistic future study of these parameters confronting real data,  we prefer  to use instead the more proper clipping method using the variable $m_{HHjj}$.  

Thus, as a general conclusion of this last part in the present work, we make the following recommendation regarding the experimental tests of BSM effective interactions given by $\delta$ and $\eta$  in $HH$ production from VBF at future colliders.  For a  unitarity safe analysis with future data at very high energy colliders, like  HL-LHC, HE-LHC, and FCC-hh  we recommend the use of a unitarization procedure with the clipping method proposed here with the variable $m_{f_3f_4HH}$  which is the proper one for on-shell $f_1 f_2 \to f_3 f_4 H_5H_6$ scattering, instead of using the variable $m_{HH}$ which is only suitable for on-shell  $VV\to HH$ scattering.

%%%%%%%%%%%%%%%%%%%%%
\section*{Acknowledgments}
D.D and M.J.H acknowledge partial financial support by the Spanish Research Agency (Agencia Estatal de Investigación) through the grant  with reference number PID2025-172338NB-I00 funded by MICIU/AEI/10.13039/501100011033 and by ERDF, EU.
% PID2022-137127NB-I00 
%funded by MCIN/AEI/10.13039/501100011033/ FEDER, UE.  
They also acknowledge financial support from the  grant IFT Centro de Excelencia Severo Ochoa with reference number CEX2020-001007-S
 funded by MCIN/AEI/10.13039/501100011033,  and from the European Union’s Horizon 2020 research and innovation programme under the Marie Sklodowska-Curie grant agreement No 860881-HIDDeN.  
 A.D.  acknowledges financial support from the project PID2025-170350NB-I00. 
 The work of R.M. is supported by CONICET and ANPCyT under project PICT-2021-00374.  ASB acknowledges the support of grants PID2023-148162NB-C21 and PID2022-137003NB-I00 from Spanish MCIN/AEI/10.13039/501100011033/ and EU FEDER.
The work of D.D.is also supported by the Spanish Ministry of Science and Innovation via an FPU grant No FPU22/03485. 

\section*{Data Availability Statement}

No Data associated in the manuscript.

%\section*{Code Availability Statement}
%
%The analysis code used in this study is available from the corresponding authors on reasonable request.

%%%%%%%%%%%%%%%%%%%%%%%%%%%%%%%%%

%%%  New Appendices (Modifications by Rober)

%%%%%%%%%%%%%%%%%%%%%%%%%%

%\newpage

%%%%%%%%%%%%%%%%%%%%%%%%%%%%%%%%%
\section*{Appendices}
\appendix
Here we collect some technical details about our computation,  including  polynomials,  transfer matrices and 4-body phase space integration.  We also collect here the numerical results for $\tilde T^{J,a} (\delta, \eta)$  in the more constraining cases and for the seven selected energies: 
$E_{12} (\rm TeV)=0.5, 1, 3, 14/3, 10, 20, 50$. 

%%%%%%   
\begin{table}[H]
    \centering
    \begin{tabular}{ccccc}
       \hline
        \hline
        \multicolumn{5}{c} {Relevant polynomials $\mathcal{P}^{J,a}$ for BSM $f_1 f_2 \to f_3 f_4 H_5 H_6$ ($J=0$ and $J=1$)}  \\
        \hline
         {($\lambda_1, \lambda_2; \lambda_3, \lambda_4, \lambda_5, \lambda_6$)} & $J$ & $a$ & $\mathcal{P}^{J,a}$  &$c_a$   \\
          ($-\frac{1}{2}, -\frac{1}{2};-\frac{1}{2}, -\frac{1}{2}, 0,0$) & 0 & 1 & $c_1$ 
          $\langle 12 \rangle \langle 56 \rangle [53][64]$ 
          & $5120\sqrt{3}\pi^3 s^{-3}$ \\
           &0 &2 &  $c_2$ $\langle 12\rangle \langle 56 \rangle [34][56]$  & $7680 \pi^3 s^{-3}$ \\
           &0 &3  &  -$c_3$ $\langle 12 \rangle \langle 53 \rangle [34][53]$  & $3840\sqrt{2}\pi^3 s^{-3}$   \\
           &0 &4 &   $c_4$ $\langle 12 \rangle \langle  54 \rangle [34][54]$  & $3840\sqrt{2}\pi^3 s^{-3}$  \\
           &0 &5 &  $c_5$ $\langle 12 \rangle  \langle 63 \rangle [34][63]$  & $3840\sqrt{2}\pi^3 s^{-3}$\\
           &0 &6 &  -$c_6$ $\langle 12 \rangle  \langle 64 \rangle [34][64]$  & $3840\sqrt{2}\pi^3 s^{-3}$ \\    
           &0 & 7 & $c_7$ $\langle 12 \rangle \langle 34 \rangle [34][34]$  & $1280\sqrt{6}\pi^3 s^{-3}$ \\   
        \hline
        ($-\frac{1}{2}, -\frac{1}{2};-\frac{1}{2}, -\frac{1}{2}, 0,0$) & 1 & 1 & 
        $2c_1$ $\langle 15 \rangle  \langle 26 \rangle [53][64]$  & $3072 \sqrt{35} \pi^3 s^{-3}$  \\
        & & &$-c_1$  $\langle 12 \rangle  \langle 56 \rangle [53][64]$ & \\
           &1 &2 &  $2c_2$ $\langle 15 \rangle  \langle 26 \rangle [34][56]$  & $2304 \sqrt{70} \pi^3 s^{-3}$  \\
        & & &$-c_2$  $\langle 12 \rangle  \langle 56 \rangle [34][56]$ & \\
           &1 &3  &  $-2c_3$ $\langle 15 \rangle \langle 23 \rangle [34][53]$  & $2304 \sqrt{35} \pi^3 s^{-3}$  \\
        & & &$+c_3$  $\langle 12 \rangle  \langle 53 \rangle [34][53]$ & \\ 
           &1 &4 & $2c_4$ $\langle 15 \rangle \langle 24 \rangle [34][54]$  & $2304 \sqrt{35} \pi^3 s^{-3}$  \\
        & & &$-c_4$  $\langle 12 \rangle  \langle 54 \rangle [34][54]$ & \\  
           &1 &5 & $-c_5$ $\langle 15 \rangle \langle 25 \rangle [53][54]$  & $4608 \sqrt{14} \pi^3 s^{-3}$  \\
           &1 &6 & $2c_6$ $ \langle 16 \rangle  \langle 23 \rangle [34][63]$  & $2304 \sqrt{35} \pi^3 s^{-3}$  \\
        & & &$-c_6$  $\langle 12 \rangle  \langle 63 \rangle [34][63]$ & \\  
           &1 & 7 & $-2c_7$ $\langle 16 \rangle \langle 24 \rangle [34][64]$  & $2304 \sqrt{35} \pi^3 s^{-3}$  \\
        & & &$+c_7$  $\langle 12 \rangle \langle 64 \rangle [34][64]$ & \\  
           &1 & 8 &  $-c_8$ $ \langle 16 \rangle \langle 26 \rangle [63][64]$  & $4608\sqrt{14}\pi^3 s^{-3}$\\
           &1 & 9 & $2c_9$ $ \langle 13 \rangle \langle 24 \rangle [34][34]$  & $2304 \sqrt{14} \pi^3 s^{-3}$  \\
        & & &$-c_9$  $\langle 12 \rangle  \langle 34 \rangle  [34][34]$ & \\   
        \hline
        \hline
 \end{tabular}
    \caption{\label{table-samehelicities} Normalized polynomials,  not yet orthogonal,  $\mathcal{P}^{J,a} $,  for the BSM case with same sign helicities,  $(- - ; - - 00)$.   The two cases $J=0$ and $J=1$ are included.  The coefficients $c_a$ fix the norm of these polynomials to one.  Here $s=(p_1+p_2)^2$,   and we assume $p_{1,2}$ incoming and $p_{3,4,5,6}$ outgoing.  Other polynomials, like $(+ + ; + + 00)$,  with $\lambda_i  \to - \lambda_i$ are obtained by the substitution $ \langle ij \rangle  \leftrightarrow [ji]$ for all $i,j$.}
    \end{table}

\newpage
\section{Appendix A}
\label{A1}  
Here we present the relevant polynomials and the transfer matrices that change,  according to \eqref{transfer},  from the ${\cal P}^{J,a}$ basis (normalized but non-orthogonal) to the 
${\cal B}^{J,a}$  orthonormal basis (normalized and orthogonal). 
The results for the $(- -; - - 0 0)$ helicity channel which  is  the most relevant one are summarized in 
\tabref{table-samehelicities}.  Notice that these matrices also apply to the $(++;++00)$ helicity channel.

The result for the transfer matrix in the $J=0$ case is: 

\begin{equation}
\mathcal{R}^{0}_{-\frac{1}{2}, -\frac{1}{2}; -\frac{1}{2}, -\frac{1}{2},0,0}= 
\left(
  \begin{array} {ccccccc} 
  0.029 & 0.063 & -0.146 & 0.144 & 0.144 & -0.146 & 0.570 \\ 
  -0.226 & -0.348 & 0.330 & -0.234 & -0.234 & 0.330 & 1.183 \\
  0. & 0. & -0.894 & 0. & 0. & 0.894 & 0. \\
  0. & 0. & 0. & -0.894 & 0.894 & 0. & 0. \\
  0.462 & 0.023 & -0.380 & -0.799 & -0.799 & -0.380 & 0.432 \\
  0.557 & 0.734 & 0.941 & 0.061 & 0.061 & 0.941 & 0.641 \\
  -1.107 & 1.391 & -0.198 & -0.694 & -0.694 & -0.198 & 0.277 \\
  \end{array}
\right)
\end{equation}
%%%%%%%%

\vspace{0.1cm}
The result for the transfer matrix in the $J=1$ case is:
  
\begin{equation}
\mathcal{R}^{1}_{-\frac{1}{2}, -\frac{1}{2}; -\frac{1}{2}, -\frac{1}{2},0,0}= 
\left(
  \begin{array} {ccccccccc} 
  -0.053 & -0.006 & 0.143 & 0.135 & 0.053 & -0.135 & -0.143 & 0.053 & 0.754 \\
  0.306 & 0.260 & -0.354 & 0.190 & -0.153 & -0.190 & 0.354 & -0.153 & 0.356 \\
  0. & 0. & 0.436 & -0.436 & 0. & -0.436 & 0.436 & 0. & 0. \\
  0.162 & -0.184 & 0.061 & -0.461 & -0.296 & 0.461 & -0.061 & -0.296 & 0.566 \\
  0. & 0. & -0.389 & -0.389 & -0.434 & -0.389 & -0.389 & 0.434 & 0. \\
  -0.569 & -0.404 & -0.812 & -0.151 & 0.172 & 0.151 & 0.812 & 0.172 & 0.613 \\
  0.236 & 0.632 & -0.060 & -0.698 & 0.604 & 0.698 & 0.060 & 0.604 & 0.440 \\
  0. & 0. & 0.648 & 0.648 & -0.725 & 0.648 & 0.648 & 0.725 & 0. \\
  -1.336 & 1.128 & 0.113 & -0.219 & -0.604 & 0.219 & -0.113 & -0.604 & 0.050 \\
  \end{array}
\right)
\end{equation}
%%%%%%%%% 

\vspace{0.1cm}
The transfer matrix in the $J=2$ case, for the four opposite sign helicity channels $(\mp\pm;\mp\pm00)$ and $(\pm\mp;\mp\pm)$, and the associated polynomials $\mathcal{P}^{J,a} $, normalized but not yet orthogonal, are collected in \eqref{Rotopphel}, Tables \ref{table-opposhelicitiesmp} and \ref{table-opposhelicitiespm}:
\begin{equation}
\mathcal{R}^{2}_{\mp\frac{1}{2}, \pm\frac{1}{2}; \mp\frac{1}{2}, \pm\frac{1}{2},0,0}=\mathcal{R}^{2}_{\pm\frac{1}{2}, \mp\frac{1}{2}; \mp\frac{1}{2}, \pm\frac{1}{2},0,0}= 
\left(
  \begin{array} {ccc} 
  0.236 & -0.236 & 0.816 \\
  0.645 & 0.645 & 0. \\
  -0.882 & 0.882 & 0.816 \\
  \end{array}
\right)
\label{Rotopphel}
\end{equation}

%%% ver PD2J2norm-pm2mp00.m y PD2J2norm-mp2mp00.m, son largos y no illuminating
\begin{table}[H]
    \centering
    \begin{tabular}{ccccc}
       \hline
        \hline
        \multicolumn{5}{c} {Relevant polynomials $\mathcal{P}^{J,a}$ for BSM $f_1 f_2 \to f_3 f_4 H_5 H_6$ ($J=2$)}  \\
        \hline
           {($\lambda_1, \lambda_2; \lambda_3, \lambda_4, \lambda_5, \lambda_6$)} & $J$ & $a$ & $\mathcal{P}^{J,a}$  &$c_a$   \\
          ($-\frac{1}{2}, \frac{1}{2}; -\frac{1}{2}, \frac{1}{2},0,0$) & 2 & 1 & $4c_1\langle14\rangle\langle25\rangle[23][25]$  & $1536\sqrt{42}\pi^3 s^{-3}$ \\
          &&& $-c_1\langle14\rangle\langle35\rangle[23][35]$ &\\
          &&& $+2c_1\langle14\rangle\langle45\rangle[24][35]$ &\\
          &&& $+c_1\langle15\rangle\langle45\rangle[25][35]$ &\\
          &&& $+c_1\langle15\rangle\langle46\rangle[25][36]$ &\\
          &&& $-2c_1\langle14\rangle\langle56\rangle[25][36]$ &\\
          &&& $-3c_1\langle14\rangle\langle45\rangle[23][45]$ &\\
          &&& $-c_1\langle14\rangle\langle56\rangle[23][56]$ &\\
           &2 &2 &  $-4c_2\langle14\rangle\langle26\rangle[23][26]$  & $1536\sqrt{42} \pi^3 s^{-3}$ \\
          &&& $+c_2\langle14\rangle\langle36\rangle[23][36]$ &\\
          &&& $-2c_2\langle14\rangle\langle46\rangle[24][36]$ &\\
          &&& $-c_2\langle15\rangle\langle46\rangle[25][36]$ &\\
          &&& $-c_2\langle14\rangle\langle56\rangle[25][36]$ &\\
          &&& $-c_2\langle16\rangle\langle46\rangle[26][36]$ &\\
          &&& $+3c_2\langle14\rangle\langle46\rangle[23][46]$ &\\
          &&& $+c_2\langle15\rangle\langle46\rangle[23][56]$ &\\
          &&& $+2c_2\langle14\rangle\langle56\rangle[23][56]$ &\\
           &2 &3  &  $-4c_3\langle14\rangle\langle24\rangle[23][24]$  & $768\sqrt{105}\pi^3 s^{-3}$   \\
          &&& $+c_3\langle14\rangle\langle34\rangle[23][34]$ &\\
          &&& $+c_3\langle14\rangle\langle45\rangle[24][35]$ &\\
          &&& $+c_3\langle14\rangle\langle46\rangle[24][36]$ &\\
          &&& $+c_3\langle14\rangle\langle45\rangle[23][45]$ &\\
          &&& $+c_3\langle14\rangle\langle46\rangle[23][46]$ &\\
        \hline
        \hline
 \end{tabular}
    \caption{\label{table-opposhelicitiesmp} Relevant normalized polynomials,  not yet orthogonal,  $\mathcal{P}^{J,a} $,  for the BSM case associated to \eqref{Rotopphel} of the opposite sign helicities channel (- + ; - + 00).  The coefficients $c_a$ fix the norm of these polynomials to one.  Here $s=(p_1+p_2)^2$,   and we assume $p_{1,2}$ incoming and $p_{3,4,5,6}$ outgoing.  Other polynomials with $\lambda_i  \to - \lambda_i$ are obtained by the substitution $ \langle ij \rangle  \leftrightarrow [ji]$ for all $i,j$.}
\end{table}

\begin{table}[H]
    \centering
    \begin{tabular}{ccccc}
       \hline
        \hline
        \multicolumn{5}{c} {Relevant polynomials   $\mathcal{P}^{J,a}$  for BSM $f_1 f_2 \to f_3 f_4 H_5 H_6$ ($J=2$, cont.)}  \\
        \hline
         {($\lambda_1, \lambda_2; \lambda_3, \lambda_4, \lambda_5, \lambda_6$)} & $J$ & $a$ & $\mathcal{P}^{J,a}$  &$c_a$   \\
          ($\frac{1}{2}, -\frac{1}{2}; -\frac{1}{2}, \frac{1}{2},0,0$) & 2 & 1 & $4c_1\langle24\rangle\langle25\rangle[13][25]$  & $1536\sqrt{42}\pi^3 s^{-3}$ \\
          &&& $-3c_1\langle24\rangle\langle35\rangle[13][35]$ &\\
          &&& $-2c_1\langle24\rangle\langle45\rangle[14][35]$ &\\
          &&& $-c_1\langle15\rangle\langle45\rangle[15][35]$ &\\
          &&& $-c_1\langle25\rangle\langle46\rangle[15][36]$ &\\
          &&& $+2c_1\langle24\rangle\langle56\rangle[15][36]$ &\\
          &&& $-c_1\langle24\rangle\langle45\rangle[13][45]$ &\\
          &&& $-3c_1\langle24\rangle\langle56\rangle[13][56]$ &\\
           &2 &2 &  $-4c_2\langle24\rangle\langle26\rangle[13][26]$  & $1536\sqrt{42} \pi^3 s^{-3}$ \\
          &&& $+3c_2\langle24\rangle\langle36\rangle[13][36]$ &\\
          &&& $+2c_2\langle24\rangle\langle46\rangle[14][36]$ &\\
          &&& $+c_2\langle25\rangle\langle46\rangle[15][36]$ &\\
          &&& $+c_2\langle24\rangle\langle56\rangle[15][36]$ &\\
          &&& $+c_2\langle26\rangle\langle46\rangle[16][36]$ &\\
          &&& $+c_2\langle24\rangle\langle46\rangle[13][46]$ &\\
          &&& $-c_2\langle25\rangle\langle46\rangle[13][56]$ &\\
          &&& $+2c_2\langle24\rangle\langle56\rangle[13][56]$ &\\
           &2 &3  &  $-4c_3\langle24\rangle\langle24\rangle[13][24]$  & $768\sqrt{105}\pi^3 s^{-3}$   \\
          &&& $+3c_3\langle24\rangle\langle34\rangle[13][34]$ &\\
          &&& $-c_3\langle24\rangle\langle45\rangle[14][35]$ &\\
          &&& $-c_3\langle24\rangle\langle46\rangle[14][36]$ &\\
          &&& $+3c_3\langle24\rangle\langle45\rangle[13][45]$ &\\
          &&& $+3c_3\langle24\rangle\langle46\rangle[13][46]$ &\\
        \hline
        \hline
 \end{tabular}
    \caption{\label{table-opposhelicitiespm} Relevant normalized polynomials,  not yet orthogonal,  $\mathcal{P}^{J,a} $,  for the BSM case associated to \eqref{Rotopphel} of the opposite sign helicities channel (+ -; - + 00).    The coefficients $c_a$ fix the norm of these polynomials to one.  Here $s=(p_1+p_2)^2$,   and we assume $p_{1,2}$ incoming and $p_{3,4,5,6}$ outgoing.  Other polynomials with $\lambda_i  \to - \lambda_i$ are obtained by the substitution $ \langle ij \rangle  \leftrightarrow [ji]$ for all $i,j$.}
\end{table}

%%%%%%

\section{Appendix B}
\label{A2}

Here we present our procedure to compute the partial wave amplitudes from the integration of non-contact amplitudes times J-basis polynomials over the 4-body phase-space.

We adopt the spherical parametrization in Eqs. (4.13)-(4.14) of Ref.~\cite{Degrande:2025uil} for the two complex-vectors $\vec{u}=(u_1,u_2,u_3,u_4)^T$ and $\vec{v}=(v_1,v_2,v_3,v_4)^T$.
They are the coefficients expanding the final spinors in terms of the initial spinors: $\vert f\rangle=u_{f-2}\vert 1\rangle+v_{f-2}\vert 2\rangle$ (with $f=3,4,5,6$).

In our case,  the massless 4-body phase-space (denoted by  $d\Phi_{\rm out}$  in section 3) is: 
\bear
&&{\rm dLips4} = (2\pi)^4d\Pi_4\delta^4(p_1+p_2-p_3-p_4-p_5-p_6) = d\theta_1 d\theta_2 d\theta_3 d\eta_1 d\eta_2 d\xi_1 d\xi_2 d\xi_3 \times \nn\\
&&\qquad  \frac{s_{12}^2}{512\pi^8} \sin^3(\eta_1) \cos (\eta_1) \sin (\eta_2) \cos (\eta_2) \sin (\theta_1) \cos (\theta_1) \sin^3(\theta_2) \cos (\theta_2) \sin^5(\theta_3) \cos (\theta_3)
\eear
where, 
\bear
d\Pi_4&=& \prod_{i=3}^{6}  \frac {d^3 p_i}{(2\pi)^3 2 E_i}, 
\eear
and the 8 integration variables are $\theta_n\in[0,\pi/2]$ ($n=1,2,3$),  $\eta_n\in[0,\pi/2]$ ($n=1,2$) and $\xi_n\in[0,2\pi]$ $(n=1,2,3)$.   Notice that these variables do not correspond to any kinematical angular variables.  Using this useful parametrization of ${\rm dLips4}$  we have checked with Mathematica the usual normalization,  i.e.,   
\bear
\int({\rm dLips4})&=& \frac{s_{12}^2}{24576 \pi^5}
\eear
For the 2-body phase space (denoted $d\Phi_{\rm in}$ in section 3) which is also need in the computation of the partial waves, we have also checked the usual normalization, i.e.,  $\int({\rm dLips2})= (1/8\pi)$. 
 
Notice also that there are additional symmetry factors $S$  that must be included in the integration process for the computation of the partial waves $T^J$,   which depend on the particular process.  These are  given in \eqref{Sfactor} for the processes of interest in this work. 

The Mathematica code provided in \cite{Degrande:2025uil} does not support amplitudes with propagators (just contact on-shell amplitudes) in the command {\tt PSIntAMPUser[amp1,amp2,n,in]}.
For our purpose, we extend the complex vectors as $\vec{u'}=(1,0,u_1,u_2,u_3,u_4)^T$ and $\vec{v'}=(0,1,v_1,v_2,v_3,v_4)^T$.  Hence all spinor products involved in the amplitudes ${\cal M}$ and polynomials $\mathcal{P}^{J,a}$ can be written in terms of the spinor products of the initial particles 1 and 2.  In this way,  the relevant Mandelstam variable in this integration process is $s_{12}$. 
\bear
\langle ij\rangle &=& {\tt ab[i,j]} = (u'_i v'_j-v'_i u'_j){\tt ab[1,2]} \nn\\
\left[ ij\right] &=& {\tt sb[i,j]} = {\tt Conjugate}[v'_i u'_j-u'_i v'_j]{\tt sb[2,1]}
\eear
In particular, $s_{ij}=-t_{ij}={\tt Abs}[u'_i v'_j-v'_i u'_j]^2s_{12}$.  In this way, the energy dependence on the $E_{12}=\sqrt{s_{12}}$ is easily tracked.

For the present work,  we performed the numerical integrations with our own  private Python code and used the VEGAS integrator~\cite{Lepage:2020tgj},  given the limited capabilities of Mathematica for numerical integration treating complicated integrands with lengthy numerators and  non-trivial denominators.  In any case,  we have also cross checked some of these integrations by VEGAS with numerical integrations by Mathematica,  getting very close results.  

%%%%%%%%%%%%%%%%%
%\newpage  
%%%%%%%%%%%

\section{Appendix C}
\label{A3}

Here we collect the numerical results for ${\tilde T^{J,a}}$ in Tables \ref{tab:udJ0delta} to \ref{tab:eeJ0eta}. 

\textcolor{white}{NECESITAMOS ESPACIO PARA QUE LAS TABLAS NO CORTEN LAS REFERENCIAS}
\vspace{200cm}
  
%%%%%%%%%%%%%
%%%%%%   
\begin{table}[!t]
    \centering
    \begin{tabular}{cccccccc}
       \hline
        \hline
        \multicolumn{8}{c} {${\tilde T}^{J=0,a}(ud \to d u HH)/\delta$,  versus $E_{12}({\rm TeV})=\sqrt{s_{12}}$.  Case: $J=0$,  for $\delta$}  \\
         \hline
         Helicities & $E_{12}=0.5$& $E_{12}=1$ &$E_{12}=3$&$E_{12}=14/3$& $E_{12}=10$& $E_{12}=20$& $E_{12}=50$ \\
         \hline
         (- -;- -00) & &  &  &  &  &  &  \\
                        a=1& -0.003 & -0.034 & -0.927 & -3.082 & -22.0 & -121 & -894 \\
                        a=2& -0.005 & -0.059 & -1.576 & -5.396 & -39.2 & -207 & -1489 \\
                        a=3& 0.002 & 0.024 & 0.632 & 2.076 & 14.3 & 76.7 & 597 \\
                        a=4& -0.002 & -0.024 & -0.623 & -2.067 & -15.1 & -78.5 & -614 \\
                        a=5& -0.002 & -0.024 & -0.633 & -2.095 & -14.9 & -73.9 & -567 \\
                        a=6& 0.002 & 0.024 & 0.639 & 2.086 & 14.2 & 75.9 & 655 \\
                        a=7& -0.002 & -0.019 & -0.551 & -1.897 & -14.0 & --74.4 & -572 \\
           \hline               
               (++;++00) & & & & & & &\\             
a=1& $<10^{-3}$ & -0.001 & -0.02 & -0.069 & -0.491 & -2.753 & -16.2 \\
a=2& $<10^{-3}$ & -0.001 & -0.036 & -0.121 & -0.901 & -5.08 & -35.4 \\
a=3& $<10^{-3}$ & 0.001 & 0.015 & 0.048 & 0.325 & 1.892 & 13.1 \\
a=4& $<10^{-3}$ & -0.001 & -0.015 & -0.049 & -0.348 & -2.004 & -16.2 \\
a=5& $<10^{-3}$ & -0.001 & -0.015 & -0.049 & -0.343 & -1.808 & -14.8 \\
a=6& $<10^{-3}$ & 0.001 & 0.014 & 0.047 & 0.327 & 1.819 & 14.3 \\
a=7& $<10^{-3}$ & $<10^{-3}$ & -0.012 & -0.041 & -0.307 & -1.809 & -14.2 \\ 
         \hline 
         \hline
 \end{tabular}
    \caption{\label{tab:udJ0delta} Results for  $u_1d_2 \to d_3u_4 H_5H_6$ with  $J=0$ for $\delta$.  WBF+ZBF in (- -;- -00) and ZBF in  (++;++00).  The channels with opposite sign helicities do not enter in $J=0$.}
    \end{table}

%%%%%%   
%\newpage
%%%%%%   
\begin{table}[!t]
    \centering
    \begin{tabular}{cccccccc}
       \hline
        \hline
    \multicolumn{8}{c} {${\tilde T}^{J=0,a}(ud \to d u HH)/\eta$,  versus $E_{12}({\rm TeV})=\sqrt{s_{12}}$.  Case: $J=0$,  for $\eta$}  \\
         \hline
         Helicities & $E_{12}=0.5$& $E_{12}=1$ &$E_{12}=3$&$E_{12}=14/3$& $E_{12}=10$& $E_{12}=20$& $E_{12}=50$ \\
         \hline
         (- -;- -00) & &  &  &  &  &  &  \\
                        a=1& -0.001 & -0.012 & -0.332 & -1.127 & -7.218 & -39.7 & -319 \\
                        a=2& -0.002 & -0.021 & -0.576 & -1.903 & -13.0 & -71.6 & -579 \\
                        a=3& 0.001 & 0.010 & 0.250 & 0.808 & 5.511 & 30.3 & 228 \\
                        a=4& -0.001 & -0.01 & -0.249 & -0.803 & -5.635 & -27.9 & -228 \\
                        a=5& -0.001 & -0.01 & -0.252 & -0.829 & -5.456 & -27.1 & -205 \\
                        a=6& 0.001 & 0.01 & 0.252 & 0.82 & 5.779 & 28.7 & 226 \\
                        a=7& -0.001 & -0.008 & -0.216 & -0.747 & -5.183 & -27.7 & -210 \\
           \hline               
               (++;++00) & & & & & & &\\             
a=1& $<10^{-3}$ & $<10^{-3}$ & -0.008 & -0.025 & -0.166 & -0.851 & -6.708 \\
a=2& $<10^{-3}$ & $<10^{-3}$ & -0.013 & -0.044 & -0.321 & -1.679 & -13.3 \\
a=3& $<10^{-3}$ & $<10^{-3}$ & 0.006 & 0.018 & 0.132 & 0.772 & 4.846 \\
a=4& $<10^{-3}$ & $<10^{-3}$ & -0.006 & -0.019 & -0.131 & -0.729 & -5.414 \\
a=5& $<10^{-3}$ & $<10^{-3}$ & -0.006 & -0.019 & -0.136 & -0.699 & -5.697 \\
a=6& $<10^{-3}$ & $<10^{-3}$ & 0.006 & 0.019 & 0.131 & 0.675 & 4.463 \\
a=7& $<10^{-3}$ & $<10^{-3}$ & -0.005 & -0.017 & -0.122 & -0.651 & -4.93 \\
         \hline 
         \hline
 \end{tabular}
    \caption{\label{tab:udJ0eta} Results for  $u_1u_2 \to u_3u_4 H_5H_6$ with  $J=0$ for $\eta$.  WBF+ZBF in (- -;- -00) and ZBF in  (++;++00).  The channels with opposite sign helicities do not enter in $J=0$.}
    \end{table} 

%\newpage
    
%%%%%%   
\begin{table}[!t]
    \centering
    \begin{tabular}{cccccccc}
       \hline
        \hline
        \multicolumn{8}{c} {${\tilde T}^{J=0,a}(uu \to u u HH)/\delta$,  versus $E_{12}({\rm TeV})=\sqrt{s_{12}}$.  Case: $J=0$,  for $\delta$}  \\
         \hline
         Helicities & $E_{12}=0.5$& $E_{12}=1$ &$E_{12}=3$&$E_{12}=14/3$& $E_{12}=10$& $E_{12}=20$& $E_{12}=50$ \\
         \hline
         (- -;- -00) & &  &  &  &  &  &  \\
a=1& -0.001 & -0.008 & -0.231 & -0.798 & -5.686 & -31.7 & -207 \\
a=2& -0.001 & -0.014 & -0.41 & -1.359 & -10.2 & -54.0 & -423 \\
a=3& 0.001 & 0.006 & 0.161 & 0.543 & 3.811 & 20.9 & 170 \\
a=4& -0.001 & -0.006 & -0.16 & -0.552 & -3.861 & -19.7 & -162 \\
a=5& -0.001 & -0.006 & -0.161 & -0.544 & -3.881 & -19.3 & -175 \\
a=6& 0.001 & 0.006 & 0.161 & 0.541 & 3.747 & 20.4 & 161 \\
a=7& $<10^{-3}$ & -0.005 & -0.14 & -0.474 & -3.464 & -20.4 & -148 \\
           \hline               
a=1& $<10^{-3}$ & -0.001 & -0.041 & -0.143 & -1.016 & -5.663 & -37.0 \\
a=2& $<10^{-3}$ & -0.003 & -0.073 & -0.243 & -1.828 & -9.652 & -75.4 \\
a=3& $<10^{-3}$ & 0.001 & 0.029 & 0.097 & 0.681 & 3.738 & 30.3 \\
a=4& $<10^{-3}$ & -0.001 & -0.029 & -0.099 & -0.69 & -3.52 & -28.9 \\
a=5& $<10^{-3}$ & -0.001 & -0.029 & -0.097 & -0.693 & -3.454 & -31.2 \\
a=6& $<10^{-3}$ & 0.001 & 0.029 & 0.097 & 0.669 & 3.655 & 28.7 \\
a=7& $<10^{-3}$ & -0.001 & -0.025 & -0.085 & -0.619 & -3.644 & -26.3 \\
         \hline 
         \hline
 \end{tabular}
    \caption{\label{tab:uuJ0delta} Results for  $u_1u_2 \to u_3u_4 H_5H_6$ with  $J=0$ for $\delta$.  Only ZBF contributes.  The channels with opposite sign helicities do not enter in $J=0$.}
    \end{table} 
%%%%%%   
%\newpage
%%%%%%   
\begin{table}[!t]
    \centering
    \begin{tabular}{cccccccc}
       \hline
        \hline
        \multicolumn{8}{c} {${\tilde T}^{J=0,a}(uu \to u u HH)/\eta$,  versus $E_{12}({\rm TeV})=\sqrt{s_{12}}$.  Case: $J=0$,  
        for $\eta$}  \\
         \hline
         Helicities & $E_{12}=0.5$& $E_{12}=1$ &$E_{12}=3$&$E_{12}=14/3$& $E_{12}=10$& $E_{12}=20$& $E_{12}=50$ \\
          \hline
         (- -;- -00) & &  &  &  &  &  &  \\
a=1& $<10^{-3}$ & -0.003 & -0.086 & -0.283 & -1.939 & -10.4 & -83 \\
a=2& $<10^{-3}$ & -0.005 & -0.144 & -0.483 & -3.747 & -18.7 & -140 \\
a=3& $<10^{-3}$ & 0.002 & 0.065 & 0.209 & 1.464 & 8.313 & 59.1 \\
a=4& $<10^{-3}$ & -0.002 & -0.064 & -0.212 & -1.506 & -7.949 & -61.9 \\
a=5& $<10^{-3}$ & -0.002 & -0.064 & -0.209 & -1.488 & -7.558 & -57.1 \\
a=6& $<10^{-3}$ & 0.002 & 0.064 & 0.216 & 1.491 & 7.861 & 57.8 \\
a=7& $<10^{-3}$ & -0.002 & -0.055 & -0.188 & -1.379 & -7.549 & -49.5 \\
           \hline               
               (++;++00) & & & & & & &\\             
a=1& $<10^{-3}$ & -0.001 & -0.015 & -0.051 & -0.346 & -1.867 & -14.7 \\
a=2& $<10^{-3}$ & -0.001 & -0.026 & -0.086 & -0.669 & -3.343 & -24.9 \\
a=3& $<10^{-3}$ & $<10^{-3}$ & 0.012 & 0.037 & 0.262 & 1.485 & 10.5 \\
a=4& $<10^{-3}$ & $<10^{-3}$ & -0.012 & -0.038 & -0.269 & -1.42 & -11.0 \\
a=5& $<10^{-3}$ & $<10^{-3}$ & -0.011 & -0.037 & -0.266 & -1.35 & -10.2 \\
a=6& $<10^{-3}$ & $<10^{-3}$ & 0.012 & 0.039 & 0.266 & 1.404 & 10.3 \\
a=7& $<10^{-3}$ & $<10^{-3}$ & -0.01 & -0.034 & -0.246 & -1.348 & -8.849 \\
         \hline 
         \hline
 \end{tabular}
    \caption{\label{tab:uuJ0eta} Results for  $u_1u_2 \to u_3u_4 H_5H_6$ with  $J=0$ for $\eta$.  Only ZBF contributes.  The channels with opposite sign helicities do not enter in $J=0$.}
    \end{table} 
%%%%%%   
%\newpage

\begin{table}[!t]
    \centering
    \begin{tabular}{cccccccc}
       \hline
        \hline
        \multicolumn{8}{c} {${\tilde T}^{J=0,a}(dd \to d d HH)/\delta$,  versus $E_{12}({\rm TeV})=\sqrt{s_{12}}$.  Case: $J=0$,  for $\delta$}  \\
         \hline
         Helicities & $E_{12}=0.5$& $E_{12}=1$ &$E_{12}=3$&$E_{12}=14/3$& $E_{12}=10$& $E_{12}=20$& $E_{12}=50$ \\
         \hline
         (- -;- -00) & &  &  &  &  &  &  \\
a=1& -0.001 & -0.012 & -0.339 & -1.17 & -8.343 & -46.5 & -304 \\
a=2& -0.002 & -0.021 & -0.602 & -1.993 & -15.014 & -79.2 & -620 \\
a=3& 0.001 & 0.009 & 0.236 & 0.796 & 5.592 & 30.7 & 250 \\
a=4& -0.001 & -0.009 & -0.235 & -0.81 & -5.666 & -28.9 & -238 \\
a=5& -0.001 & -0.009 & -0.236 & -0.798 & -5.695 & -28.3 & -257 \\
a=6& 0.001 & 0.009 & 0.236 & 0.793 & 5.497 & 30.0 & 236 \\
a=7& -0.001 & -0.007 & -0.206 & -0.695 & -5.083 & -29.9 & -217 \\
           \hline               
               (++;++00) & & & & & & &\\             
a=1& $<10^{-3}$ & $<10^{-3}$ & -0.01 & -0.036 & -0.254 & -1.416 & -9.254 \\
a=2& $<10^{-3}$ & -0.001 & -0.018 & -0.061 & -0.457 & -2.413 & -18.8 \\
a=3& $<10^{-3}$ & $<10^{-3}$ & 0.007 & 0.024 & 0.17 & 0.935 & 7.595 \\
a=4& $<10^{-3}$ & $<10^{-3}$ & -0.007 & -0.025 & -0.172 & -0.88 & -7.242 \\
a=5& $<10^{-3}$ & $<10^{-3}$ & -0.007 & -0.024 & -0.173 & -0.863 & -7.817 \\
a=6& $<10^{-3}$ & $<10^{-3}$ & 0.007 & 0.024 & 0.167 & 0.914 & 7.184 \\
a=7& $<10^{-3}$ & $<10^{-3}$ & -0.006 & -0.021 & -0.155 & -0.911 & -6.599 \\
         \hline 
         \hline
 \end{tabular}
    \caption{\label{tab:ddJ0delta} Results for  $d_1d_2 \to d_3d_4 H_5H_6$ with  $J=0$ for $\delta$.  Only ZBF contributes.  The channels with opposite sign helicities do not enter in $J=0$.}
    \end{table} 
%%%%%%   
%\newpage
%%%%%%   
\begin{table}[!t]
    \centering
    \begin{tabular}{cccccccc}
       \hline
        \hline
        \multicolumn{8}{c} {${\tilde T}^{J=0,a}(dd \to dd HH)/\eta$,  versus $E_{12}({\rm TeV})=\sqrt{s_{12}}$.  Case: $J=0$,  
        for $\eta$}  \\
         \hline
         Helicities & $E_{12}=0.5$& $E_{12}=1$ &$E_{12}=3$&$E_{12}=14/3$& $E_{12}=10$& $E_{12}=20$& $E_{12}=50$ \\
          \hline
         (- -;- -00) & &  &  &  &  &  &  \\
a=1& $<10^{-3}$ & -0.004 & -0.126 & -0.415 & -2.846 & -15.3 & -121 \\
a=2& -0.001 & -0.008 & -0.212 & -0.709 & -5.498 & -27.4 & -205 \\
a=3& $<10^{-3}$ & 0.004 & 0.095 & 0.307 & 2.148 & 12.1 & 86.7 \\
a=4& $<10^{-3}$ & -0.004 & -0.095 & -0.311 & -2.21 & -11.6 & -90.8 \\
a=5& $<10^{-3}$ & -0.004 & -0.093 & -0.306 & -2.184 & -11.0 & -83.8 \\
a=6& $<10^{-3}$ & 0.004 & 0.094 & 0.317 & 2.188 & 11.5 & 84.9 \\
a=7& $<10^{-3}$ & -0.003 & -0.081 & -0.276 & -2.023 & -11.0 & -72.6 \\
           \hline               
               (++;++00) & & & & & & &\\             
a=1& $<10^{-3}$ & $<10^{-3}$ & -0.004 & -0.013 & -0.087 & -0.467 & -3.69 \\
a=2& $<10^{-3}$ & $<10^{-3}$ & -0.006 & -0.022 & -0.167 & -0.836 & -6.231 \\
a=3& $<10^{-3}$ & $<10^{-3}$ & 0.003 & 0.009 & 0.065 & 0.371 & 2.64 \\
a=4& $<10^{-3}$ & $<10^{-3}$ & -0.003 & -0.009 & -0.067 & -0.355 & -2.766 \\
a=5& $<10^{-3}$ & $<10^{-3}$ & -0.003 & -0.009 & -0.066 & -0.338 & -2.552 \\
a=6& $<10^{-3}$ & $<10^{-3}$ & 0.003 & 0.01 & 0.067 & 0.351 & 2.585 \\
a=7& $<10^{-3}$ & $<10^{-3}$ & -0.002 & -0.008 & -0.062 & -0.337 & -2.212 \\
         \hline 
         \hline
 \end{tabular}
    \caption{\label{tab:ddJ0eta} Results for  $d_1d_2 \to d_3d_4 H_5H_6$ with  $J=0$ for $\eta$.  Only ZBF contributes.  The channels with opposite sign helicities do not enter in $J=0$.}
    \end{table} 
%%%%%%   
%\newpage
%%%%%%   
\begin{table}[!t]
    \centering
    \begin{tabular}{cccccccc}
       \hline
        \hline
        \multicolumn{8}{c} {${\tilde T}^{J=0,a}(e {\bar e} \to e {\bar e}HH)/\delta$,  versus $E_{12}({\rm TeV})=\sqrt{s_{12}}$.  Case: $J=0$,  
        for $\delta$}  \\
         \hline
         Helicities & $E_{12}=0.5$& $E_{12}=1$ &$E_{12}=3$&$E_{12}=14/3$& $E_{12}=10$& $E_{12}=20$& $E_{12}=50$ \\
          \hline
         (- -;- -00) & &  &  &  &  &  &  \\
a=1& $<10^{-3}$ & 0.004 & 0.117 & 0.413 & 2.939 & 16.2 & 116 \\
a=2& 0.001 & 0.007 & 0.207 & 0.684 & 5.19 & 25.5 & 224 \\
a=3& $<10^{-3}$ & -0.003 & -0.08 & -0.274 & -1.99 & -10.3 & -96.3 \\
a=4& $<10^{-3}$ & 0.003 & 0.079 & 0.28 & 1.913 & 8.487 & 71.4 \\
a=5& $<10^{-3}$ & 0.003 & 0.08 & 0.272 & 1.963 & 9.21 & 91.9 \\
a=6& $<10^{-3}$ & -0.003 & -0.08 & -0.279 & -1.914 & -10.2 & -80.5 \\
a=7& $<10^{-3}$ & 0.002 & 0.071 & 0.243 & 1.745 & 10.2 & 67.7 \\
         \hline 
         \hline
 \end{tabular}
    \caption{\label{tab:eeJ0delta} Results for  $e_1 {\bar e}_2 \to e_3 {\bar e}_4 H_5H_6$ with $J=0$ for $\delta$. Same results for (++;++00) as explained in the text.  Only ZBF contributes.  The channels with opposite sign helicities do not enter in $J=0$.}
    \end{table} 
%%%%%%   
%\newpage
%%%%%%   
\begin{table}[!t]
    \centering
    \begin{tabular}{cccccccc}
       \hline
        \hline
        \multicolumn{8}{c} {${\tilde T}^{J=0,a}(e {\bar e} \to e {\bar e}HH)/\eta$,  versus $E_{12}({\rm TeV})=\sqrt{s_{12}}$.  Case: $J=0$,  
        for $\eta$}  \\
         \hline
         Helicities & $E_{12}=0.5$& $E_{12}=1$ &$E_{12}=3$&$E_{12}=14/3$& $E_{12}=10$& $E_{12}=20$& $E_{12}=50$ \\
          \hline
         (- -;- -00) & &  &  &  &  &  &  \\
a=1& $<10^{-3}$ & 0.002 & 0.043 & 0.144 & 1.007 & 5.689 & 45.0 \\
a=2& $<10^{-3}$ & 0.003 & 0.072 & 0.239 & 1.952 & 9.316 & 64.5 \\
a=3& $<10^{-3}$ & -0.001 & -0.032 & -0.107 & -0.724 & -3.993 & -31.9 \\
a=4& $<10^{-3}$ & 0.001 & 0.032 & 0.104 & 0.772 & 3.866 & 31.6 \\
a=5& $<10^{-3}$ & 0.001 & 0.031 & 0.101 & 0.728 & 3.645 & 25.2 \\
a=6& $<10^{-3}$ & -0.001 & -0.033 & -0.107 & -0.756 & -4.08 & -32.9 \\
a=7& $<10^{-3}$ & 0.001 & 0.027 & 0.092 & 0.694 & 3.905 & 21.9 \\
         \hline 
         \hline
 \end{tabular}
    \caption{\label{tab:eeJ0eta} Results for  $e_1 {\bar e}_2 \to e_3 {\bar e}_4 H_5H_6$ with  $J=0$ for $\eta$.  Same results for (++;++00) as explained in the text.  Only ZBF contributes.  The channels with opposite sign helicities do not enter in $J=0$.}
    \end{table} 
%%%%%% 

%%%%%%%%%%%%%%%
\textcolor{white}{NECESITAMOS ESPACIO PARA QUE LAS TABLAS NO CORTEN LAS REFERENCIAS}
\vspace{200cm}

\textcolor{white}{NECESITAMOS ESPACIO PARA QUE LAS TABLAS NO CORTEN LAS REFERENCIAS}
\vspace{200cm}

\textcolor{white}{NECESITAMOS ESPACIO PARA QUE LAS TABLAS NO CORTEN LAS REFERENCIAS}
\vspace{200cm}

\textcolor{white}{NECESITAMOS ESPACIO PARA QUE LAS TABLAS NO CORTEN LAS REFERENCIAS}
\vspace{200cm}

\textcolor{white}{NECESITAMOS ESPACIO PARA QUE LAS TABLAS NO CORTEN LAS REFERENCIAS}
\vspace{200cm}

\textcolor{white}{NECESITAMOS ESPACIO PARA QUE LAS TABLAS NO CORTEN LAS REFERENCIAS}
\vspace{200cm}

\textcolor{white}{NECESITAMOS ESPACIO PARA QUE LAS TABLAS NO CORTEN LAS REFERENCIAS}

\bibliography{DDHM}

\end{document}